%% file: drafts_2025_10_24_submitted_WP.tex
\documentclass[a4paper,12pt]{article}
\usepackage{natbib}
\usepackage[margin=2.5cm]{geometry}
\usepackage{setspace}
\usepackage{floatrow}
\usepackage{url}
\usepackage[T1]{fontenc}
\usepackage{microtype}
\usepackage{lmodern} 
\usepackage{amsmath,amsfonts,amssymb}
\usepackage{bm}
\usepackage{graphicx}
\usepackage{ltxtable}
\usepackage{threeparttable}
\usepackage{booktabs}
\usepackage{multirow}
\usepackage{subcaption}
\usepackage{adjustbox}
\usepackage[official]{eurosym}
\usepackage[dvipsnames]{xcolor}
\usepackage{placeins}
\usepackage{tikz}
\usepackage{appendix}
\usepackage{hyperref}
\usetikzlibrary{arrows.meta,positioning,calc}

\definecolor{ccombined}{RGB}{103,0,31}
\definecolor{chigh}{RGB}{178,24,43}
\definecolor{clow}{RGB}{214,96,77}
\definecolor{bblue}{RGB}{55,126,184}

\tikzset{%
  >={Latex[width=2mm,length=2mm]},
            base/.style = {rectangle, draw=white,
                           minimum width=4cm, minimum height=1cm,
                            font=\sffamily},
  activityStarts/.style = {base, text centered, fill=gray!70},
       startstop/.style = {base, text centered},
    activityRuns/.style = {base, text centered},
         process/.style = {base, minimum width=2.5cm, align=left, fill=, fill=gray!30,
                           font=\sffamily
                           },
}

\newcommand{\secstageupdatepred}{This figure shows the belief-update effects of the information treatment. We compute the aggregate causal effect as a linear combination of second-stage coefficients from regression~\eqref{eq:second_stage} on the posterior belief weighted by the average belief shift in the treatment group informing about all four occupations (see equation~\eqref{eq:aggregation}). Means and standard deviations of the outcomes for the untreated control group are reported in the labels on the vertical axis. Error bars represent 95\% confidence intervals, with standard errors calculated via the delta method to account for the joint estimation.~}

\newcommand{\secstageupdatepredsub}{This figure shows subgroup-specific causal effects of the information treatment. We compute the aggregate causal effect as a linear combination of second-stage coefficients from regression~\eqref{eq:second_stage} on the posterior belief weighted by the subgroup-specific average belief shift in the treatment group for all four occupations (see equation~\eqref{eq:aggregation}). Error bars represent 95\% confidence intervals, with standard errors calculated via the delta method to account for the joint estimation.~}

\newcommand{\redformIV}{The second-stage regression is given in equation~\eqref{eq:second_stage}. The reduced form regression is $\text{Y}_{i} = \pi_{0} + \sum_{o}\pi_{1o} \cdot \text{prior}_{io} + \sum_{j} \pi_{2j} \cdot D_{ij} + \sum_{o}\sum_{j} \pi_{3oj} \cdot (D_{ij} \times \text{prior}_{io}) + \nu_{i}.$ Effects (captured by coefficient $\pi_{1o}$, see Appendix~\ref{appendix:theory}) are evaluated at the sample-average belief updating across occupations (analogous to equation~\eqref{eq:aggregation}); points are estimates and bars are 95\% confidence intervals. Units are percentage points.~} 

\hypersetup{
    pdfmenubar=true,        
    pdffitwindow=false,     
    pdfstartview={FitH},    
  pdftitle={Beliefs about Bots: How Employers Plan for AI in White-Collar Work},    
  pdfauthor={Eduard Brüll, Samuel Mäurer, Davud Rostam-Afschar},     
    pdfsubject={Bayesian Updating Survey Experiment},   
    pdfkeywords={Automation, Labor Demand, White Collar Jobs, Human Capital Investment, Technological Displacement}, 
    pdfnewwindow=true,      
    colorlinks=true,       
    linkcolor=black,          
    citecolor=black,        
    filecolor=black,      
    urlcolor=black,           
}

\newcommand{\customfigure}[6]{%
    \begin{figure}[#6]
        \caption{#1}
        \label{#2}
        \centering
        #3 
        \begin{threeparttable}
            \begin{tablenotes}
                \footnotesize 
                \item \emph{Note:} #4
                \item \emph{Source:} #5
            \end{tablenotes}
        \end{threeparttable}
    \end{figure}
}

\def\sym#1{\ifmmode^{#1}\else\(^{#1}\)\fi}
\begin{document}
\pagebreak

\title{Beliefs about Bots:\\ How Employers Plan for AI in White-Collar Work\thanks{We thank Melanie Arntz, Efi Adamopoulou (discussant), Dietmar Fehr, Gabriel Leite Mariante (discussant), Manuel Menkhoff, Henning Hermes, Cäcilia vom Baur, and the German Federal Chamber of Tax Advisors for insightful comments and discussions, as well as participants at various seminars and conferences for helpful comments. The study was pre-registered in the AEA RCT registry (RCT ID: AEARCTR-0014788). This experiment was reviewed by the Ethics Committee of the University of Mannheim (EC-8/2024). We declare that we have no interests, financial or otherwise, that relate to the research described in this paper. We are grateful to the Deutsche Forschungsgemeinschaft (DFG, German Research Foundation) for financial support through CRC \textit{TRR 266 Accounting for Transparency} (Project-ID 403041268).} 
}

\author{
\normalsize\begin{tabular}{c}
Eduard Brüll\footnote{ZEW Leibniz Centre for European Economic Research, 68161 Mannheim, Germany (e-mail: eduard.bruell@zew.de),},
Samuel Mäurer\footnote{University of Mannheim, 68131 Mannheim, Germany, Germany (e-mail: smaeurer@mail.uni-mannheim.de),},
Davud Rostam-Afschar\footnote{University of Mannheim, 68131 Mannheim, Germany; GLO; IZA; NeST (e-mail: rostam-afschar@uni-mannheim.de),}\\
{\date{\begin{small}\today\end{small}\\
}
}
\end{tabular}
}
\maketitle

\begin{abstract}\noindent
We provide experimental evidence on how employers adjust expectations to automation risk in high-skill, white-collar work. Using a randomized information intervention among tax advisors in Germany, we show that firms systematically underestimate automatability. Information provision raises risk perceptions, especially for routine-intensive roles. Yet, it leaves short-run hiring plans unchanged. Instead, updated beliefs increase productivity and financial expectations with minor wage adjustments, implying within-firm inequality like limited rent-sharing. Employers also anticipate new tasks in legal tech, compliance, and AI interaction, and report higher training and adoption intentions.



\vfill
\end{abstract}

\pagebreak


\section{Introduction}
\label{sec: Introduction}

Each new wave of digital innovation drives growth while transforming the world of work. Emerging technologies such as generative artificial intelligence (GenAI) are being adopted at an accelerating pace \citep[e.g.,][]{HumlumVestergaard2025, Menkhoff2025}. By automating ever more complex cognitive tasks, these technologies raise pressing questions about their effects on employment, wage inequality, and the demand for skills.

Historically, automation has primarily affected blue-collar jobs in manufacturing and manual labor-intensive sectors, where robotics and mechanization replaced routine tasks \citep[e.g.,][]{AcemogluRestrepo2020}. However, the latest wave of AI-driven automation differs markedly: highly educated, high-income white-collar occupations are now among the most exposed \citep[e.g.,][]{Eloundou2023}. Recent studies suggest that AI assistance can significantly enhance worker productivity in professional settings. For instance, \citet{NoyZhang2023} show that AI-assisted writers complete tasks faster and produce higher-quality output, while \citet{BrynjolfssonLiRaymond2023} document a 14 \% productivity increase among customer service agents using generative AI, with the largest gains among less-experienced workers. At the same time, \citet{Felten2023} highlight that occupations relying on communication, analysis, and creative abilities---once considered resistant to automation---are now highly exposed to AI-driven disruption.

Despite the growing literature on automation's effects, most existing studies focus on workers rather than employers. Little is known about how firms perceive and respond to automation risks, although evidence on this is crucial because employer expectations directly shape labor demand and technology adoption. To investigate this, we study tax consulting---a white-collar, high-skill setting where generative AI is particularly relevant not only because much of the work is structured and repetitive, but also because it relies heavily on language-based legal tasks. A key advantage of this setting is the availability of granular occupational categories at the five-digit level, which enables us to distinguish between tax clerks, certified tax assistants, tax advisors, and auditors. This granularity reduces bias from occupation-task mismatches and allows us to capture heterogeneity in automation exposure across different expertise levels, a factor relevant for the direction of the effect automation has on employment and wages \citep{Autor2025}. In this setting, we exogenously shift employers beliefs about automation rates using a randomized information intervention. Our survey, which targeted the entire population of tax advisors listed in Germany's official register, allows us to analyze how belief updates about automation potential influence employment and wage setting plans. Additionally, we examine firm-level outcomes such as revenue, profit, and cost expectations to assess whether employers view automation as an opportunity for efficiency gains or a disruptive force of workforce reductions.

This study provides new insights into how automation reshapes firm decision-making and labor market dynamics. In particular, we examine how automation transforms the organization of tasks and how firms respond by adjusting their employee training strategies. We also provide evidence on other outcomes such as firms' attitudes towards automation and AI adoption.

Our results reveal several striking patterns. First, employers systematically underestimate automation potential for their profession. Initial beliefs about the share of tasks that could be automated within the next decade are significantly lower than expert assessments. After receiving objective information on automation potential from the research institute of the German Federal Employment Agency (IAB), respondents revise their beliefs upward, particularly for lower-skilled roles such as tax clerks and certified tax assistants. Belief updating is weaker for higher-skilled occupations like auditors and tax advisors, suggesting that firms perceive greater reluctance towards automation at the top of the professional hierarchy.

Second, despite substantially higher revenue and profit growth expectations, wage growth expectations remain negligibly small, suggesting that firms intend to retain productivity gains rather than pass them on to employees. We argue that these extra profits come from productivity gains by tax clerks who serve more clients, because prices are largely fixed in German tax advisory. At the same time, firms expect cost savings on average but these are not significant and heterogeneous, perhaps because anticipated investments in new technologies or upskilling initiatives could even increase costs.

We find no evidence that firms immediately revise their hiring or firing plans, indicating that automation-induced workforce displacement are not a primary concern in the near term. Firms which regularly use AI, even reduce dismissal plans but also hiring plans. The reported increase in profit growth is driven by higher expected revenues per hour in lower-skill occupations. Together with modest wage growth expectations this may help explain why these jobs persist, even though they could, in principle, be fully automated.

Firms exposed to new information on automation not only reassess existing job roles but also anticipate new tasks emerging as a consequence of AI adoption. In particular, employers expect increased demand for legal tech expertise, compliance monitoring, and AI interaction skills such as prompt engineering. This aligns with a growing recognition that generative AI does not merely replace existing jobs but also reshapes job content and skill requirements. Arntz et al.\ \citeyearpar{ArntzGregoryZierahn2016,ArntzGregoryZierahn2017}, for instance, highlight that automation affects tasks within jobs rather than eliminating whole occupations.

Fourth, employers exposed to automation information are significantly more likely to report plans for further training and upskilling investments for their staff. Our results show that belief updating about automation not only shifts expectations about task content but also increases the likelihood that firms intend to invest in specialized digital skills, legal tech, and AI-related training. At the same time, treated firms report a greater openness toward adopting AI solutions and perceive automation more as an opportunity than as a threat. 

The results we find on perceptions, expectations, and intentions translate into actions. Firms that plan to adopt AI tools also actively search for AI solutions. \cite{Menkhoff2025} shows that this is a general pattern and that also for other services, manufacturing, and retail firms stated adoption plans eventually materialize. Consistent with the null effects on hiring and firing shares, we can also show that there is no change in actual vacancy postings of treated respondents, based on record-linked data to Germany’s largest vacancy aggregator.

Our study contributes to three strands of research. First, we extend the literature on automation and labor markets. Much of this literature focuses on the effects of automation in manufacturing and manual labor-intensive sectors. For example, \citet{AcemogluRestrepo2020} find that increased robot adoption in US. manufacturing is associated with significant reductions in employment and wages, with localized displacement effects that are not fully offset by gains in other sectors. \cite{AghionEtAl2022} survey the recent literature and emphasize that automation's effects on labor demand are heterogeneous and depend on task composition and firm context. For instance, \citet{Bessen2020} demonstrate that firms adopting automation technologies often save labor while maintaining wage growth. More recently, attention has shifted toward AI and white-collar work. Evidence suggests that AI adoption may exacerbate wage inequality by disproportionately benefiting high-wage workers \citep{Bonfiglioli2023}, while also reshaping task structures: \citet{GathmannGrimmWinker2024} show that AI reduces demand for abstract tasks but raises demand for high-level routine ones. Closest to our setting, \citet{HumlumVestergaard2025} provide large-scale evidence from Denmark showing that AI chatbot adoption leads to occupational switching and task restructuring, but has minimal short-run effects on wages or hours, with precisely estimated null effects of less than 2\% two years after ChatGPT's release.

Similarly to \cite{HumlumVestergaard2025}, we find negligibly small wage growth effects of automation, underscoring that fears of short-run (3 years ahead) white-collar displacement may be overstated. Firms do not meaningfully adjust wage expectations and also not cost expectations even though they anticipate higher revenue and profits. The setting of tax advisory in Germany allows us to narrow down the mechanism behind the increased revenues and profit expectations. Rather than to charge more per service, firms must serve more clients in the tax advisory industry, because prices are fixed by a market-wide fee schedule. Since the labor market for lower skilled jobs in tax advisory in Germany has been tight for years, substantial wage growth could have been expected. The absence of substantial wage increases suggests that efficiency gains from generative AI are already internalized in firm expectations but are not being shared with workers.

Second, our study takes a novel employer-centered perspective, which allows us to detect potential adjustments in hiring plans, wage strategies, and skill investment decisions. This allows a better understanding of the decision making process. This is particularly valuable because the existing literature, while rich in documenting the impacts of automation at the worker and firm levels, often examines outcomes only after automation technologies have been implemented. For example, \citet{Bessen2023} examine worker-level outcomes following firm-level automation expenditures, finding significant impacts on worker displacement and cumulative wage losses. Similarly, \citet{AcemogluAutorHazellRestrepo2022} analyze the adoption of AI using vacancy-level data, demonstrating shifts in hiring patterns and skill requirements at AI-exposed establishments between 2010 and 2018. However, their analysis does not extend to the most recent wave of generative AI adoption, leaving open questions about how firms anticipate and adapt to these transformative technologies. Although, these approaches are invaluable for understanding post-adoption consequences, they do not shed light on how firms plan for or adapt to automation before investments are made. Our study instead captures firms' anticipatory responses, showing that automation beliefs influence business expectations and investment strategies.

Our study methodologically builds upon a growing literature employing information interventions to examine how expert-assessments or other factual information signals can correct misperceptions and influence economic preferences and behaviors \citep[e.g.,][]{Fehr2024, HaalandRothWohlfart2023,Coibion2018,WiswallZafar2015,BursztynEdererFermanYuchtman2014}. A growing number of studies in labor economics now employ survey experiments to analyze how beliefs shape expectations and choices. For example, \citet{JaegerEtAl2021} show that workers anchor their expectations about outside options on current wages and adjust search and bargaining plans when informed about the actual wage distribution. \citet{CortesPanPilossophReubenZafar2023} demonstrate experimentally that gender differences in job search behavior contribute to the earnings gap.

Within this framework, recent work has turned to automation specifically: \cite{Agarwal2023} find that professional radiologists underweight AI predictions and treat them as independent from their own information, \citet{Jeffrey2021} shows that the framing of automation influences policy preferences, \citet{Lergetporer2023} demonstrate that workers underestimate the automatability of their occupations and update their training intentions when informed, and \citet{GolinRauh2022} highlight that exposure to information about automation potential increases support for redistribution but has limited effects on retraining intentions. Relative to these studies, we shift the perspective from workers to employers, examining how belief updating about automation potential influences firm expectations, wage strategies, and training investments in a white-collar industry.

The remainder of this paper is organized as follows. Section \ref{sec:data} describes the survey, experimental setup, and estimation strategy. Section \ref{sec:results} presents the results, while Section \ref{sec:conclusion} concludes. 


\section{Survey and Experimental Setup}
\label{sec:data}

\subsection{The GBP Tax Advisor Survey}
Our analysis draws on a specialized survey of tax advisors  conducted between November 2024 and April 2025. As part of the German Business Panel (GBP), this survey targeted all professionals listed in Germany’s official register of licensed tax advisors (\textit{Steuerberater}), using over 80,000 email addresses. Since the official register is both mandatory and exhaustive, it covers all individuals and firms authorized to practice as tax advisors in Germany.

In Germany, tax advisors are classified as \textit{Freiberufler} (liberal professionals), a designation that differentiates them from traditional firms. Freiberufler, include tax advisors, lawyers, and doctors, and constitute a highly organized and economically significant part of the service sector: as of January 2023, about 1.47 million self-employed professionals in these fields accounted for roughly 3\% of the workforce, employing around 4.2 million people and generating over 10\% of Germany’s GDP (German Federal Association of Liberal Professions).\footnote{Professional chambers oversee admission, standards, and representation for liberal professions.} Tax advisors operate independently under distinct occupation-specific legal frameworks (e.g., Steuerberatungsgesetz) that regulate market entry and conduct. They frequently employ significant numbers of white-collar workers, playing a vital role in the labor market. However, standard firm-level datasets, such as those used in business or employer-employee panel studies, typically exclude \textit{Freiberufler}, creating a significant data gap. Using the mandatory register for fielding allows us to bridge this gap, offering direct insights into this unique professional group and their responses to automation in a new custom survey.

\subsection{Experimental Setup}

The key part of our survey is a randomized information intervention designed to examine how the firms of the tax advisory industry respond to updated information on the automation potential of their workforce. The experiment follows the sequence, visually represented in Figure \ref{fig:Experimental_Design}.

Before the intervention, we collect respondents' employment levels and their prior beliefs about the automatability of four tax-related occupations: tax clerks, certified tax assistants, tax advisors, and auditors. Participants estimate the percentage of core activities within each occupation that they believe can be automated as of 2024.

Following this, respondents are randomly assigned to one of three treatment groups or a control group, each with equal probability. The treatment groups receive information about the automatability of each occupation, based on occupation-level estimates from the IAB Job-Futuromat.

\paragraph{The IAB Job Futuromat.} The IAB Job-Futuromat is a tool developed by the Institute for Employment Research (IAB), a research division of Germany's Federal Employment Agency. It provides a systematic assessment of how digital technologies impact various occupations by evaluating the degree to which specific tasks within those roles can be automated. It covers approximately 4,000 occupations and is based on expert-driven task analyses, making it one of the most detailed and policy-relevant resources on labor automation.

The automatability scores in the Job-Futuromat are built on BERUFENET data, an expert database maintained by the German Federal Employment Agency, which documents occupational tasks, required skills, and competencies for career guidance and job placement. The methodology behind BERUFENET, as described by \citet{Dengler2018}, follows a task-based approach similar to O*NET in the U.S., systematically mapping occupations to their core tasks and assessing their substitutability by automation.\footnote{The automatability of an occupation is obtained by first dividing the number of automatable core tasks of an occupation by all its core tasks and multiplying the result by 100. This is known as the substitution potential. Only the core tasks of this occupation are taken into account in the calculation. It is assumed that each core task is performed with the same frequency and therefore has the same influence on the calculation of automatability of an occupation $o$.
$$
\text{automatability}_o = \frac{\text{Number of core tasks that can be automated}_o}{\text{Number of all core tasks}_o} \times 100.$$
The automatability of a task indicates whether this task could be performed fully automatically by a computer or a computer-controlled machine with the current technology. We accessed the Job-Futuromat in November 2024.} This expert-driven approach offers a robust alternative to survey-based task measurements, ensuring that occupational analyses remain consistent and reliable over time.

\input{gph/experimental_design}

Previous studies have drawn on the Job-Futuromat to examine labor market dynamics. Research using Job-Futuromat data has revealed that occupations with higher substitutability potential tend to experience lower employment growth on average \citep[e.g.][]{Dengler2018}.\footnote{However, some highly automatable professions have still seen employment growth, indicating that factors beyond technological feasibility, such as economic demand, regulatory environments, and skill shortages, play a crucial role in the adoption of automation.} The Job-Futuromat has also been used in experimental settings: \citet{Lergetporer2023} study how workers adjust their training and upskilling demand when they learn about their occupation’s automatability, and \citet{Resnjanskij2023JPE} find that disadvantaged adolescents guided by mentors tend to aspire to occupations with lower automation risk. 

\begin{table}[h]
    \centering
    \caption{Automation Potential of Tax Occupations According to the Job-Futuromat.}
    \label{tab:job_futuromat}
    \begin{tabular}{l c}
        \toprule
        \textbf{Occupation} & \textbf{Automation Potential} \\
        \midrule
        Tax Clerk (Steuerfachangestellter) & 100\% \\
        Certified Tax Assistant (Steuerfachwirt) & 80\% \\
        Tax Advisor (Steuerberater) & 62\% \\
        Auditor (Wirtschaftsprüfer) & 57\% \\
        \bottomrule    
    \end{tabular}    
\footnotesize\\[2ex]\textit{Note}: See~Figure~\ref{fig:expert_signal} for screenshots of the IAB Job-Futuromat.
\end{table}

The automation potential estimates for the four tax-related occupations considered in this study are strikingly high (see Table \ref{tab:job_futuromat}). According to the Job-Futuromat, tax clerks face full automation potential (100\%), while certified tax assistants also exhibit a high substitutability potential (80\%). Even among higher-skilled roles, tax advisors (62\%), and auditors (57\%) show considerable exposure to automation. The full substitutability of tax clerks is plausible given that their core activities (bookkeeping, payroll, VAT filing, and record management) are highly standardized, rule-based, and already directly supported by existing accounting and tax software prior to modern generative AI.\footnote{Also note that the Job-Futuromat measures technological feasibility, not realized automation, and thus should not be read as a deterministic prediction of job loss. High substitution potentials reflect the substitutability of occupational tasks, but adoption of automation technology in practice depends on regulatory frameworks, organizational change, and demand-side conditions. Within this framework, however, the classification of tax clerks at 100\% is methodologically coherent, since virtually all of their BERUFENET-listed core tasks have direct software analogs.}

\paragraph{The Information Treatment.} To examine how individuals respond to expert-provided automation assessment, we implement an information treatment that randomly assigns respondents into one of four groups including a passive control group. Respondents assigned to one of the three treatment arms receive an animated visualization comparing their own estimates of the automation potential in their occupation to expert assessments from the Institute for Employment Research.

\customfigure
    {Example Screenshot of the Information Treatment} 
    {fig:screenshot}              
    {\includegraphics[width=\textwidth]{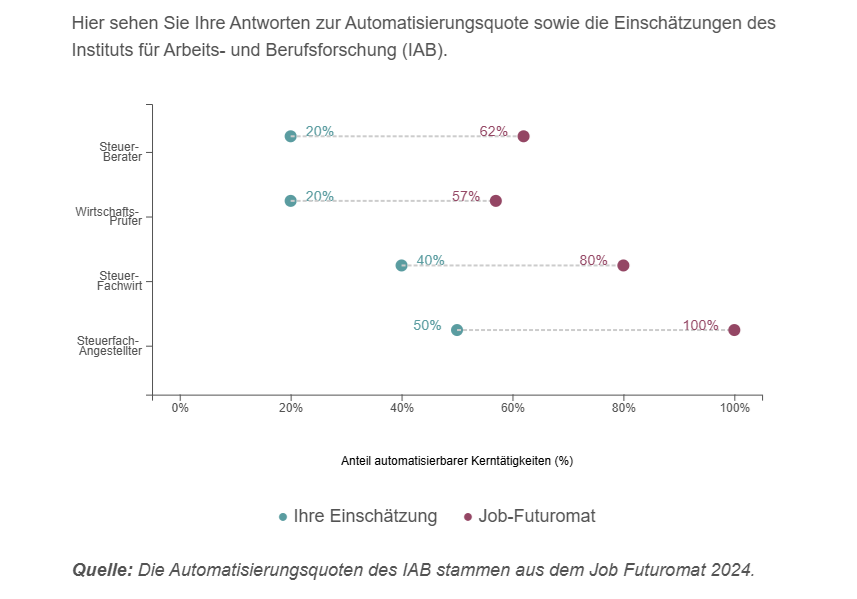}} 
    {This figure presents a screenshot of the combined treatment animation.}     
    {German Business Panel Screenshot.}            
    {H} 

The treatment heading states (see Figure~\ref{fig:screenshot} for an example screenshot for the ``combined treatment'' arm): \textit{"Here you can see your answers on the automation rate, along with the assessments of the Institute for Employment Research (IAB)"}.\footnote{The visualization consists of a dumbbell plot (see Figure~\ref{fig:screenshot}) where each occupation is represented by a horizontal line connecting two color-coded points. The animation unfolds smoothly, starting with respondents' own beliefs and then progressively revealing the objective IAB values, visually emphasizing the gap between the two.\footnote{The visualization is implemented using \texttt{d3.js}, a JavaScript library for producing dynamic, interactive data visualizations in web browsers \citep{Bostock2011}.} The animation design follows best practices in visual perception research, using motion to guide attention while avoiding excessive cognitive load. For the control group only a static plot is displayed, showing the own beliefs of the respondents graphically. 
}

\begin{enumerate}
    \item \textbf{Control Group:}  Only respondents' own beliefs are displayed.
    \item \textbf{Lower-Skilled Treatment:} Respondents' beliefs are compared with expert assessments for tax clerks and certified tax assistants.
    \item \textbf{Higher-Skilled Treatment:} Respondents' beliefs are compared with expert assessments for tax advisors and auditors.
    \item \textbf{Combined Treatment:} A comprehensive visualization comparing prior beliefs and expert assessments for all four listed occupations.
\end{enumerate}

After the experiment, we ask whether respondents want to update their beliefs to elicit a posterior for all four occupations.\footnote{The questions were asked on a new screen from which the respondents could not go back to their prior beliefs. All respondents were asked to enter a new estimate regardless of whether they changed their assessment. This approach allows to distinguish between those who consciously maintain their belief and those who revise it. It prevents anchoring on prior responses or mechanical copying.} We then proceed with several questions on hiring and firing, revenue, profit, cost, and wage expectations as well as perceived automation potential of tasks and emergence of new tasks due to automation. 

\subsection{Data Quality and Plausibility Checks}\label{sec:data_quality}

Ensuring the reliability and representativeness of our survey data is crucial for deriving meaningful insights about tax advisory firms.

\customfigure
    {Survey and Register Data} 
    {fig:appointment_year}              
    {\includegraphics[width=\textwidth]{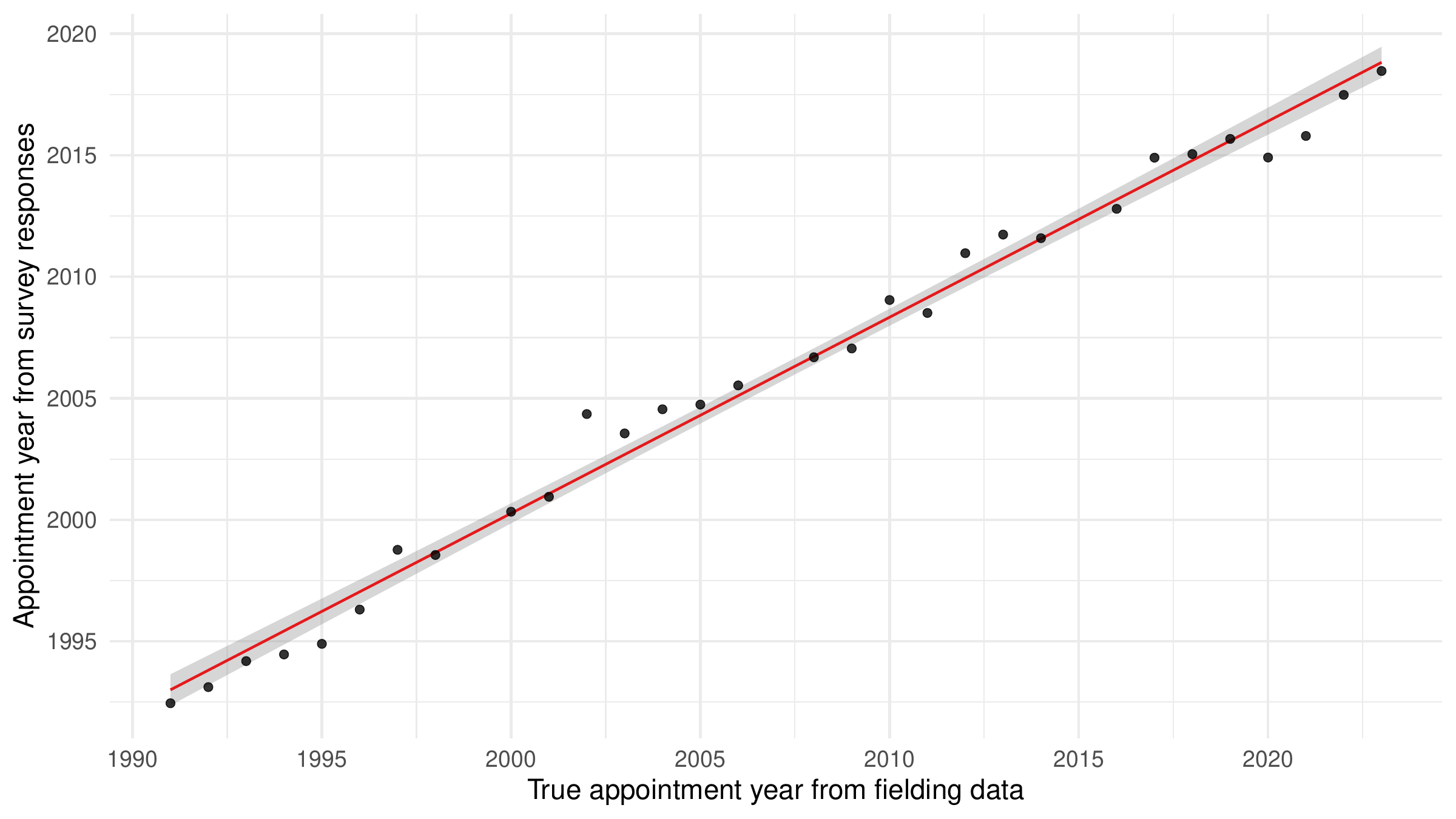}} 
    {This figure presents a binned scatter plot comparing self-reported appointment years from the survey with official register data. Each point represents the average self-reported appointment year within 30 equally sized bins of the true appointment year from the register data. A linear fit in red, demonstrates a strong positive correlation, indicating high consistency between self-reported and official records. Deviations are most pronounced among respondents with early appointment years, likely reflecting inactive professionals who retain their designation.}     
    {German Business Panel Tax Advisor Survey 2025 and German Registers of Tax Advisors.}            
    {H} 

\paragraph{Consistency with Register Data.} A critical test of our dataset's accuracy is the matching between self-reported and official register data. To this end, we compare survey answers on the appointment year as tax advisor to the official register entry for each respondent. We check this in a binned scatter-plot in Figure \ref{fig:appointment_year}, which reveals a strong positive correlation, reflecting the reliability of responses. The red trend line and confidence interval suggest that, for the majority of respondents, self-reported data closely matches the fielding data from the register. Extreme deviations are rare and especially present for the oldest respondents in our sample (i.e. the early appointment years), who are unlikely to still be economically active as tax advisors. To prevent inconsistencies, we restricted the data to cases where the reported appointment years deviated by no more than five years from the register. 

\customfigure
    {Representativity of the Survey} 
    {fig:represent}              
    {\includegraphics[width=0.9\textwidth]{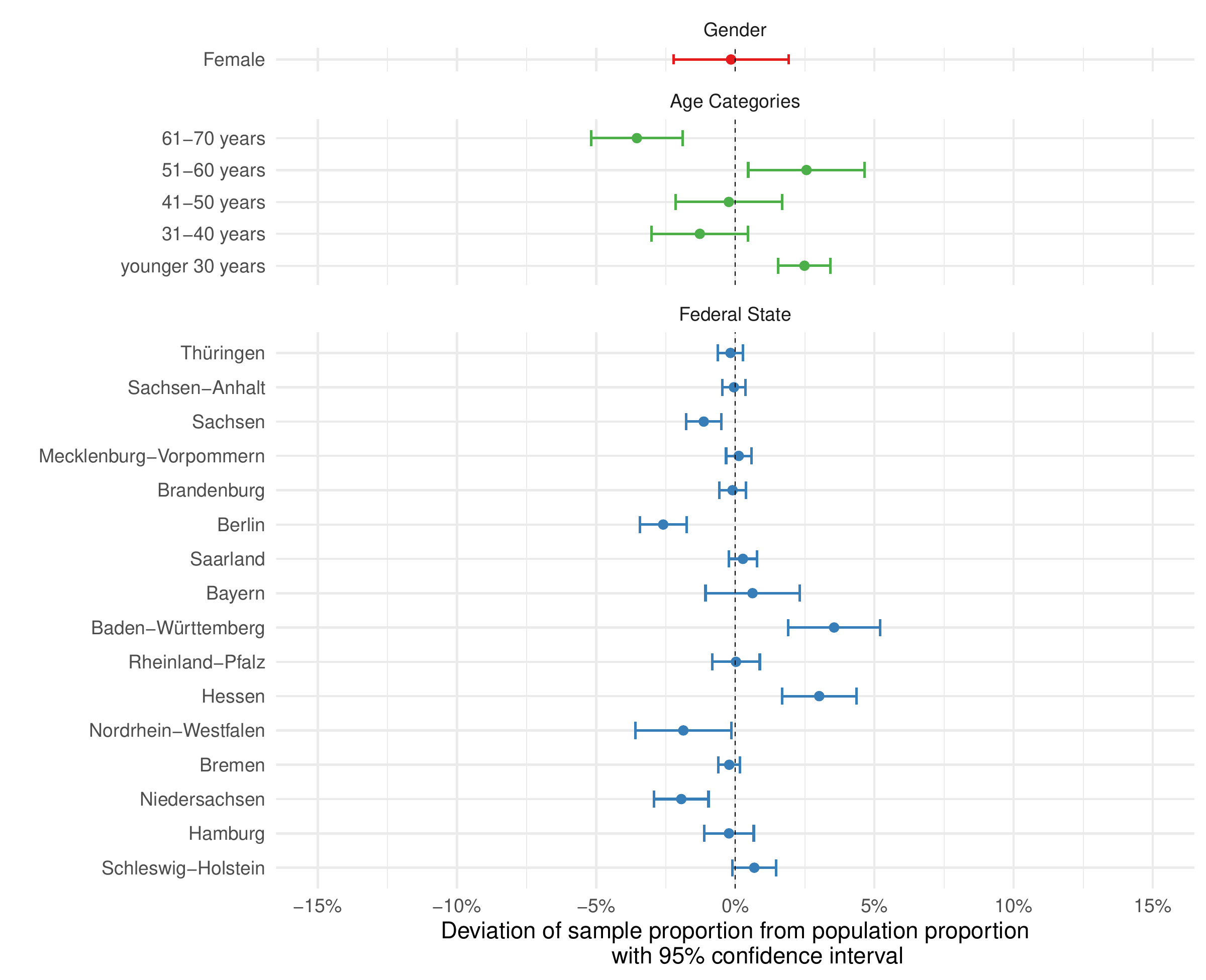}} 
    {This figure compares key demographic characteristics of survey respondents with population benchmarks from the official statistics of the Chamber of Tax Advisors, with whiskers representing 95\% confidence intervals. The plotted coefficients represent differences in respondent proportions relative to the population across categories such as gender, age groups, and federal states.}     
    {German Business Panel Tax Advisor Survey 2025 and Official Statistics of the Chamber of Tax Advisors.}    
    {H} 

\paragraph{Sample Representativeness Across Key Demographics.} We compared sample proportions with population benchmarks from the official statistics of the chamber of tax advisors. Figure~\ref{fig:represent} illustrates deviations across gender, age categories, and federal states. The results demonstrate that the survey largely captures the target population, with most deviations falling within acceptable ranges. These differences primarily stem from the nature of the official register and our focus on economically active tax firms. For instance, since the titles of tax advisor and auditor are lifelong, many older professionals retain their designation despite no longer being active. 

We filtered and cleaned the survey data to retain only active, independent tax advisory firms by excluding respondents outside the target occupation (tax or law professors, retirees), firms with implausible or extreme revenue or employment figures, and a few cases with clearly erroneous responses (see Appendix~\ref{appendix:filtering}). After this, the final sample consists of 1,736 observations.

We have also obtained balance sheet data for firms who completed the survey, if available, using data from the Bureau van Dijk Orbis database. Although the Orbis data are only observed irregularly and sometimes up to six years prior to the survey, they show high agreement for the measures of the number of employees and of log revenue as reported in Orbis and in the survey (see Figure~\ref{fig:compare_orbis_survey} in the supplementary appendix). This further shows that stated and actual values correspond very well.

\subsection{Descriptive Statistics and Covariate Balance}

\paragraph{Respondent Characteristics.}
The average respondent is around 51 years old, with a strong representation of self-employed professionals (75\%). Most respondents are tax advisors (98\%), with a minority working in related auditing roles. Female representation stands at 32\%, reflecting broader industry demographics.

\paragraph{Revenue and Employment Distribution.} Figure \ref{fig:firm_distribution} visualizes the distribution of firm revenue (left) and total employees (right) in more detail for our main target group of smaller tax firms with less than 150 employees and revenues below 10 million Euros. The majority of firms have revenue below 2.5 million euros and employ fewer than 50 individuals, though some large firms, primarily large multinational auditing firms, contribute to a long right tail in both distributions. The median employment is 9 employees. Revenue statistics exhibit a similar pattern, with a median revenue of approximately 1 million euros but a mean exceeding 3 million euros.

\paragraph{Self-reported AI usage.} We also elicited self-reported AI usage. Figure \ref{fig:aiuse} in the Appendix shows the current use of generative AI across firms of different sizes. Among the smallest firms (0-3 employees), more than two thirds report never using generative AI, and only about 16\% use it often or always. In contrast, larger firms (with more than 11 employees) show markedly higher adoption rates, with only 39\% never using AI and nearly 24\% reporting frequent use (often or always).

\paragraph{Covariate Balance.} The covariate balance plot in Figure \ref{fig:balance} verifies the success of the randomization process. Mean differences between treatment arms and the control group remain small across all key firm characteristics, with confidence intervals largely overlapping zero. This ensures that any treatment effects observed in later analyses are not driven by pre-existing differences in firm size, revenue, or regional distribution. The balance in employment and revenue distributions further underscores the robustness of the experimental design.


\section{Results}

\label{sec:results}
Before analyzing how information influences tax advisors' expectations and decision-making, we first examine their prior beliefs about automation potential across different occupations.

Figure~\ref{fig:prior_dist} displays the distribution of participants’ prior beliefs about the share of automatable tasks in 2024 for tax clerks, certified tax assistants, tax advisors, and auditors. The vertical dashed lines indicate expert assessments from the Institute for Employment Research (IAB), providing a benchmark against which subjective expectations can be compared.

The ranking of occupations in participants’ priors closely mirrors that of the expert assessments: tax clerks are expected to face the highest degree of automation potential, followed by certified tax assistants, while auditors and tax advisors are perceived as less exposed. Across all occupations, however, the majority of respondents underestimate automation potential relative to the expert assessment, with only a small minority assigning higher probabilities.

\customfigure
    {Distribution of Prior Beliefs} 
    {fig:prior_dist}              
    {\includegraphics[width=\textwidth]{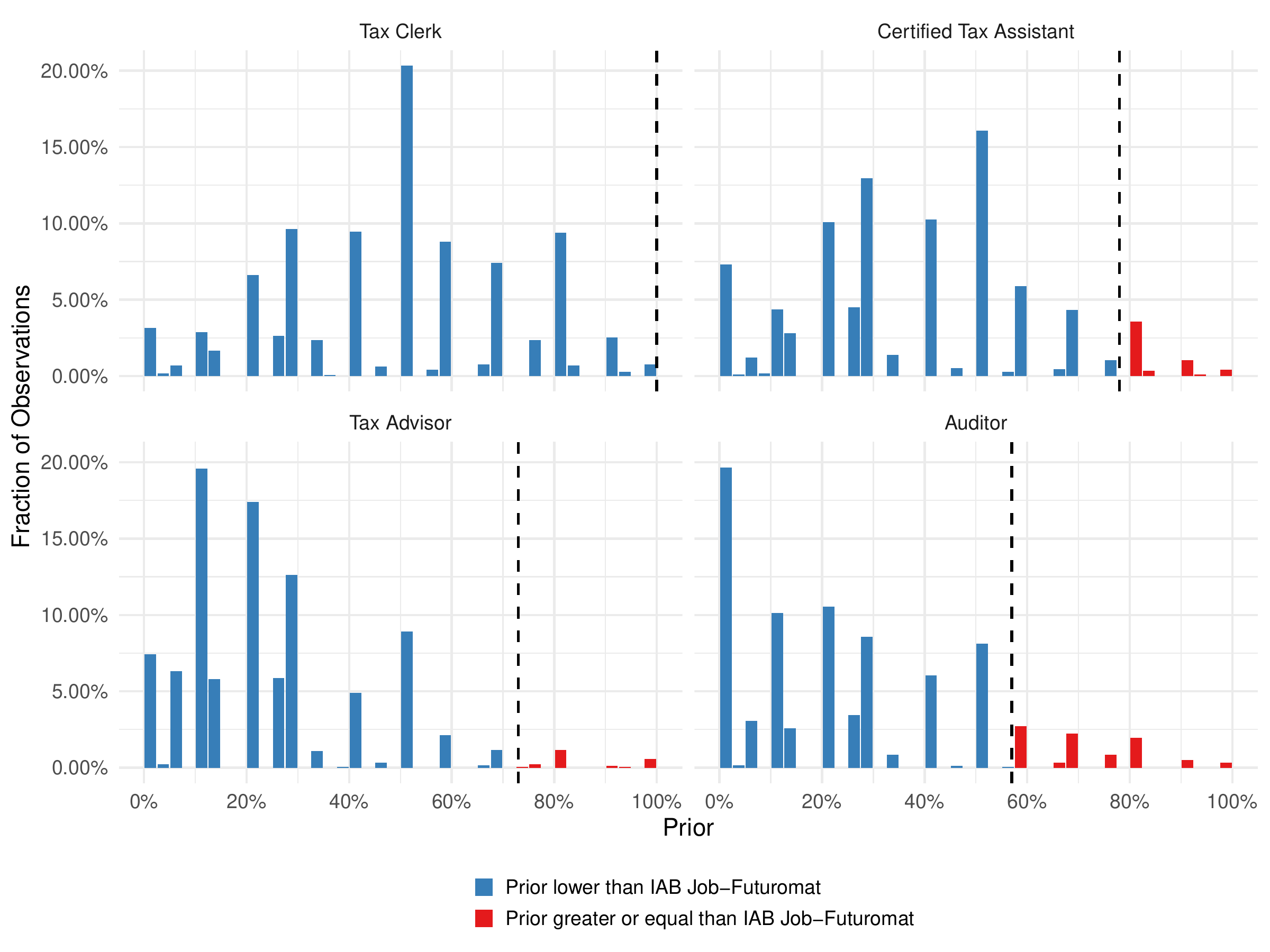}} 
    {This figure displays the distribution of prior beliefs about job automatability for four tax-related occupations: Tax clerk, certified tax assistant, tax advisor, and auditor. The horizontal axis represents the subjective assessment (prior belief) of the share of core tasks in a given occupation that can be automated as of 2024, while the vertical axis indicates the fraction of respondents reporting each probability level. The dashed vertical lines denote the automatability estimates from the IAB Job-Futuromat signal. The shading differentiates between respondents whose prior beliefs are below (blue) or at least as high (red) as the benchmark estimate.
    A large majority of respondents report priors below the IAB Job-Futuromat benchmark: 99.3\% for tax clerks, 93.9\% for certified tax assistants, 97.7\% for tax advisors, and 89.1\% for auditors.}     
    {German Business Panel Tax Advisor Survey 2025.}    
    {H} 

\subsection{Information Provision and Belief Updating}

To examine how tax firms revise their beliefs about automatability in response to new information, we compare prior and posterior beliefs across treatment and control groups.

Figure~\ref{fig:updating_by_occ} plots the average belief shift (posterior minus prior) by occupation and treatment arm. As expected, belief changes in the control group are minimal and centered around zero, suggesting that expectations remain stable in the absence of new information. In contrast, all three treatment arms exhibit clear upward revisions, particularly for lower-skilled occupations such as tax clerks and certified tax assistants. This indicates that the information signals prompt respondents to reassess the automation potential of specific job roles.

\customfigure
{Belief Updating by Occupation and Treatment Arm}
{fig:updating_by_occ}
{\includegraphics[width=0.9\textwidth]{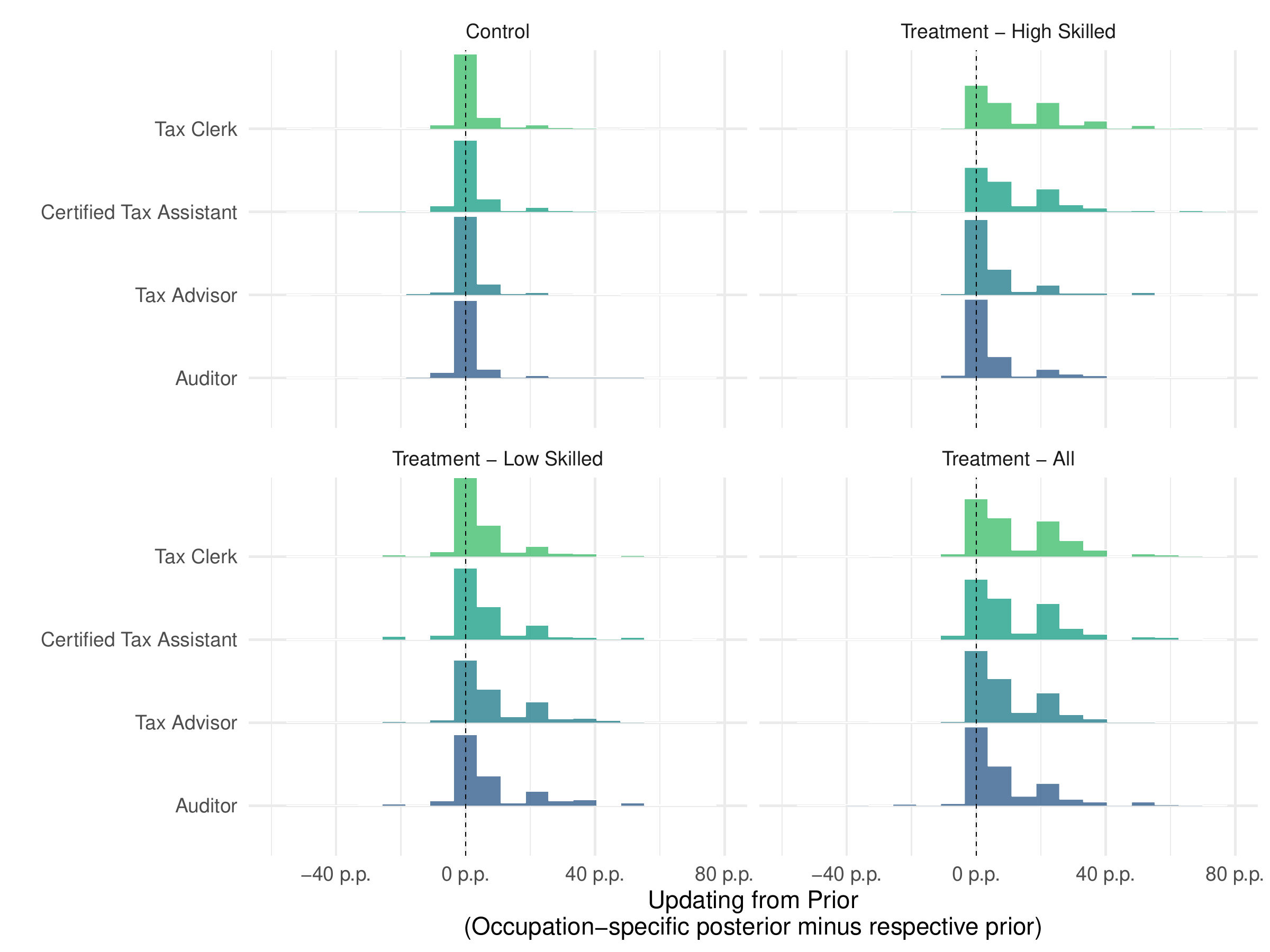}}
{Each panel shows posterior minus prior beliefs about the share of automatable tasks, broken down by occupation and treatment arm. Shifts to the right reflect belief updating toward higher perceived automatability.}
{German Business Panel Tax Advisor Survey 2025.}
{H}

Importantly, belief updating is most pronounced for lower-skilled occupations such as tax clerks and certified tax assistants, while higher-skilled roles like tax advisors and auditors show smaller shifts. Belief changes are not limited to the occupation specifically mentioned in the information treatment. For example, respondents receiving information about tax advisors and auditors also revise their beliefs about tax clerks or certified tax assistants upward, suggesting cross-occupational learning. This pattern is consistent with respondents generalizing the automation message to tax-specific roles, indicating a broader reinterpretation of occupational risk once exposed to the information.\footnote{From a methodological point of view, our experiment is designed to allow quantifying cross-learning. Figure~\ref{fig:updating_by_occ} shows that when confronted with information on low-skilled occupations, respondents revise beliefs also about high-skilled, perhaps even more strongly (see bottom-left graph) and vice-versa (see top-right graph). This suggests that it is important to elicit beliefs of related occupations, even if no signal has been provided about them in information provision experiments.}

Our empirical strategy exploits the information-provision experiment to identify the causal effect of automatability news on firms’ perceptions, expectations, and plans \citep[see][for a general discussion of the design and interpretation of information-provision experiments]{HaalandRothWohlfart2023}. Appendix~\ref{appendix:theory} details how this causal effect can be identified within a 2SLS framework based on Bayesian belief updating, and discusses the corresponding implications for the first-stage and second-stage regression models.

\paragraph{Bayesian Learning Framework for Belief Updating.}
We formalize belief updating using a Bayesian learning framework in which respondents combine new information from the treatment with their prior beliefs about the automatability of specific occupations \citep[e.g.][]{CoibionGorodnichenkoWeber2025}. Because we elicited detailed priors and posteriors for four occupations, we estimate a fully saturated model that allows belief updating to vary flexibly across occupations and treatment arms.\footnote{Equation~\eqref{eq:belief_update} omits priors for other occupations for brevity. See Table~\ref{tab:first_stage} for the full specification.}

\begin{equation}
\text{posterior}_{io} = \alpha_o + \beta_o \cdot \text{prior}_{io} + \sum_{j} \delta_{oj} \cdot D_{ij} + \sum_{j} \gamma_{oj} \cdot (D_{ij} \times \text{prior}_{io}) + \varepsilon_{io},
\label{eq:belief_update}
\end{equation}

where $i$ indexes respondents, $o \in \{\text{Tax Clerk}, \text{Tax Assistant}, \text{Tax Advisor}, \text{Auditor}\}$ denotes the occupation, and $j \in \{\text{Low-skill}, \text{High-skill}, \text{Combined}\}$ indexes the three treatment arms.  
The dummy variable $D_{ij}$ equals one if respondent~$i$ received treatment~$j$, and zero otherwise. The control group serves as a baseline. This specification allows both the direct treatment effect ($\delta_{oj}$) and the slope with respect to prior beliefs ($\gamma_{oj}$) to vary by occupation and treatment.  
$\gamma_{oj}$ indicates whether individuals with stronger priors discount new information more heavily, the coefficient $\beta_o$ captures overall persistence in beliefs in the control group. The framework allows for heterogeneous learning in line with Bayesian updating, where individuals rationally weigh new information relative to their existing beliefs depending on its perceived novelty or informativeness \citep{Sims2003}. A key implication of this set-up is that $\gamma_{oj}/\beta_{o}$ serves as a measure of the learning rate. Since $\beta_{o}$ is close to unity empirically, posterior beliefs equal prior beliefs, such that there is no updating for the control group and $\gamma_{oj}/\beta_{o}\approx\gamma_{oj}$.

\paragraph{Scatterplots showing Bayesian Updating.} We first visually examine whether belief updating follows this Bayesian learning process, whereby respondents combine their prior beliefs with the signal received in the information treatment, starting with a simple scatterplot to present the main intuition. Figure \ref{fig:scatter_update_main} plots individual-level prior and posterior beliefs for tax clerks, with separate trends for the control and treatment groups.

\customfigure
    {Prior and Posterior Beliefs for Tax Clerks} 
    {fig:scatter_update_main}              
    {\includegraphics[width=\textwidth]{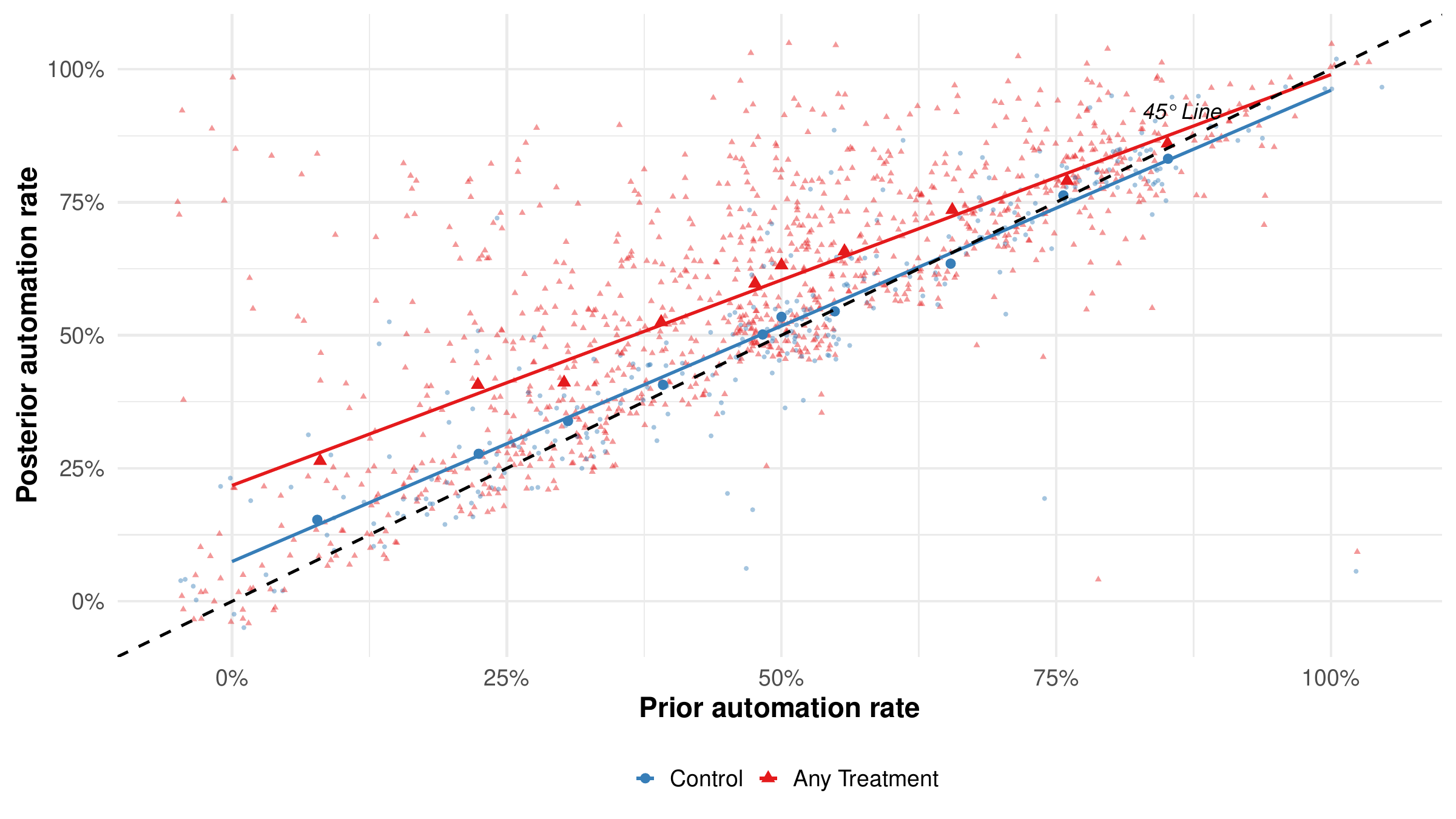}} 
    {This figure illustrates belief updating about automation rates based on regression equation~\eqref{eq:belief_update} for tax clerks. The horizontal axis represents respondents' prior beliefs, and the vertical axis shows their posterior beliefs about automation rates. Light blue dots represent individual values in the control group, while red triangles denote those in any treatment arm. Larger, darker markers indicate averages for 10 quantile bins within each group. The dashed 45 degree line represents no belief updating.}     
    {German Business Panel Tax Advisor Survey 2025.}            
    {H} 

The control group (blue dots) largely aligns to the 45 degree line, suggesting that in the absence of new information, respondents' beliefs remain stable. In contrast, those being in any of the three treatment arm and receiving an information treatment (red triangles) tend to shift upward, suggesting that structured information interventions lead employers to revise upward their beliefs of how automatable the task of tax clerks are.

Similar adjustments can also be observed across other tax occupations, as shown in Figure \ref{fig:updating_any_all} in the Appendix. This adjustment is most pronounced for lower-skilled roles, such as tax clerks and certified tax assistants, while belief updating is weaker for higher-skilled roles, such as tax advisors and auditors. However, the extent of these adjustments varies systematically with the strength of prior beliefs.

\customfigure
    {Learning Rates} 
    {fig:learning_rates}              
    {\includegraphics[width=\textwidth]{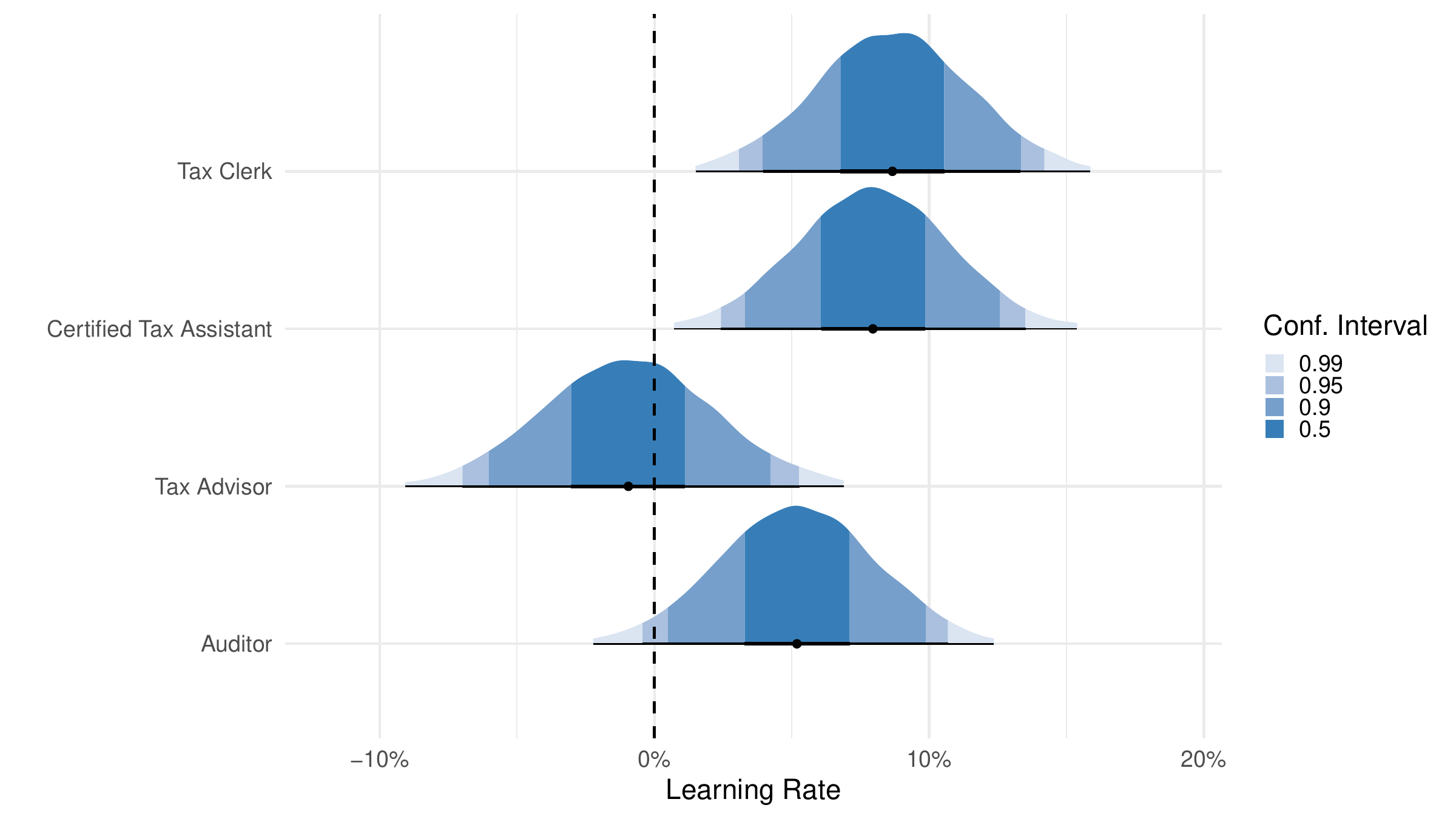}} 
    {This figure presents the estimated learning rates for different occupations in the tax advisory sector. Learning rates are derived from a Bayesian updating framework based on regression equation~\eqref{eq:belief_update}, where belief shifts in automation potential are modeled as a function of prior expectations and information treatment exposure. The individual $\gamma_o$ slope parameters from equation~\eqref{eq:belief_update} shown in Figure~\ref{fig:updating_any_all} are aggregated with the delta method using the proportions of the treatment arm (roughly 0.25 for each arm). The density plots visualize the distribution of estimated learning rates across occupations, with shading indicating different confidence intervals (50\%, 90\%, 95\%, and 99\%). A higher learning rate suggests greater responsiveness to new information about automation rates.}     
    {German Business Panel Tax Advisor Survey 2025.}            
    {H} 

\paragraph{Empirical Evidence of Belief Adjustment.} Next, we present results for the occupation-specific (Table~\ref{tab:stage1_by_occ}) and the full specification from equation \eqref{eq:belief_update} for all four occupations (Table~\ref{tab:first_stage}). We focus our discussion on Figure~\ref{fig:updating_any_all}, which is based on the full specification, to facilitate an interpretation that takes the cross-terms into account. The information treatments prompt substantial belief updating among employers regarding the automatability of tax occupations. Across all models, the coefficient on prior beliefs ($\beta_o$ in the specification without cross-terms) is close to one, indicating that, in the absence of new information, employers largely rely on their existing expectations when forming posterior beliefs about automation potential, perhaps revising them slightly downward. However, the negative and statistically significant interaction terms ($\gamma_{oj}$) for tax clerks and tax assistants together with the significantly higher coefficients on the treatment dummies show that the posterior beliefs are moved upwards and the slopes become flatter. This shows that, when exposed to the information treatment, employers discount their prior beliefs more heavily and are more likely to revise their expectations for these lower-skilled roles.
For tax advisors, belief updating appears limited, as reflected in small and statistically insignificant interaction terms across all treatment arms. For auditors, by contrast, we find a negative and statistically significant interaction effects indicating information updating. 

\paragraph{Learning Rates and Responsiveness to Information.} The results that employers are more open to updating their views about automation potential for lower-skilled positions, while remaining more anchored in their priors for higher-skilled roles, potentially reflects stronger convictions or perceived job complexity at higher levels. 

Figure~\ref{fig:learning_rates} visualizes the overall learning rate for each occupation, derived from aggregating the interaction coefficients across treatment arms using the delta method, and drawing from the estimated distribution to visualize confidence regions. The figure shows that learning rates are highest for tax clerks and certified tax assistants, where firms are most responsive to new information, and lowest for tax advisors, where beliefs remain largely unchanged. The confidence intervals confirm that belief updating is statistically significant for the lower-skilled roles, marginally significant for auditors, while the learning rate for tax advisors is not distinguishable from zero, highlighting the limited effect of the information treatment on beliefs about this occupation.\footnote{The raw first stage estimates are shown in Table \ref{tab:first_stage} in the Appendix.}

\subsection{Instrumenting Automation Beliefs}

While the results above demonstrate that employers systematically revise their beliefs about the automatability of their workforce in response to new information, shifting expectations about automation does not necessarily translate into changes in firm behavior. Whether belief updates lead to revisions in concrete decision-making, such as hiring, wage setting, or investment strategies, remains an open question.

To identify the causal effect of automation beliefs on these firm-level outcomes, we exploit the experimental variation from the information treatment as an instrument. Specifically, we estimate a two-stage least squares (2SLS) model, where posterior beliefs about automation rates, endogenously determined by priors and treatment assignment, serve as  instrumented predictors of firm behavior. In Appendix~\ref{Reduced Form Estimates}, we show reduced form estimates that capture the intent-to-treat effects and compare them to the 2SLS estimates. The 2SLS approach isolates the exogenous variation in belief shifts, precluding bias from confounders and, more importantly in a survey context, is consistent under specific forms of measurement error. Appendix~\ref{appendix:theory} shows that the main coefficient of interest should be similar in the the reduced form and the 2SLS approach in absence omitted variables, simultaneity problems and measurement error, which is true in our case (see Appendix~\ref{Reduced Form Estimates}). The second-stage regression is specified as follows:

\begin{equation}
Y_i = \beta_0 + \sum_{o} \beta_{1o} \widehat{\text{posterior}}_{io}  + \mathbf{X}_i'\gamma + \eta_i,
\label{eq:second_stage}
\end{equation}
where $Y_i$ denotes the outcome of interest for firm $i$ (e.g., revenue expectations or employment plans), $\widehat{\text{posterior}}_{io}$ are the fitted values from the first-stage regressions (see Table~\ref{tab:first_stage}) for each occupation $o$, and $\mathbf{X}_i$ is a vector of predetermined firm and individual-level control variables. Since the experiment achieved good balance in observable characteristics across treatment arms due to randomization, results stay virtually unchanged if no controls are included. 

Typically, the estimates are interpreted in terms of the standard deviation of posterior beliefs within each treatment arm, since this serves as a proxy for information diffusion or belief dispersion. However, we take a more conservative approach and summarize average effects based on mean belief updating. In our data, standard deviations in posteriors are typically about twice as large as the mean shift, so relying on averages typically leads to lower treatment effect ranges that are in line with actual belief shifts.\footnote{Because the information treatment generates joint variation in beliefs across all four occupations, it is not meaningful to interpret the $\beta_{1o}$ coefficients as ceteris paribus effects.}

Concretely, we summarize the overall impact of belief updating by  calculating the impact at an average belief updating for all occupations in the combined treatment using a linear combination of the second-stage coefficients:
\begin{equation}
\sum_{o=1}^{4} \hat{\beta}_{1o} \cdot \overline{\text{Updating}}_{o,\text{Treatment-Combined}}.
\label{eq:aggregation}
\end{equation}

To account for joint estimation uncertainty, we apply the delta method to derive standard errors and confidence intervals for this linear combination. The mean belief shift caused by the combined information treatment is substantial across all four occupations: on average, beliefs about the automatable task share increased by 13.1 for tax clerks, 12.2 for certified tax assistants, 9.1 percentage points for tax advisors, and 9.3 for auditors.
 
This approach offers two key advantages. First, it yields a single, interpretable quantity that captures the expected effect  of the information treatment at representative levels of belief updating. Second, it avoids problematic ceteris paribus interpretations that arise when individual  posteriors are treated as if they could vary independently, despite being jointly determined by the treatment. 

\subsection{Employment Plans and Productivity Expectations}

The coefficient plot in Figure \ref{fig:coefplot_main_outcomes} illustrates the effects of belief updating about automation on firm-level employment and financial expectations. Estimates are derived from second-stage IV regressions using updated automation beliefs instrumented by our randomized information intervention (see Appendix~\ref{Reduced Form Estimates}).

\customfigure
    {Coefficient Plot for Firm-Level Employment Plans and Financial Expectations} 
    {fig:coefplot_main_outcomes}              
    {\includegraphics[width=\textwidth]{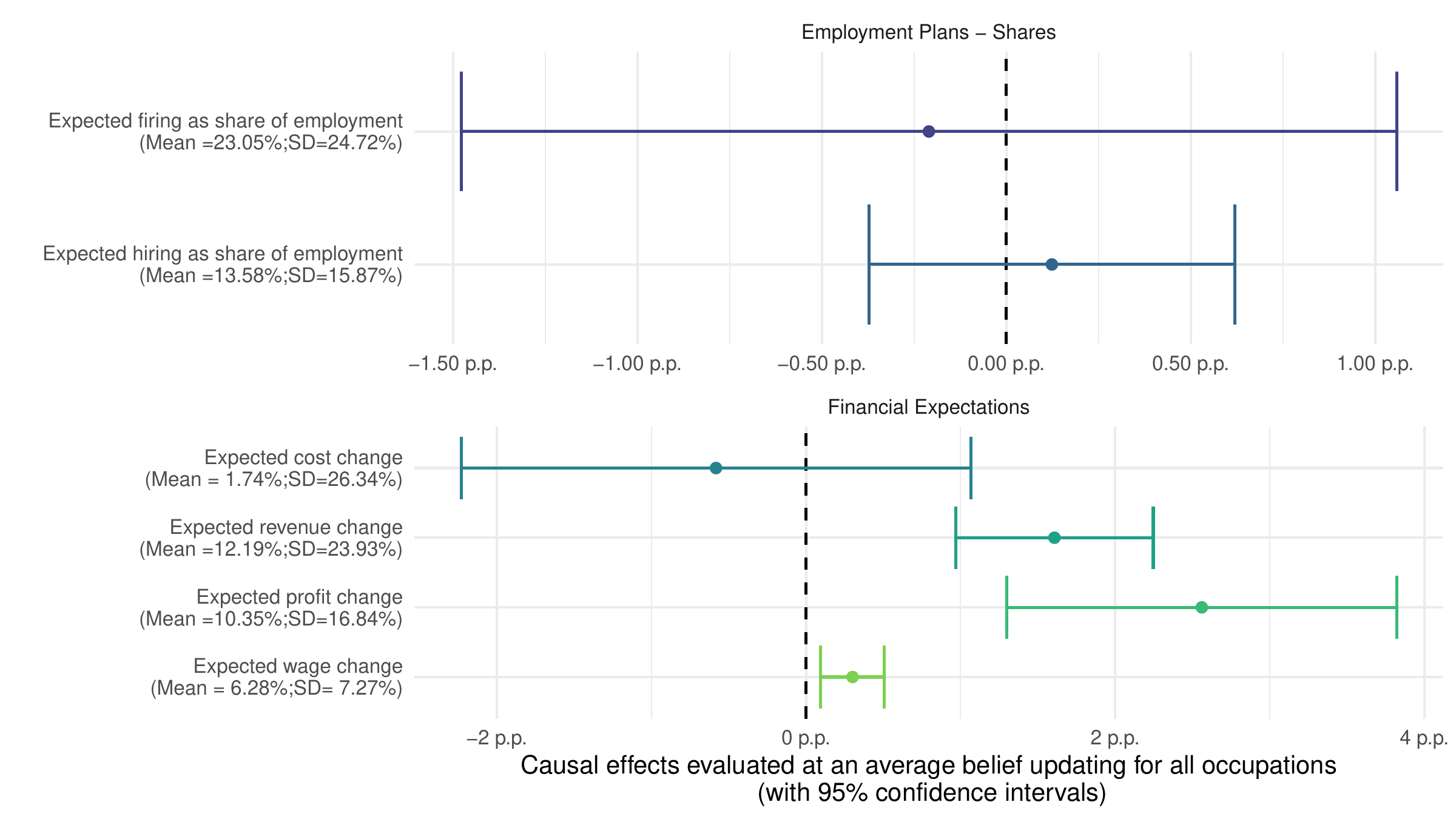}} 
    {\secstageupdatepred The outcome variables are hiring and firing plans, cost, revenue, and profit change expectations as well as wage change expectations.}     
    {German Business Panel Tax Advisor Survey 2025.}            
    {H} 

In the upper panel (Employment Plans), we see that the effects of updated automation beliefs on both firing and hiring plans as a share of employment are statistically indistinguishable from zero. Firing plans refer to employer-initiated separations and do not include voluntary turnover or retirements. The estimated coefficients are small, and the confidence intervals span both positive and negative values. The baseline for expected firing in the control group is 23.05\% (SD: 24.72\%) and for expected hiring is 13.58\% (SD: 15.87\%). It is important to note that the average firm in the sample is small, with fewer than 10 employees, so these large percentage shares reflect relatively small absolute adjustments planned over the next three years. The results must also be interpreted in light a labor market tightness (vacancies/unemployed) of more than 3.5 in 2023 for tax clerks (2 for tax advisors), where employer-initiated separations are rare. Compared to this substantial underlying variation, the treatment effects  of the information intervention on employment plans are negligible.\footnote{In open text fields, respondents explain why they are hiring tax clerks, even if their tasks are fully automatable. One frequently stated reason is that tax advising is a “people’s business”, where clients want personalized explanations, confidentiality, and human interaction. A classic way to do this is shoebox accounting: the informal practice of storing all receipts and financial records in a box and handing them to an tax advisor at year’s end for sorting and bookkeeping.}

The lower panel (Financial Expectations) shows the effects of updated automation beliefs on firms' cost, productivity measured in revenue, profit, and wage expectations changes. The estimated effect on expected cost change is negative and statistically insignificant (point estimate = $-0.6$ percentage points, p = $0.49$), suggesting that firms do not anticipate significant cost increases due to AI. By contrast, there are statistically significant positive effects on both expected revenue and profit: the intervention increases expected revenue by $1.6$ percentage points ($p < 0.001$) and expected profit by $2.6$ percentage points ($p < 0.001$) over the baseline means. Although these magnitudes may appear modest, they are notable relative to the mean and standard deviation of the respective outcomes, especially for profits.

Although firms in our sample revise their beliefs upward for revenue and profit expectations, it is not obvious that productivity gains from AI would translate systematically into higher profits in the long run. Whether firms can retain these gains depends on market structure and regulation. In most markets, competitive pressure or consumer demand might limit firms’ ability to capture the full surplus. In tax advisory services, this is shaped by a regulated fee schedule (Steuerberatervergütungsverordnung) and professional rules that restrict competition (e.g., limits on advertising, restricted entry). The fee schedule is a government-mandated schedule that fixes prices for many standard services, common also in other white collar occupations, e.g., architects and engineers  \citep{Rostam2019}. With prices semi-fixed, higher productivity mainly allows firms to serve more clients per hour rather than to charge more per service.\footnote{Moreover, because firms are legally required to use licensed tax advisors for many filings, direct substitution of advisory services by AI tools is unlikely. Historically, the large-scale vertical replacement of tax advisors by software has not materialized. \citet{goolsbee2004turbotax} shows that early tax software such as TurboTax displaced some routine preparation work, especially among affluent, educated households that overlapped with paid preparers’ clientele. Yet complex cases continued to rely on professional advisors, and selective adoption limited the scope of substitution. In the German context, this substitution is further constrained by law: certain filings, such as reports and taxes for limited liability companies, must legally be prepared by licensed advisors.}

Interestingly, the effect on expected wage change is also positive and statistically significant, but much smaller in magnitude (just $0.3$ percentage points, $p = 0.004$). This effect is minor compared to the baseline mean of 6.28\% (and  standard deviation of 7.27\%), implying that firms do not plan to pass on the anticipated financial gains from automation to employees in the form of higher wages, at least not on a comparable scale. At the same time, the narrow confidence interval for wage expectations  reinforces the precision of this economically negligible effect.

These findings align with the research on rent-sharing, that shows that firms often do not share productivity gains with employees, especially when such gains are derived from automation or technological advancements. For instance, \citet{Kline2019} found that workers capture only a fraction of the surplus generated by patents, with significant disparities based on tenure and position within the firm. Similarly, \citet{Cho2022} observed that rent-sharing within firms is uneven, favoring higher-earning employees.\footnote{This might be in part because the wage schedule for most tax occupations is fixed, although discretionary, informal rewards for for exceptional client management, accuracy, or firm profits can occur. This is in line with research by \citet{Franceschelli2010}, who show that productivity gains under performance pay schemes translate more directly into higher wages compared to fixed-wage schemes, where productivity improvements have a more limited effect on employee compensation, even when both types of workers achieve similar productivity increases.} Evidence for architects and engineers reported by \cite{Rostam2019} shows that an unexpected 10\% increase of their otherwise fixed prices raised employers’ personal net income by 8 \%, with no corresponding wage gains for employees.

Moreover, these findings are consistent with recent evidence from Denmark. \citet{HumlumVestergaard2025} study chatbot adoption among workers in AI-exposed occupations using linked survey and administrative data. Despite widespread usage, time savings, and firm-level encouragement, they find no measurable changes in hours worked or earnings, with precise confidence intervals that rule out even modest effects. Only a small share of the productivity gains (estimated at 3-7\%) is passed through to wages. Our results echo this pattern: while firms revise their beliefs about automation and anticipate financial benefits, they do not expect to share these gains proportionally with workers. This highlights how belief formation and firm expectations shape the early-stage impact of AI.

\subsection{Occupation-Level Productivity Expectations}

To further examine how belief updating affects firms' economic expectations, Figure \ref{fig:coefplot_occupation_specific} breaks down the expected productivity expectation, measured as revenue growth attributed to automation-induced belief updates by different tax-advisory occupations. We again use the same IV strategy and present the aggregated effects for average belief updating. 

The estimates reveal that across all occupations, higher posterior beliefs about automatability are associated with significantly higher expected productivity, measured as expected growth of revenue per hour. This suggests that firms anticipating greater automation in these roles expect efficiency gains to translate into revenue increases. However, the magnitude of this effect varies across occupations and decrease with higher expertise requirements.

\customfigure
    {Coefficient Plot for Occupation-Level Hourly Revenues} 
    {fig:coefplot_occupation_specific}              
    {\includegraphics[width=\textwidth]{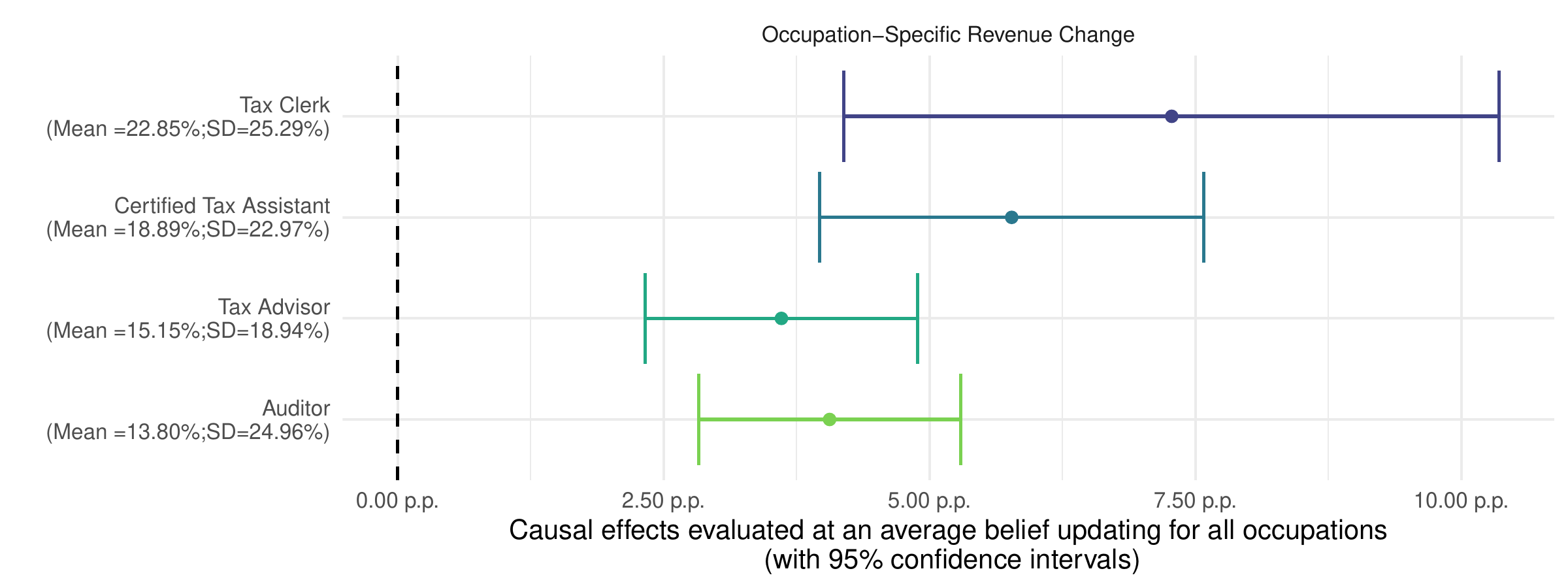}} 
    {\secstageupdatepred The outcome variables are perceived occupation-specific revenue changes per hour.}     
    {German Business Panel Tax Advisor Survey 2025.}            
    {H} 

The largest estimated effect appears for tax clerks, where a representative treatment-induced update in automation beliefs is associated with an expected increase in hourly revenues of 7.3 percentage points (95\% CI: 4.2 to 10.4). Certified tax assistants exhibit a somewhat smaller effect at 5.8 percentage points (95\% CI: 4.0 to 7.6). For higher-skilled roles, the estimated effects are more modest, with expected increases of 3.6 (95\% CI: 2.3 to 4.9) and 4.1 (95\% CI: 2.8 to 5.3) percentage points for tax advisors and auditors, respectively. However, the confidence intervals for these estimates overlap, particularly between the two lower-skilled and two higher-skilled occupations, indicating that the differences in point estimates may not be statistically significant. Yet, for the high-skilled occupations for which we observed lower learning rates, less productivity increases are expected. This may raise questions about barriers of automation due to hierarchy and bureaucracy. The baseline revenue growth expectations, however, also decrease with higher expertise requirements, therefore the relative effects are remarkably similar (7.3/27.9$\ \approx \ $26\%,26\%,22\%,28\%).

\subsection{Automation Potential and New Tasks}

\customfigure
    {Coefficient Plot for Automation Potential} 
    {fig:coefplot_dummies_ap}              
    {\includegraphics[width=\textwidth]{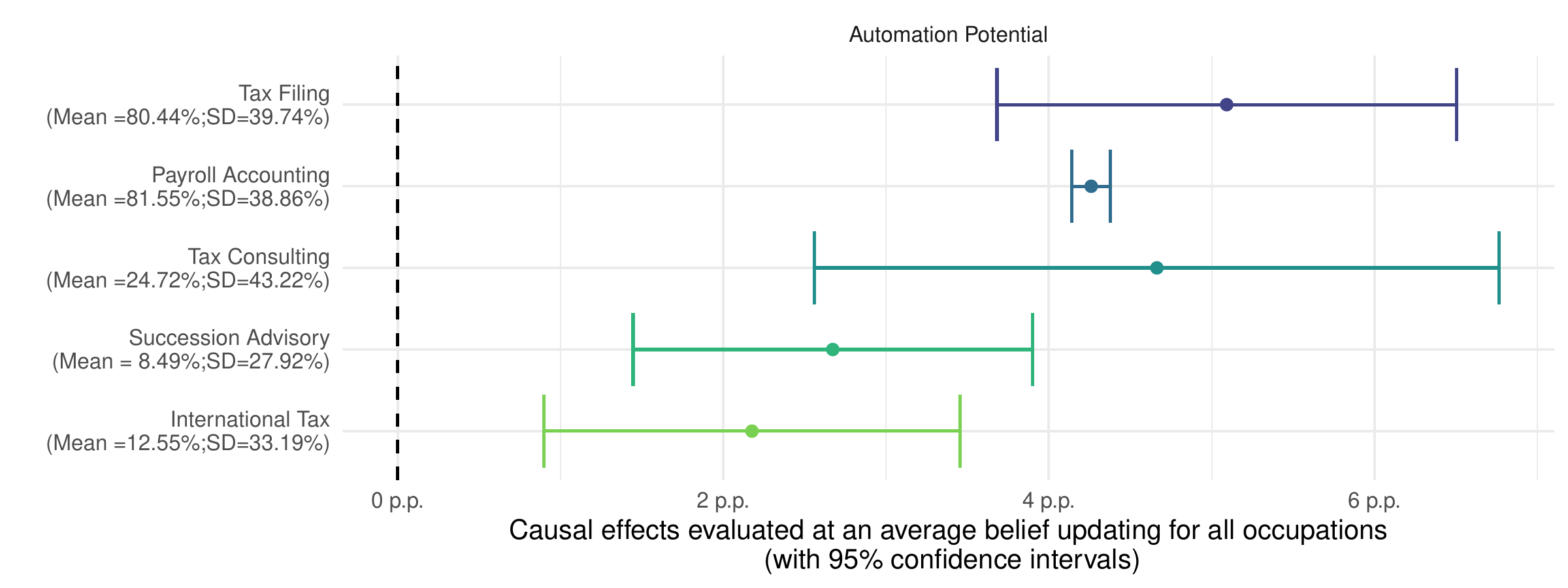}} 
    {\secstageupdatepred The outcome variables are perceived automation potential at the task level. Each point shows the change in probability that respondents assign automation potential to a specific task (tax filing, payroll accounting, tax consulting, succession advisory, or international tax advisory)}     
    {German Business Panel Tax Advisor Survey 2025.}            
    {H} 

The evidence presented in Figure \ref{fig:coefplot_dummies_ap} reveals how the overall information shock translates into perceived automation potential at the task level.

Here, the dependent variables are binary indicators equal to one if a respondent believes that a given task, such as \textit{tax filing}, \textit{payroll accounting}, \textit{tax consulting}, \textit{succession advisory}, or \textit{international tax advisory}, has automation potential. Consistent with our earlier findings, respondents exposed to the information treatment report significantly higher automation potential for lower-skilled tasks, particularly those traditionally performed by tax clerks and certified tax assistants such as tax filing and payroll accounting.  

Treated respondents report significantly higher probabilities that routine tasks such as tax filing and payroll accounting can be automated. The estimated increases for these tasks are around four to five percentage points, indicating that the information treatment leads to a marked shift in perceptions of automation potential in the most routine-intensive areas. For more complex or advisory tasks like tax consulting, succession advisory, and international tax advisory, the effects are smaller and estimated less precisely, albeit still significantly positive. This pattern demonstrates that firms' beliefs about which aspects of their work are automatable are particularly responsive in routine, lower-skilled tasks.

However, the fact that treated respondents revise their beliefs even for high-skilled tasks suggests that AI-based tools and automation solutions are beginning to shape expectations beyond purely routine work. While automation expectations remain strongest for procedural and compliance-related work, firms do not entirely discount the potential for AI-driven automation even in high-skilled advisory roles.

\customfigure
    {Coefficient Plot for the Emergence of New Tasks} 
    {fig:coefplot_new_tasks}              
    {\includegraphics[width=\textwidth]{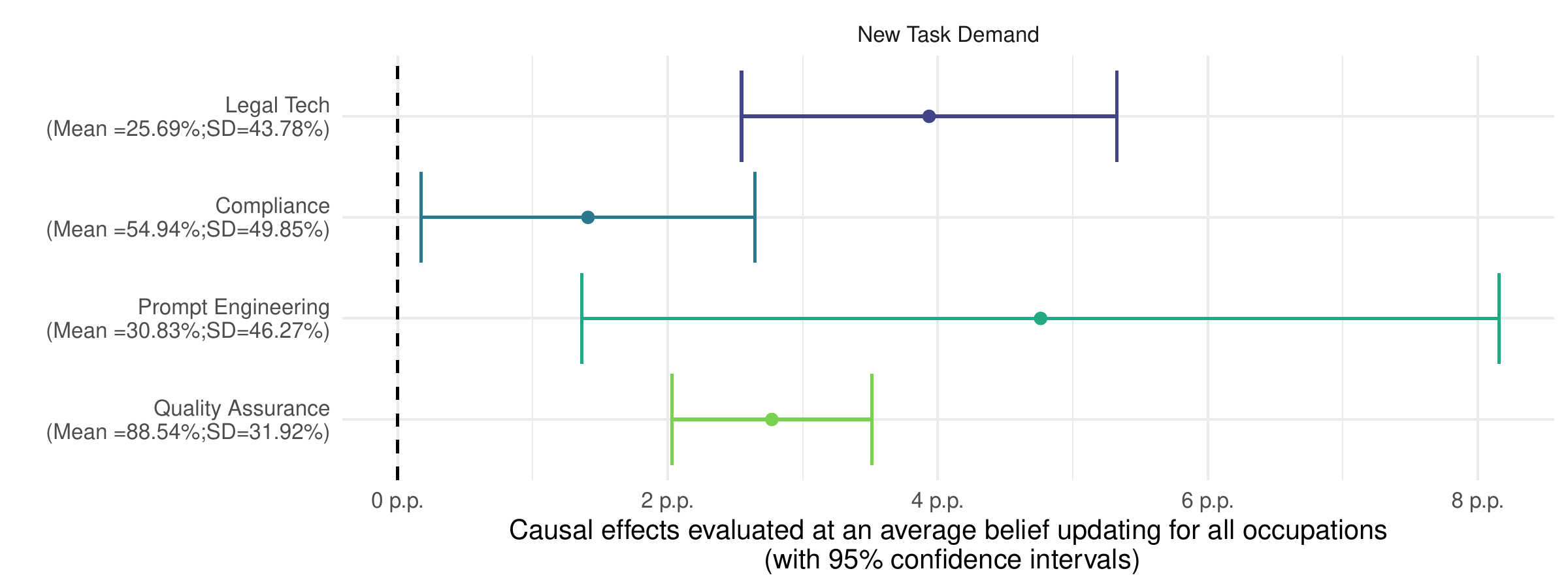}} 
    {\secstageupdatepred The outcome variables are expectations for the emergence of new tasks, including the probability that respondents report increased relevance of legal tech, compliance, prompt engineering, and quality assurance.}     
    {German Business Panel Tax Advisor Survey 2025.}            
    {H} 

Beyond revising their expectations about the automation potential of existing tasks, respondents also anticipate the emergence of new tasks as a consequence of automation. This raises the question of how firms expect job roles to evolve in response to automation and whether they foresee a net displacement of tasks or an expansion into new responsibilities that complement AI-driven workflows.

To explore this, Figure \ref{fig:coefplot_new_tasks} highlights anticipated changes in demand for new tasks emerging alongside automation. Here, the dependent variables are binary indicators for whether respondents expect \textit{legal tech}, \textit{compliance}, \textit{prompt engineering}, or \textit{quality assurance} to become relevant as part of their evolving job responsibilities.

The results show that higher automation beliefs significantly increase the likelihood of respondents considering new tasks relevant, though the effect sizes vary across task types. 

The largest effects are observed for prompt engineering, involving direct interaction and optimization of AI outputs, and legal tech, with estimated increases of roughly $4.8$ and $3.9$ percentage points, respectively. While still a relatively new concept in professional tax work, this suggests that some tax professionals are beginning to anticipate the growing role of AI interaction and optimization as part of their job, reflecting broader labor market trends where demand for AI-related skills has increased across diverse occupations, often accompanied by wage premiums \citep{AlekseevaAzarGineSamilaTaska2021}.

Quality assurance also sees a significant positive effect of $2.8$ percentage points, despite already being regarded as highly important even for individuals who do not update their priors (with  88.54\% of the respondents reporting that it is important).

For compliance, the effect is smaller, about $1.4$ percentage points, but still statistically significant. This pattern suggests that respondents foresee a shift not just toward the automation of existing, routine work, but also toward an expansion of new, tech-oriented or oversight-oriented responsibilities. The means reported in the figure labels further illustrate that while some of these tasks (like quality assurance and compliance) are already widely relevant, others (such as legal tech and prompt engineering) are slated for significant growth in importance.

These differences underscore varying degrees of perceived complementarity between AI and human expertise. Legal tech and compliance are seen as domains likely to require upskilling in legal automation and regulatory monitoring, while prompt engineering signals an early recognition of AI interaction skills as a novel job component. Quality assurance, by contrast, remains a core responsibility, likely shifting focus toward mitigating AI errors rather than creating entirely new workflows.

Rather than replacing professionals, automation appears to be driving a transition toward augmented work, where human oversight and AI-driven processes increasingly coexist. This aligns with the perspective that automation often complements, rather than substitutes for, human labor-creating new tasks where human expertise remains critical \citep[e.g.,][]{Autor2025, AcemogluRestrepo2019, Brynjolfsson2014}. \citet{Brynjolfsson2014} emphasize that while routine tasks are automated, new roles arise requiring advanced cognitive and interactive skills, reinforcing the idea that technology reshapes job content rather than simply eliminating work. As \citet{Autor2025} argue, the labor market consequences of automation depend on how task automation reshapes occupational expertise requirements: when automation removes inexpert tasks, expertise demands and wages rise but employment falls, whereas eliminating expert tasks lowers expertise demands and wages but expands employment.

\subsection{Training from the Employer's Perspective }
Automation is widely seen as a catalyst for organizational change, prompting firms to adapt not only their processes but also their approach to workforce development. Previous studies have shown that automation potential can drive both upskilling and changes in job mobility. 

\customfigure
{Effect of Information Treatment on Planned Training Investments} 
{fig:coefplot_training} 
{\includegraphics[width=\textwidth]{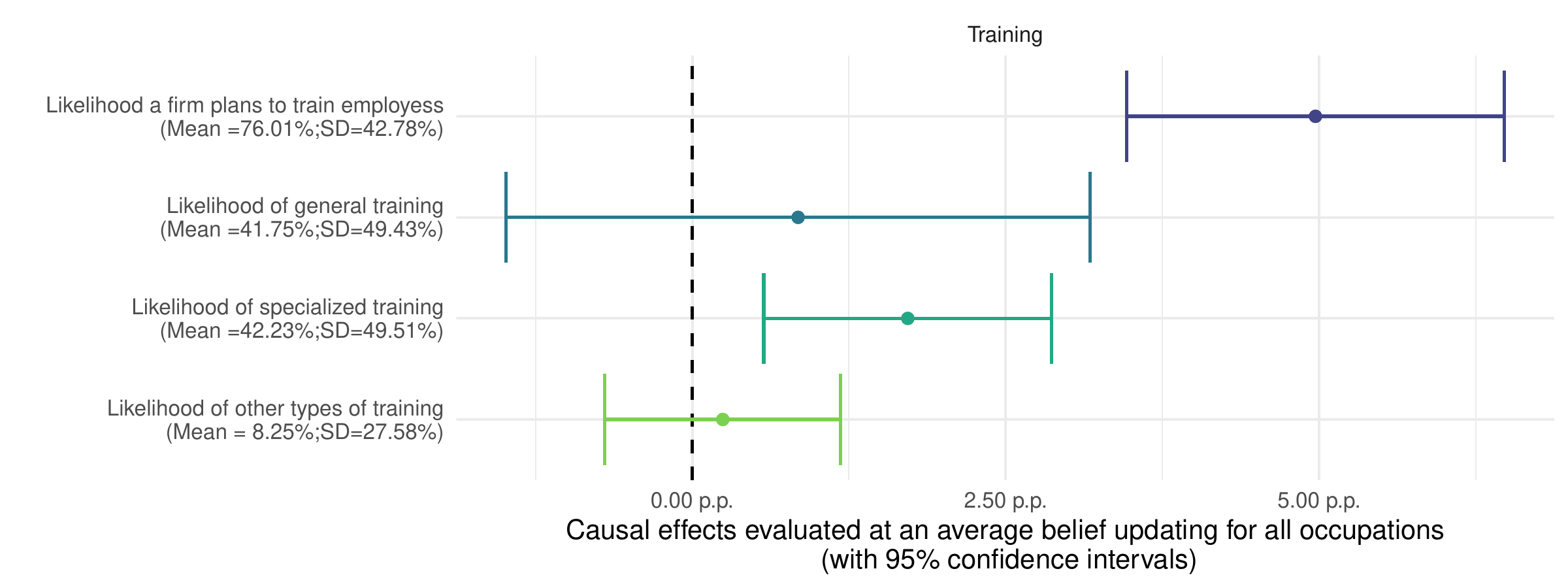}} 
{\secstageupdatepred The outcome variables are the likelihoods that firms plan automation-related training investments.}
{German Business Panel Tax Advisor Survey 2025.} 
{H} 

For instance, \citet{Blanas2019} argue that workers confronted with automation either move into lower-paid, less automatable occupations or acquire new skills to complement technology and access higher-paid roles. Similarly, \citet{Lergetporer2023} find that providing information about automation risks increases employees' willingness to participate in further training, particularly among those in highly automatable occupations, with effects of around five percentage points. However, they also highlight that misperceptions about which tasks are truly automatable may explain why participation in training remains low among less-skilled groups.

Building on this literature and our previous results that show that employers anticipate the emergence of new tasks (e.g., legal tech and prompt engineering) due to automation, we examine firms' plans for employee  training (or upskilling) in response to the information intervention. 

As shown in Figure~\ref{fig:coefplot_training}, exposure to automation information significantly increases the probability that firms plan further training for their staff, raising the likelihood by about 5 percentage points from an already high baseline of 76\%. This effect is concentrated in specialized training, such as technical courses or certifications, which increases by about 1.7 percentage points and is statistically significant. By contrast, the treatment does not meaningfully affect plans for general training (e.g., part-time study) or other types of training, where baseline levels are much lower and effects are not statistically significant.

\customfigure
{Effect of Automation Information on Automation Attitudes, and Career Plans} 
{fig:coefplot_attitudes} 
{\includegraphics[width=\textwidth]{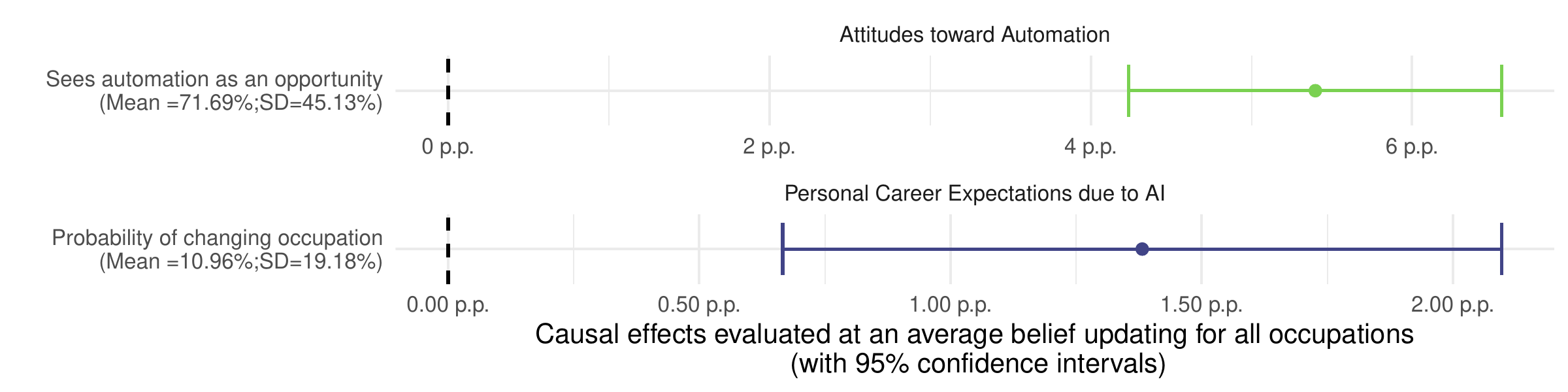}} 
{\secstageupdatepred The outcome variables are:  (i) perceiving automation as an opportunity, and (ii) the self-assessed probability of changing occupation.} 
{German Business Panel Tax Advisor Survey 2025.} 
{H} 

Similarly, our treatment impacts attitudes towards automation.  To this end, Figure \ref{fig:coefplot_attitudes} complements our earlier firm-level results by documenting how the information shock reshapes individual mind-sets and planned behavior.

We use the probability of seeing automation as an opportunity as a measure of general sentiment. This figure rises by $5.4$ p.p. (about $7.5$\% relative to its $71.69$\% mean). This shift is noteworthy providing evidence about a different perception, compared to prior evidence from Germany suggesting that automation is predominantly perceived as a threat to employment \citep{Arntz2022}.

Finally, the treatment increases the self-reported likelihood of changing the respondent's own occupation. Tax advisors in Germany often pursue the auditor certification after several years of practice to offer more comprehensive client support. Auditors assume greater responsibility in economic matters and engage in a broader range of activities compared to tax advisors.\footnote{In our survey, respondents indicated moving into politics, consulting, or acting as CFO frequently. Given the average age of 51, some also mentioned retirement or crafts as next career step.} The treatment changed the likelihood of changing one's occupation by $1.38$ p.p. on a 0-100\% scale. Although modest in absolute terms, this is a 12.6\% relative rise over the 10.96\% baseline, suggesting that a minority of professionals begin to contemplate career moves once automation risks are made salient. The pattern aligns with cross-country evidence suggesting that automation concerns stimulate skill investment and career planning \citep{InnocentiEtAl2022}, yet it also reveals that most respondents refrain from making drastic career changes; instead, they prefer upskilling and tool adoption within their current field.

\subsection{Intentions and Actions}\label{sec:actions}

To quantify the extend to which intentions stated in the survey translate into real-world actions, we measure how the information provision experiment affects the plans to adopt AI solutions and have included an information acquisition task about AI solutions. Figure~\ref{fig:coefplot_actions} reports the results.

For plans to adopt AI solutions we find large responses, which rise by $9.4$ p.p. from a baseline mean of 71.15\%. This corresponds to a $13\%$ increase relative to the mean, underscoring that belief updating will translate into actual AI use for affected firms. 

For individuals unfamiliar with tax-specific AI tools,\footnote{These are typically retrieval-augmented chatbots that accurately cite court decisions and tax laws.} we offered the option to receive a link to an informational page and measured subsequent click behavior as our outcome. In this subgroup, we observe a small but statistically significant increase of $2.4$ percentage points. The muted effect size is consistent with a ceiling effect, given the already high baseline click rate of 80.88\%.

Finally, we test whether the observed belief updating and AI adoption intentions have any effect on actual vacancy posting. For this purpose, we link survey responses to vacancy postings recorded in Germany’s largest vacancy aggregator, operated by the Federal Employment Agency and drawing on multiple job platforms.\footnote{We elicited linkage consent to other data from the survey respondents and obtain nearly complete agreement for linking the survey to external sources. Record linkage was performed using names and addresses as identifiers, with index-based record-linkage that relies on a token-identification-potential search rule \citep[see][for a detailed description]{doherr2023searchengine}. The link was implemented with the \texttt{MatchMakeR} R package, which has been applied and validated in related work; see \citet[][App. D.2]{BruellEtAl2025} for a short description.} We track postings over the four months from April to August 2025, which allows us to examine near-term adjustments in labor demand following the intervention.

\customfigure
{Effect of Automation Information on AI-Adoption Intentions and Actions} 
{fig:coefplot_actions} 
{\includegraphics[width=\textwidth]{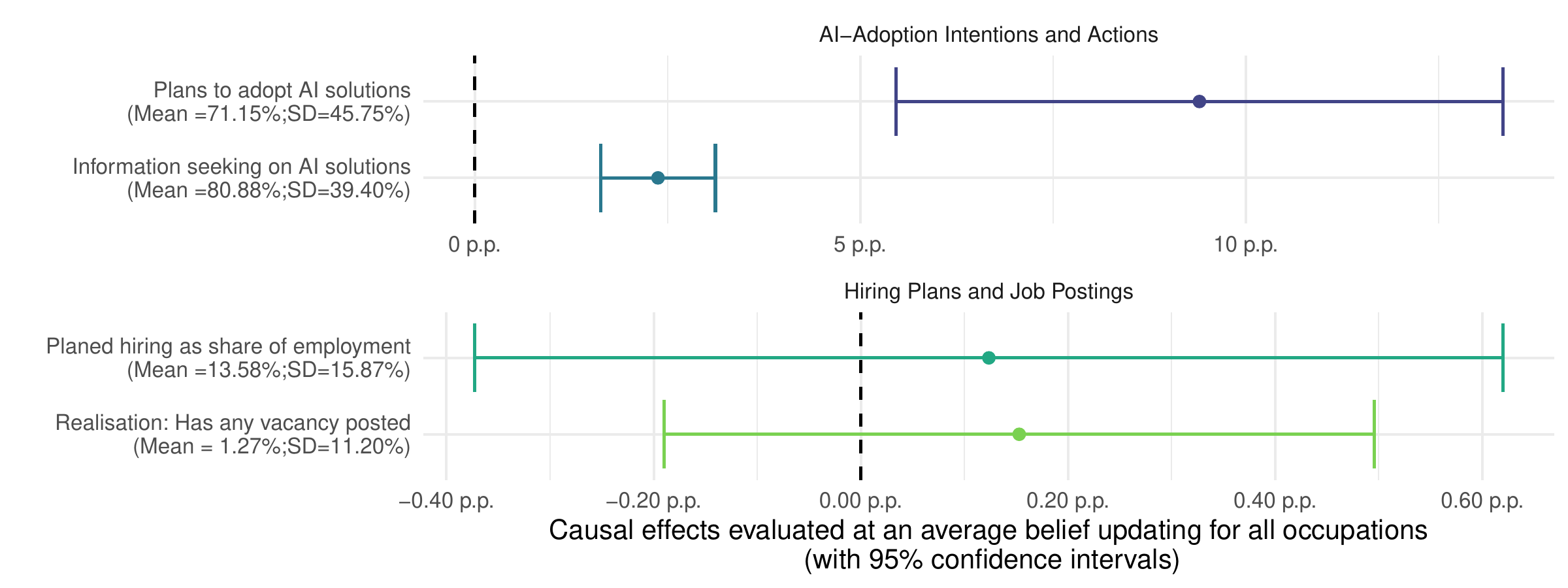}} 
{\secstageupdatepred The outcome variables are: (i) plans to adopt AI solutions, (ii) active information seeking about such tools within those individuals who have not heard of taxation-specific AI solutions prior to the survey and (iii) planned hiring and (iv) an indicator of having posted a vacancy in the months from April to August 2025.} 
{German Business Panel Tax Advisor Survey 2025.} 
{H} 

Consistent with the survey-based null effects for hiring and firing shares, we find no evidence of a treatment effect on vacancy posting. The point estimate is close to zero at 0.15 percentage points (s.e. 0.18 p.p.), relative to a baseline posting probability of 1.3\%. The confidence interval [-0.19 p.p.,+0.50 p.p.] excludes even moderate effects, indicating that firms do not respond to updated automation beliefs with immediate changes in job advertising. This external confirmation underscores that while employers revise beliefs and adoption intentions, these shifts do not translate into observable hiring outcomes.

\subsection{Robustness}\label{sec:robustness}
We perform a series of tests to assess the robustness of the estimates. As a first check, we show in Appendix~\ref{appendix:theory} the conditions under which the Bayesian-updating model and the reduced-form regressions identifies the same causal estimates as 2SLS. In Appendix~\ref{Reduced Form Estimates} we empirically compare 2SLS estimates with corresponding reduced form estimates. Across all outcomes, the reduced-form estimates line up closely with the 2SLS estimates that scale by the treatment-induced change in beliefs. Both approaches show essentially no effect on expected hiring or firing, but positive effects on expected revenue and profits and a very small, precisely estimated effect on wages when evaluated at the sample-average belief update. By occupation, both estimators imply larger expected revenue gains where tasks are more routine (tax clerks, certified assistants) and smaller, though still positive, for higher-skill roles (tax advisors, auditors). For task automation, they agree that belief updates raise the perceived automatability of routine tasks (tax filing, payroll) more than advisory tasks.

Additional sensitivity analyses, reported in Appendix~\ref{appendix:robusntness}, examine legal form (sole proprietors vs.~other legal forms), survey completion duration, and response timing (comparing respondents who answered within one week to the initial invitation with those who replied after receiving a reminder). The results are consistent across subgroups, indicating that our main findings are robust to differences in response behavior and firm-type composition.

Taken together, the empirical results are in line with the theoretical restrictions for both the reduced-form and 2SLS estimates in this setting and survive the robustness checks.

\subsection{Heterogeneity}\label{sec:heterogeneity}

\cite{Humlum2025} document that higher-income and more experienced workers adopt generative AI tools earlier and more intensively, whereas women and lower-earning workers lag behind. To assess whether main effects of belief updating on revenue and employment plans differ across relevant margins, we re-estimate the aggregated causal effects separately by (i) respondents' regular use of generative AI, (ii) firm size (total employment above/below the median), and (iii) respondent age (above/below the median). 

The top panel of Figure~\ref{fig:coefplot_heterogeneity_ai} shows that AI-using firms expect weaker employment growth but also less dismissals following the treatment. This indicates that AI-intensive firms expect efficiency gains that allow them to scale without proportional increases in labor input. 
\customfigure
{Effect Heterogeneity by Baseline AI Use} 
{fig:coefplot_heterogeneity_ai} 
{\includegraphics[width=\textwidth]{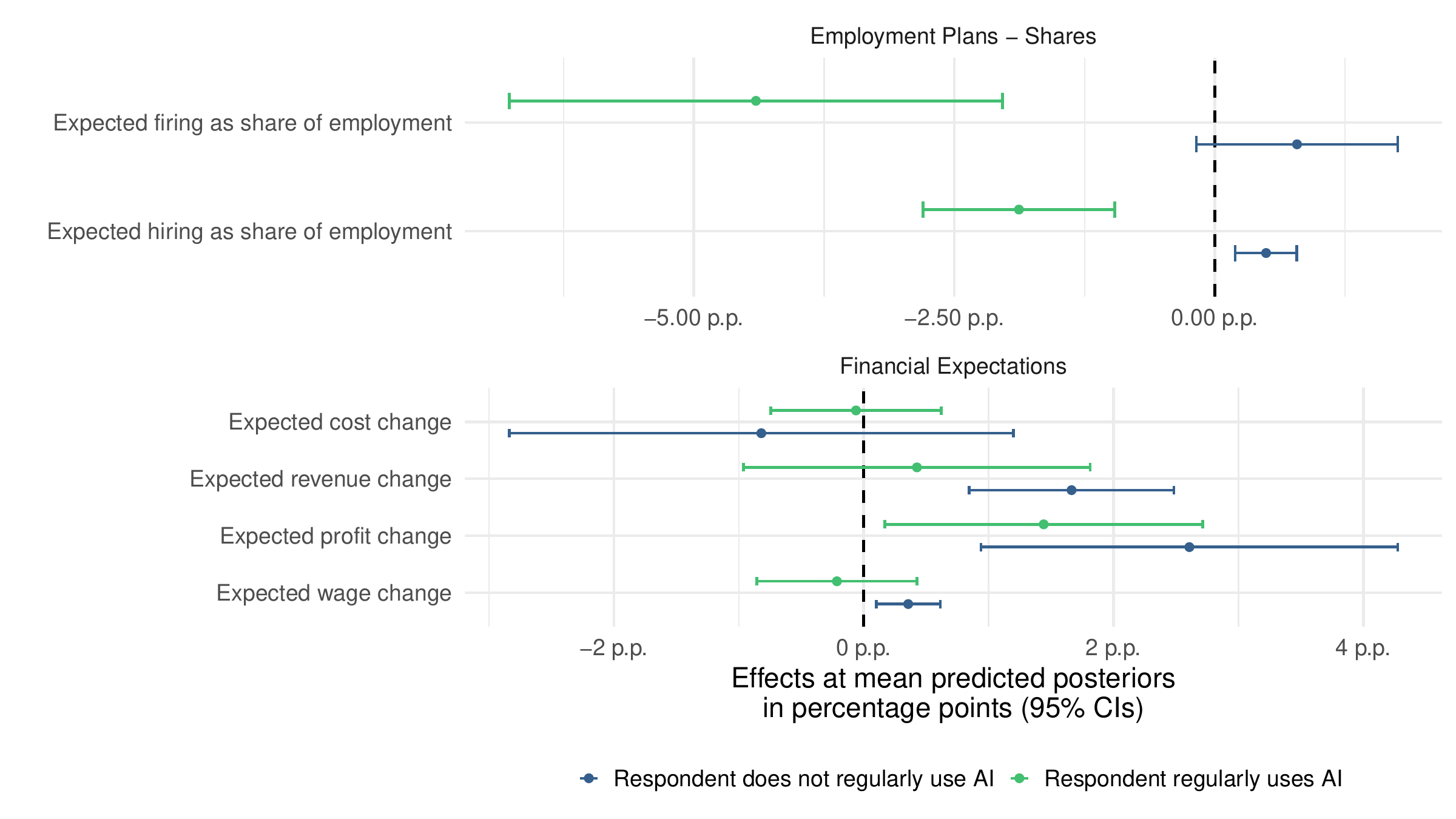}} 
{\secstageupdatepredsub Outcome variables are i) employment and ii) financial expectations, separately for respondents who (do not) regularly use AI.} 
{German Business Panel Tax Advisor Survey 2025.} 
{H} 

In contrast, firms that do not regularly use AI show effects close to zero, suggesting that their employment expectations are largely unaffected. Overall, while not substantial, the pattern is suggestive of “catch-up” effects: firms that have not yet integrated AI perceive greater scope for productivity improvements when prompted by new information.

Figure~\ref{fig:coefplot_heterogeneity_ai} shows that respondents who do \emph{not} yet use AI regularly exhibit larger point estimates for forward-looking financial expectations (bottom panel). This is consistent with the signal being more salient for less informed individuals. However, confidence intervals for both regular and non-regular users overlap substantially, indicating that most differences are not statistically significant.

\customfigure
{Effect Heterogeneity by Firm Size} 
{fig:coefplot_heterogeneity_fs} 
{\includegraphics[width=\textwidth]{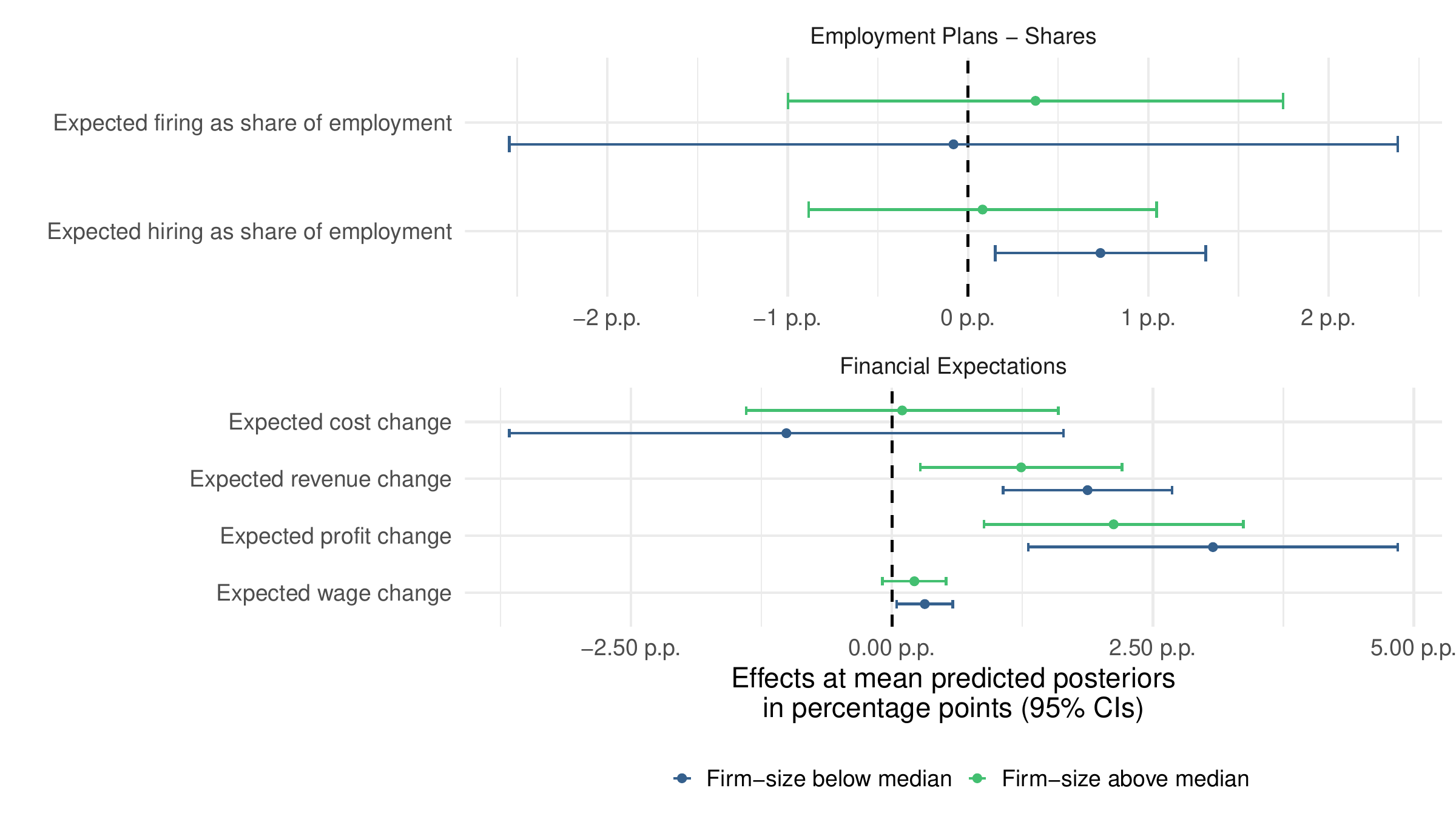}} 
{\secstageupdatepredsub Outcome variables are i) employment and ii) financial expectations, separately for firms below and above the median number of employees.} 
{German Business Panel Tax Advisor Survey 2025.} 
{H} 

Figure ~\ref{fig:coefplot_heterogeneity_fs} shows broadly similar effects of belief updating for small and large firms. In both groups, higher automation beliefs translate into higher expected revenue and profits, and confidence intervals overlap substantially, indicating that the between-group differences are not statistically distinguishable. Only smaller firms display a statistically significant within-group wage effect ($+0.32$ p.p.), but the wage coefficients themselves are also not significantly different from each other. Overall, the data are consistent with positive financial expectations across the size distribution rather than sharply divergent responses.

\customfigure
{Effect Heterogeneity by Respondent Age} 
{fig:coefplot_heterogeneity_age} 
{\includegraphics[width=\textwidth]{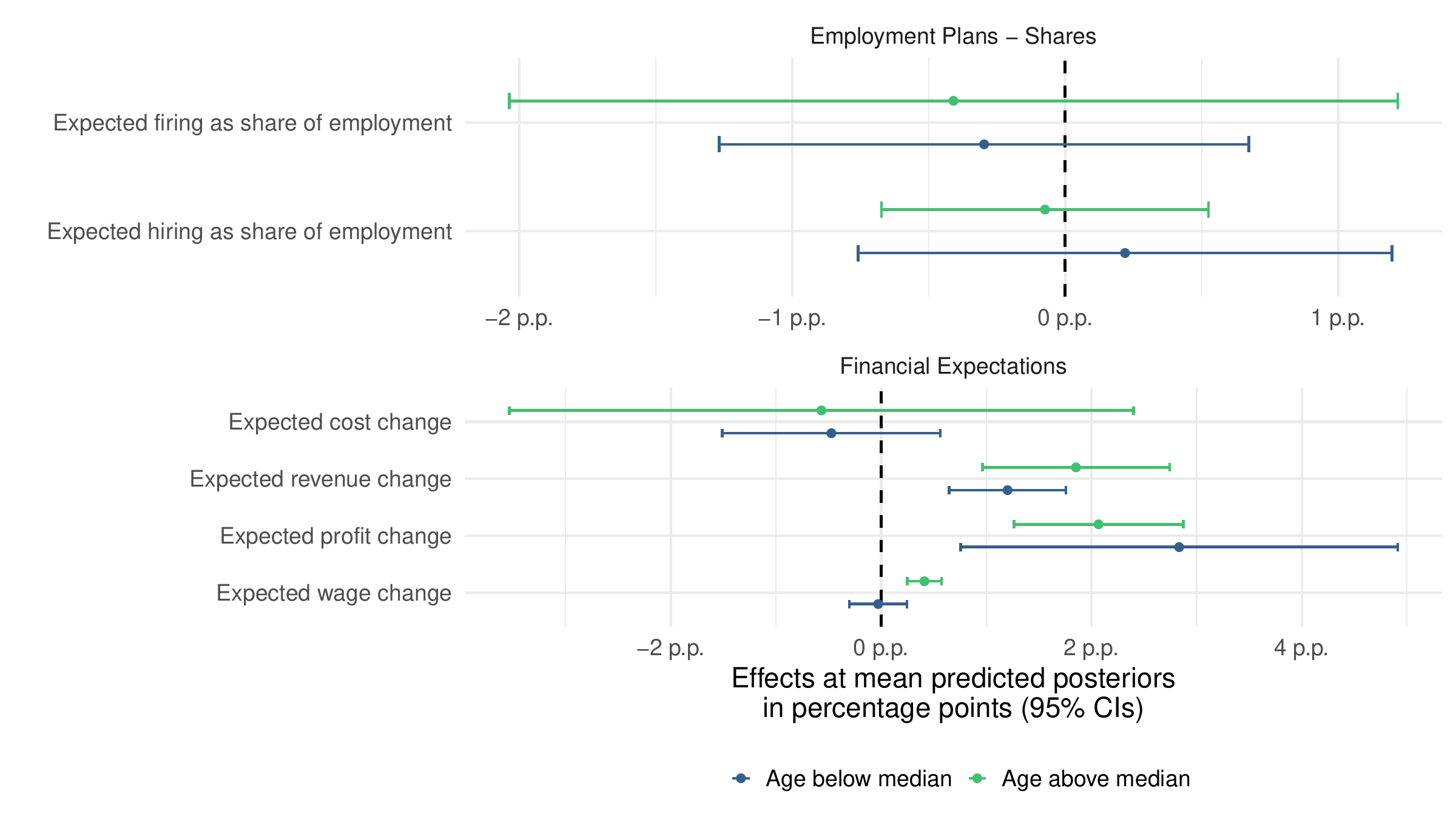}} 
{\secstageupdatepredsub Outcome variables are i) employment and ii) financial expectations, separately for respondents below and above the median age.} 
{German Business Panel Tax Advisor Survey 2025.} 
{H} 

Figure \ref{fig:coefplot_heterogeneity_age} shows similar revenue and profit effects for younger and older respondents. The only clear between-group contrast concerns wages: older respondents anticipate a modest but statistically significant wage increase ($+0.41$ p.p.), whereas the younger group's point estimate is near zero. This is the only difference between the two age-groups that borders   statistical significance.

Taken together, our heterogeneity analyses reveal a remarkably consistent pattern: the effects of automation belief updating on firms' employment and revenue expectations are largely uniform across key subgroups. Whether we split the sample by baseline firm size, or respondent age, the main results remain robust-belief shifts translate into improved financial outlooks but do not trigger immediate changes in hiring or firing plans. While there are small differences in magnitude, particularly with less regular AI users or older respondents, these are neither substantial nor statistically significant for most outcomes. Experience with AI through AI adoption, however, seems to reduce employee turnover.

\section{Conclusions}
\label{sec:conclusion}
Our findings reveal several novel insights into how tax advisory firms perceive and respond to the changed automation expectations. First, we observe a significant gap between initial employer beliefs and expert assessments regarding automation potential. While firms tend to underestimate automation potential, our information intervention successfully prompts belief updating, particularly for lower-skilled occupations like tax clerks and certified tax assistants. For higher-skilled roles such as tax advisors and auditors, belief adjustments are more limited, suggesting that firms perceive greater barriers to automation with higher expertise requirements at the top of the professional hierarchy.

Despite updating their beliefs about automation, firms do not revise their hiring or firing plans, indicating that automation-induced workforce reductions are not a primary concern in the near term. If firms already use AI regularly, there even \textit{decrease} dismissal and hiring plans somewhat. However, firms that update their automation expectations anticipate higher revenue and profit growth, consistent with the notion that efficiency gains from automation may enhance firm performance rather than lead to immediate labor displacement. Interestingly, wage expectations remain largely negligible, implying that anticipated productivity gains are not expected to translate into higher employee compensation in the short run.

Moreover, our results highlight that automation is not perceived as a labor-replacing force but as a driver of job transformation. Firms exposed to updated automation information expect new tasks to emerge, particularly in areas like legal tech, compliance, and AI interaction roles such as prompt engineering. This suggests that while automation reshapes job content, it also creates opportunities for skill development and specialization rather than rendering professional roles obsolete.

These findings contribute to the broader economic literature on automation and labor markets in three key ways. First, they extend existing research beyond manufacturing and manual labor-intensive industries, showing how generative AI might influence a broad spectrum of white-collar occupations. Second, by taking an employer-centered perspective, our study captures anticipatory responses to automation, offering a forward-looking view of technological adaptation and insights into the decision making process of firms more generally. Finally, our results underscore the outreaching impact of automation on task composition, showing that rather than reducing overall employment, automation may shift the skill demands of the workforce in ways that require continued investment in complementary human capital.
\newpage
\phantomsection
\addcontentsline{toc}{section}{References}
\bibliography{literature}
\newpage
\phantomsection
\addcontentsline{toc}{section}{Appendix}
\begin{appendix}
\renewcommand{\thesubsection}{\Alph{subsection}}
\renewcommand{\thetable}{\Alph{subsection}.\arabic{table}}
\renewcommand{\thefigure}{\Alph{subsection}.\arabic{figure}}
\setcounter{subsection}{0} \renewcommand{\thesubsection}{\Alph{subsection}}
\setcounter{figure}{0} \renewcommand{\thefigure}{\Alph{subsection}.\arabic{figure}}
\setcounter{table}{0} \renewcommand{\thetable}{\Alph{subsection}.\arabic{table}}
\label{sec: Appendix}
\subsection{Additional Figures and Tables}

\customfigure
    {Expert Signal on Automatability by Occupation} 
    {fig:expert_signal} 
    {
    \begin{adjustbox}{center}
        \begin{tabular}{cc}
         (a) Tax Clerk & (b) Certified Tax Assistant \\
          \includegraphics[scale=0.33]{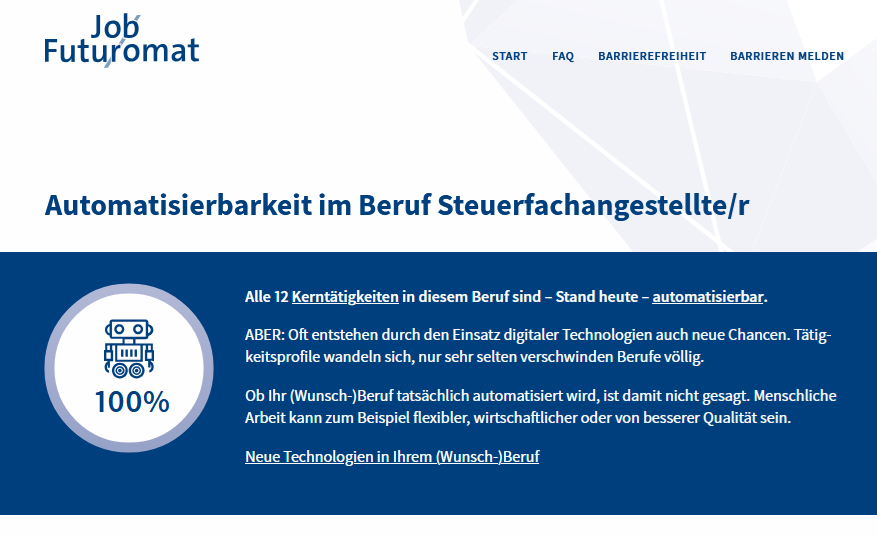} &
         \includegraphics[scale=0.33]{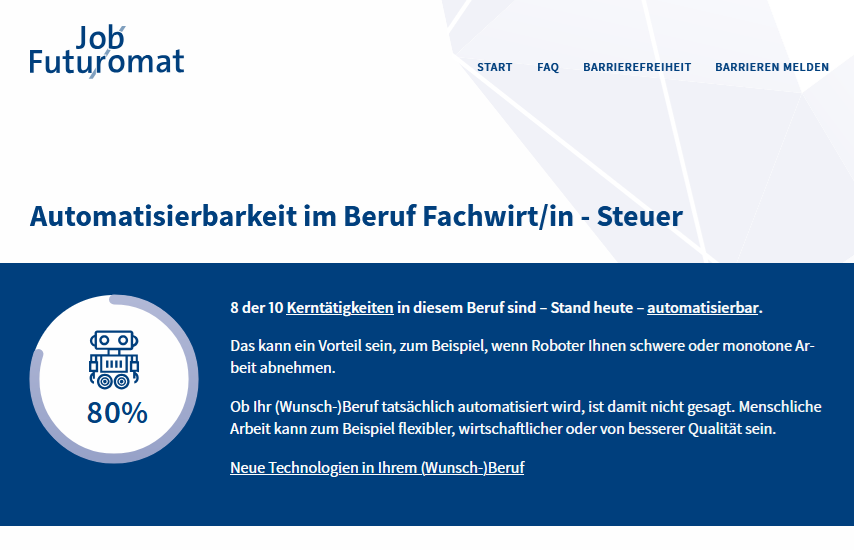}    \\
          (c) Tax Advisor & (d) Auditor \\
          \includegraphics[scale=0.33]{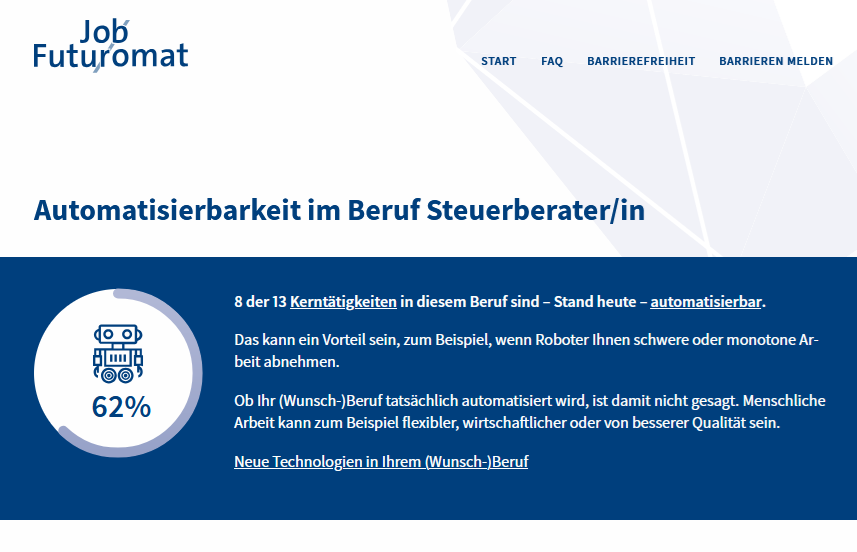} &
         \includegraphics[scale=0.33]{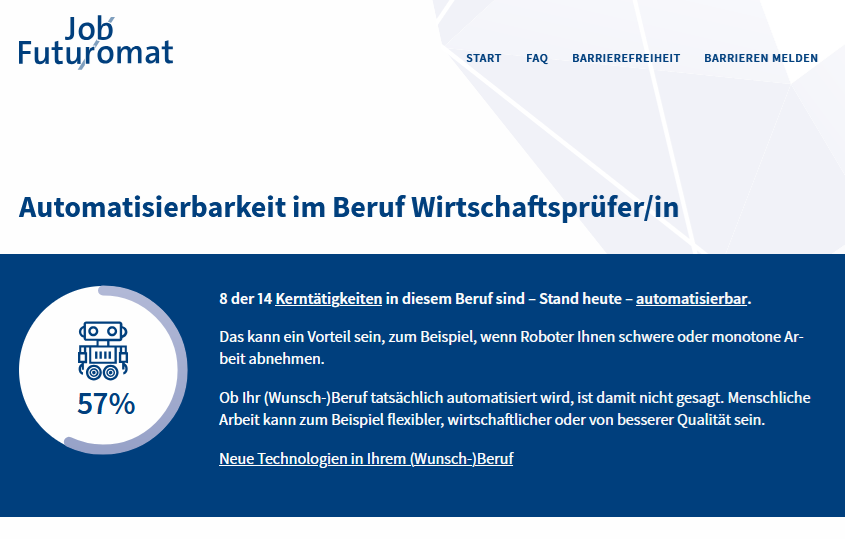}  \\
        \end{tabular}
    \end{adjustbox}
    } 
    {This figure illustrates expert assessments of the automatability of tasks in selected tax-related occupations, based on data from Job-Futuromat, a tool provided by the German Federal Employment Agency. Each panel displays the share of tasks in the given occupation that can be automated with current technologies. The automatability estimates range from 57\% for auditors to 100\% for tax clerks, indicating significant heterogeneity even within a single professional domain.} 
    {Institute for Employment Research (IAB) \href{https://job-futuromat.iab.de/en}{https://job-futuromat.iab.de/en}, 2025.} 
    {H} 

\customfigure
    {Distribution of Firm Revenue and Employment} 
    {fig:firm_distribution} 
    {\includegraphics[width=0.9\textwidth]{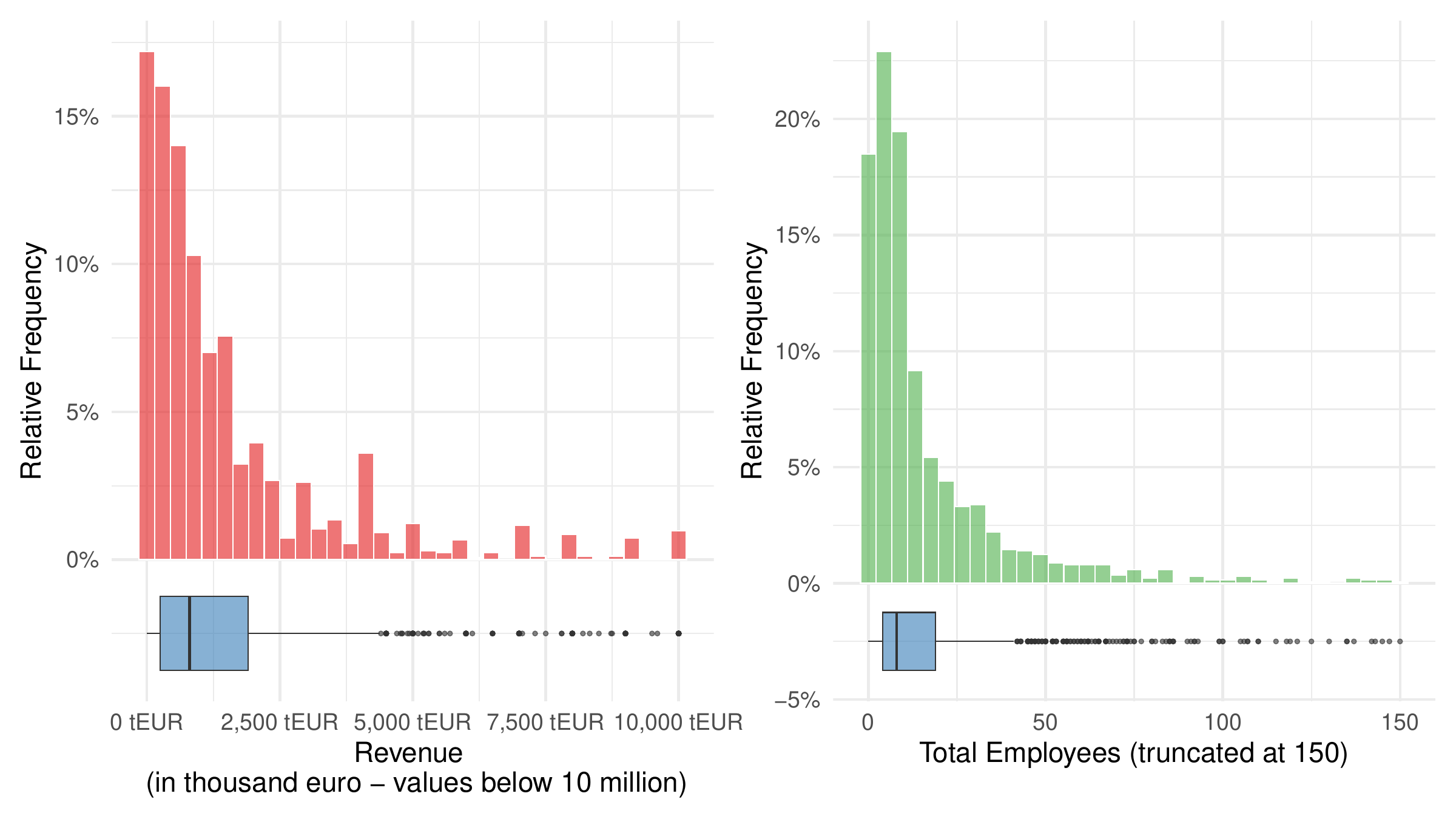}} 
    {This figure presents the distribution of firm revenue (left) and total employees (right) among survey respondents. The revenue distribution is displayed in thousand euros and excludes firms with revenues above 10 million euros. The employee distribution is truncated at 150 employees. Histograms illustrate the relative frequency of firms within each range, while boxplots provide additional insight into the spread and presence of outliers. The distributions confirm the presence of a highly skewed firm size distribution, with most firms being relatively small but a subset of large firms contributing to long right tails.} 
    {German Business Panel Tax Advisor Survey 2025.} 
    {H} 

\customfigure
    {Covariate Balance across Treatment Arms} 
    {fig:balance}              
    {\includegraphics[width=0.9\textwidth]{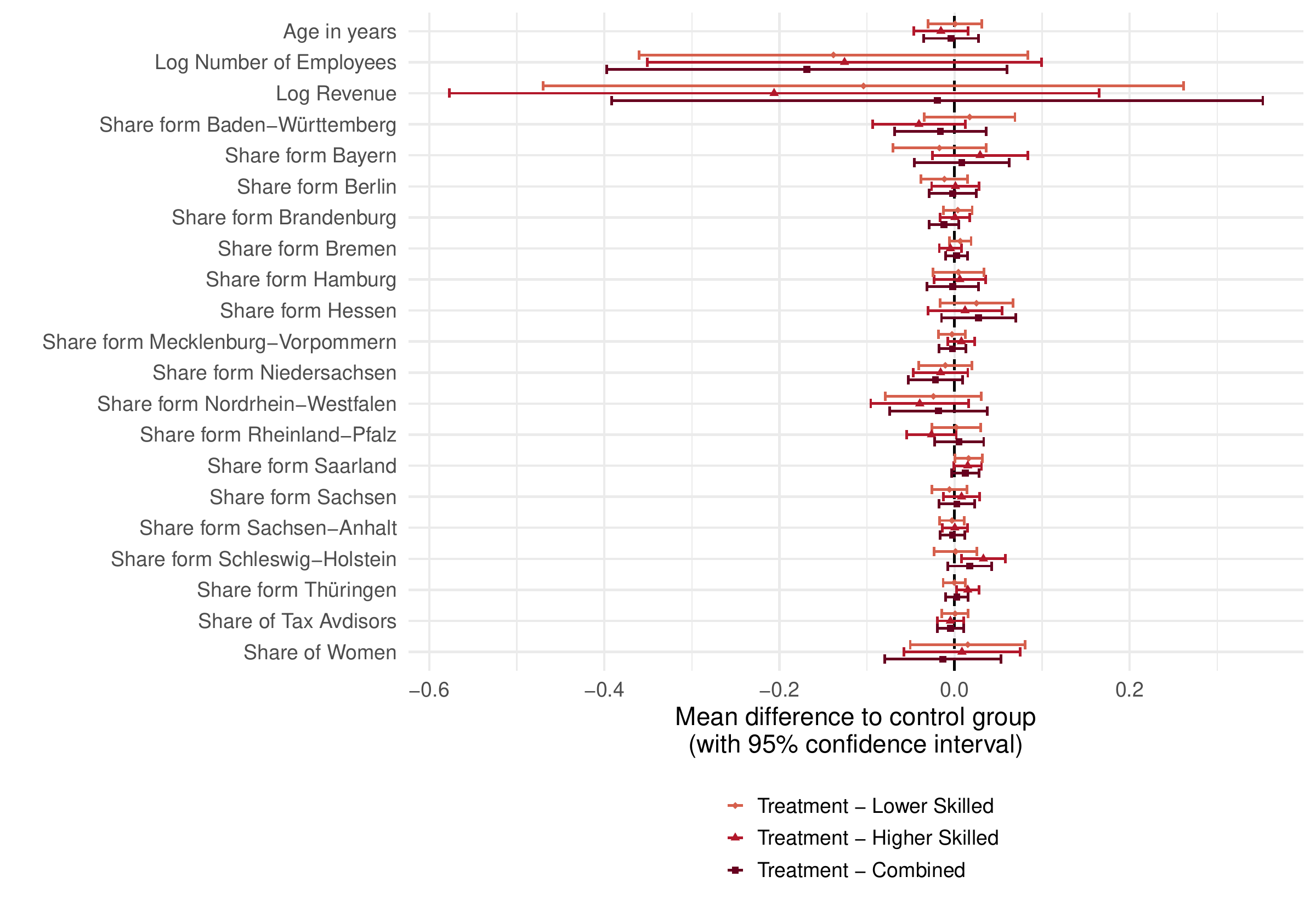}} 
    {This figure displays the mean differences in covariates between each treatment arm and the control group, with whiskers representing 95\% confidence intervals. The results indicate no systematic imbalances across key firm characteristics, including firm size, revenue, regional composition, and workforce demographics, confirming that the randomization was successful.}     
    {German Business Panel Tax Advisor Survey 2025.}            
    {H} 

\customfigure
    {Current Generative AI Use by Firm Size} 
    {fig:aiuse}              
    {\includegraphics[width=0.9\textwidth]{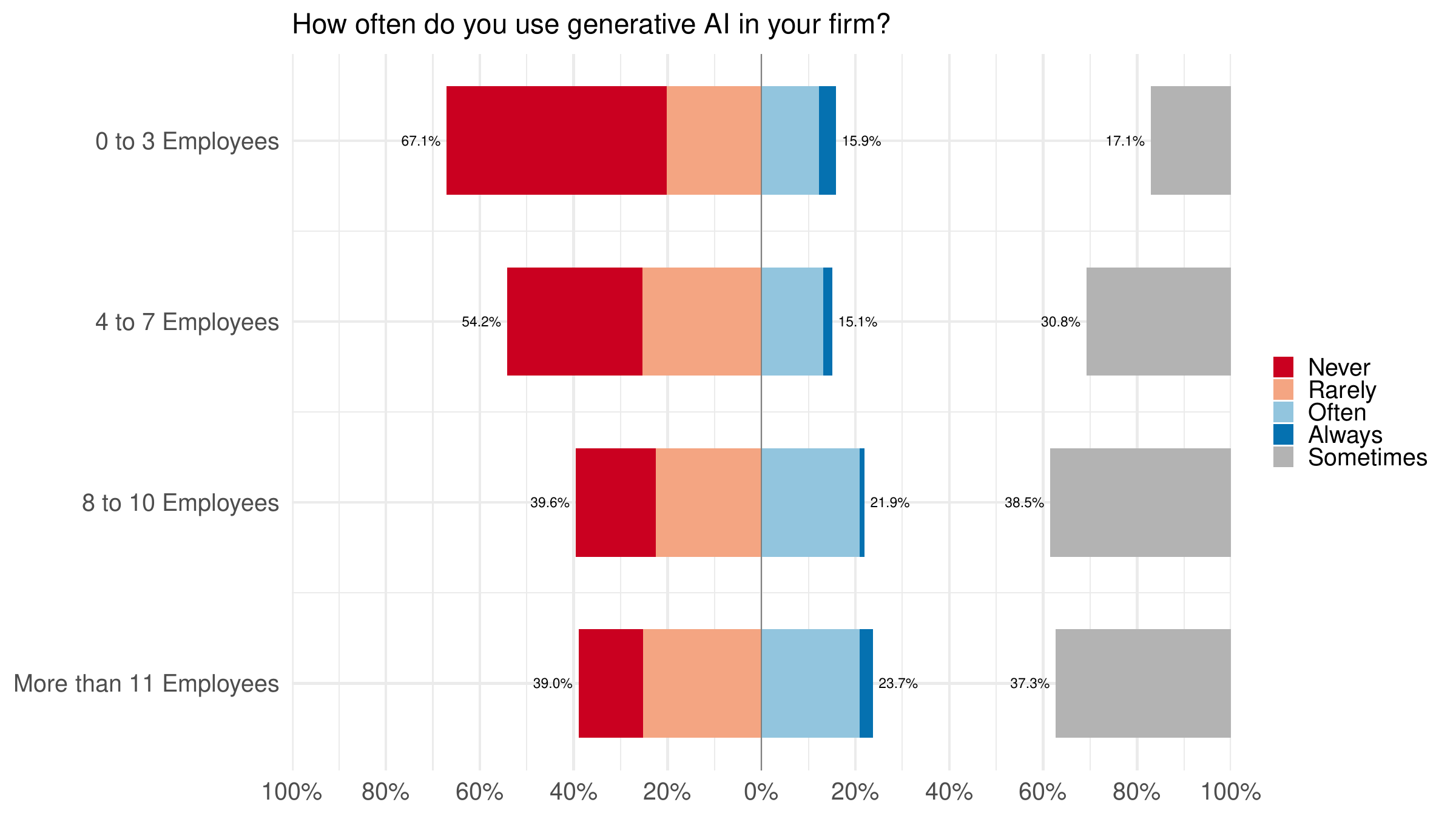}} 
    {This figure shows the frequency of generative AI use across firms of different sizes: 0-3, 4-7, 8-10, and 11-144 employees. Categories include "Never," "Rarely," "Sometimes," "Often," and "Always."}     
    {German Business Panel Tax Advisor Survey 2025.}            
    {H} 

\customfigure
    {Belief Updating by Occupation} 
    {fig:updating_any_all}              
    {\includegraphics[width=0.9\textwidth]{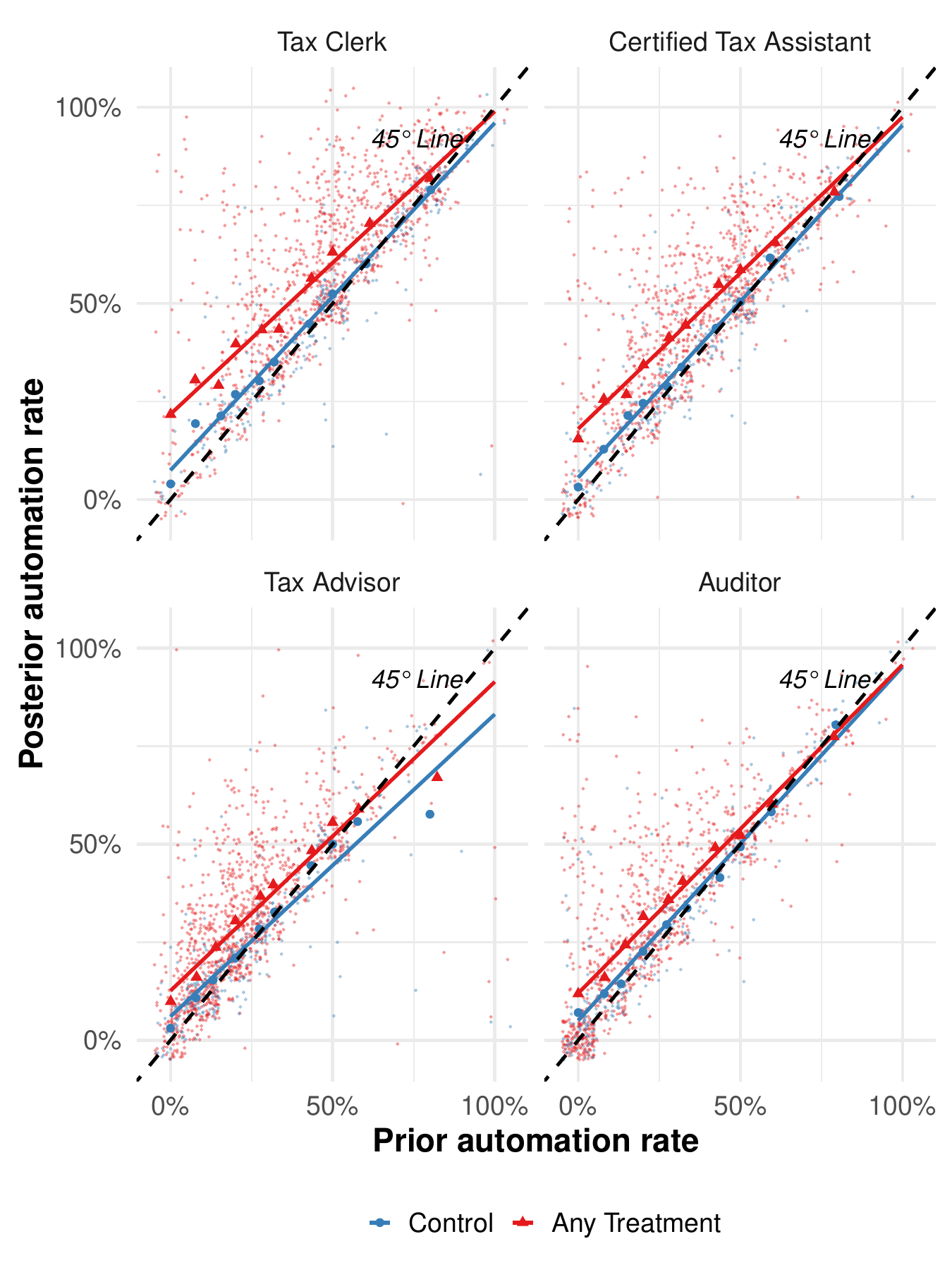}} 
    {This figure illustrates belief updating about automation rates for all four elicited occupations based on regression equation~\eqref{eq:belief_update}, where belief shifts in automation potential are modeled as a function of prior expectations and information treatment exposure. The individual $\beta_o$ and $\gamma_o$ parameters represent the slopes of the red lines. The horizontal axis represents respondents' prior beliefs, and the vertical axis shows their posterior beliefs about automation rates. Light blue dots represent individual values in the control group, while red triangles denote those who receive any treatment. Larger, darker markers indicate averages for 10 quantile bins within each group. The dashed 45-degree line represents no belief updating.}     
    {German Business Panel Tax Advisor Survey 2025.}            
    {H} 

\input{tab/tab_first_stage_by_occ}
\input{tab/tab_first_stage}

\FloatBarrier
\newpage
\subsection{Identification of Effect of Automatability Beliefs on Outcomes}
\label{appendix:theory}

Our information intervention shifts respondents' beliefs about the automation potential of specific occupations. To interpret these belief shifts as causal in a two-stage least squares framework, we rely on a simple Bayesian learning model and derive a Wald estimator that recovers the causal effect of beliefs on outcomes. We show why this is the case, using a single binary treatment and single prior/posterior pair for notational simplicity. However, the logic extends directly to multiple treatment arms and multiple prior/posterior belief pairs. Each treatment arm provides a different signal (e.g., about high-skilled or low-skilled occupations), leading to updates within participants' beliefs that can be written as a convex combination of the prior belief and the signal. The aim is to identify the causal effect \( \beta_1 \).

\paragraph{Bayesian Belief Updating.}
Let respondents $i$ begin with a prior belief, \( \text{prior}_i \), about the automatability of an occupation. The information treatment provides a signal, \( \text{signal}_i \), an expert assessment from the Job-Futuromat. If respondents update rationally in a Bayesian manner, the posterior belief, \( \text{posterior}_i \), reflects a convex combination of prior and signal:
\[
\text{posterior}_i = (1 - a_i)\,\text{prior}_i + a_i\,\text{signal}_i,
\]
where \( a=\mathbb{E}[a_i], a_i \in [0,1] \) denotes the learning rate: the weight placed on the new information. This can be rewritten as $\text{update}_i = a_i\,(\text{signal}_i-\text{prior}_i)$.

\paragraph{First Stage.}
To estimate belief updating empirically, we regress the posterior on the prior, treatment, and their interaction:
\begin{equation}\label{eq: first stage}
\text{posterior}_i = \alpha_0 + \alpha_1\,\text{prior}_i + \delta\,D_i + \gamma\,D_i \times \text{prior}_i + \epsilon_i,
\end{equation}

where \( D_i \) is the treatment indicator. If \( D_i = 0 \) respondents do not receive a signal and thus have no reason to change their beliefs. Their posteriors equal their priors on average, implying \( \alpha_0 = 0 \) and \( \alpha_1 = 1 \) for the pure control group. Under these assumptions and \( D_i =1 \), we can compare coefficients:
\[
\delta = a \cdot \mathbb{E}[\text{signal}_i], \qquad  \gamma = -a.
\]

\paragraph{Reduced Form.}
We use the exogenous variation in posteriors induced by the information treatment to estimate the causal effect of belief changes on outcomes. We write the second stage as:
\begin{equation}\label{eq: second stage}
\text{outcome}_i = \beta_0 + \beta_1\,\text{posterior}_i + \nu_i,
\end{equation}
where \( \text{posterior}_i \) is the posterior belief formed after treatment.

Inserting equation~\eqref{eq: first stage} into~\eqref{eq: second stage} gives the reduced form
\begin{equation}\label{eq: reduced form}
      \text{outcome}_i
      = \pi_0
        + \pi_1\,\text{prior}_i
        + \pi_2\,D_i
        + \pi_3\,(\text{prior}_i\times D_i)
        + \upsilon_i.
\end{equation}

Testable restrictions are $\hat{\pi}_3=-\hat{\pi}_2 / \mathbb{E}[\text{signal}_i]$ and $\hat{\pi}_1/\hat{\alpha}_1=\hat{\pi}_2/\hat{\delta}=\hat{\pi}_3/\hat{\gamma}$. With the pure control group parameters \( \alpha_0 = 0 \) and \( \alpha_1 = 1 \), equation~\eqref{eq: reduced form} also identifies $\beta_1$, since $\pi_0 = \beta_0 + \beta_1\,\alpha_0, \pi_1 = \beta_1\,\alpha_1$.

\paragraph{Wald Estimator.}
To construct the Wald estimator, we consider differences in expectations across treatment for individuals $i$ in the treatment group and control respondents $j$. The average priors in both groups are statistically equal, such that $\mathbb{E}[\text{prior}_i\mid D_i=1]=\mathbb{E}[\text{prior}_j\mid D_j=0]$, since groups are randomized:
\begin{align}
\Delta D &= \mathbb{E}[\text{posterior}_i \mid D_i=1] - \mathbb{E}[\text{posterior}_j \mid D_j=0] \nonumber\\
&= a\mathbb{E}[\text{signal}_i \mid D_i=1] -a\mathbb{E}[\text{prior}_i\mid D_i=1] \nonumber\\
&= \delta + \gamma \cdot \mathbb{E}[\text{prior}_i]=\mathbb{E}[\text{update}_i],\nonumber
\end{align}
\begin{align}
\Delta Y &= \mathbb{E}[\text{outcome}_i \mid D_i=1] - \mathbb{E}[\text{outcome}_j \mid D_j=0]\nonumber\\
&= \beta_1 (\delta + \gamma \cdot \mathbb{E}[\text{prior}_i])=\beta_1\;\mathbb{E}[\text{update}_i].\nonumber
\end{align}

Taking the ratio gives the Wald grouping estimator:
\[
\beta_{WG} = \frac{\Delta Y}{\Delta D} = \beta_1.
\]

While posterior beliefs and outcome expectations may both be influenced by baseline characteristics such as ability or experience, we do not need to control for them explicitly. This is because we observe each respondent's prior belief, \( \text{prior}_i \), immediately before treatment. That is, any effect of baseline characteristics on posteriors is mediated through the prior. Conditional on \( \text{prior}_i \), the treatment-induced change in posterior beliefs is exogenous, since it only depends on the exogenously given signal.

\paragraph{Interpretation and Identification.}
This approach identifies the causal effect of belief changes on outcome expectations (\( \beta_1 \)), even when posterior beliefs and outcomes are both correlated with unobserved baseline characteristics such as ability. The key is that we observe the prior belief \( \text{prior}_i \) immediately before treatment, and can condition on it directly in the reduced form. In the Wald estimator, the prior drops out. Since the treatment is randomly assigned and the prior captures pre-treatment heterogeneity, the variation in posteriors induced by the treatment is exogenous. No additional controls are needed in either stage. Identification further requires that the information signal is constant as in our application (or has the a constant expectation across treated respondents). The exclusion restriction requires that the treatment affects outcomes only through its effect on posterior beliefs, which is unlikely to be violated, since information that affects treatment participants differently is not likely to occur within the typically short completion time of the survey. If the distributions of the prior and posterior are different, for example from asking different questions for each, measurement error can bias first-stage and reduced from estimates strongly.

Supportive evidence for the validity of the approach is if the estimate for $\beta_1$ from the reduced from corresponds to the estimate of $\beta_1$ from the IV approach.

\FloatBarrier
\newpage 
\subsection{Reduced Form Estimates\label{Reduced Form Estimates}}

\customfigure
    {Comparing Reduced Form and IV Estimates: Main Outcomes} 
    {fig:rf_coef_main}                                       
    {\includegraphics[width=0.9\textwidth]{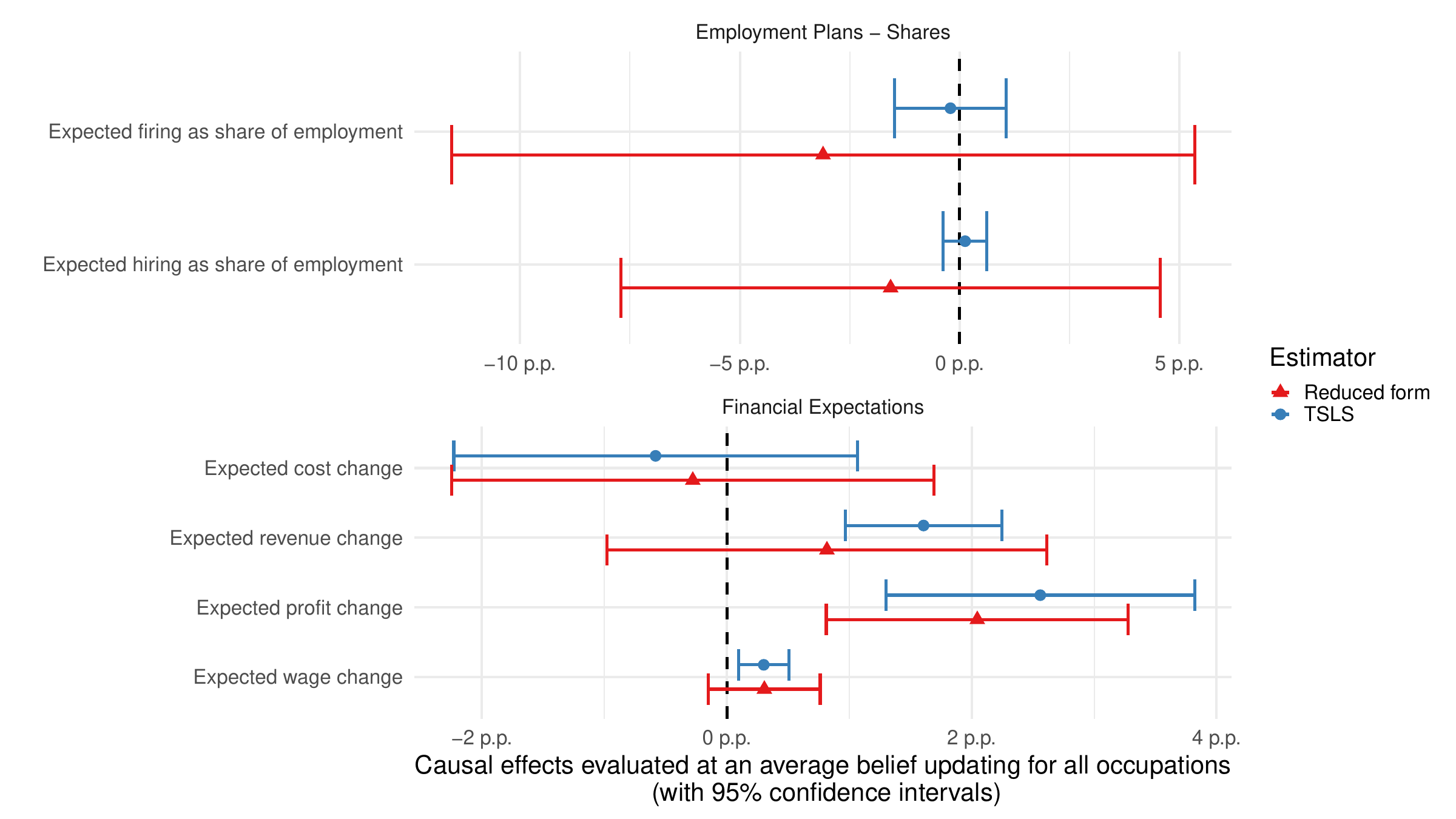}} 
    {Coefficient plots contrasting reduced-form and 2SLS (IV) effects of information-induced belief updating on employment and financial expectations. \redformIV} 
    {German Business Panel Tax Advisor Survey 2025.}          
    {H} 

\customfigure
    {Comparing Reduced Form and IV Estimates: Outcomes by Occupation} 
    {fig:rf_coef_rev_by_occ}                                
    {\includegraphics[width=0.9\textwidth]{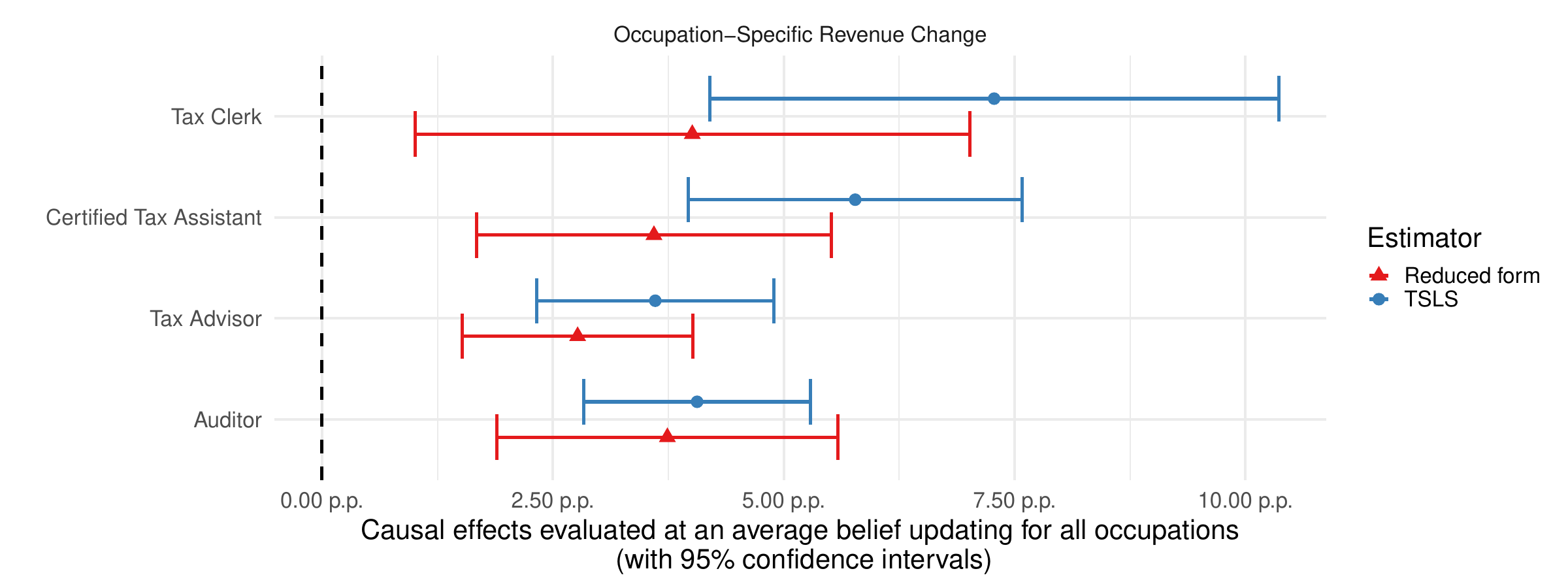}} 
    {Reduced-form vs. 2SLS (IV) estimates on revenue per hour by occupation group. \redformIV} 
    {German Business Panel Tax Advisor Survey 2025.}         
    {H}

\customfigure
    {Comparing Reduced Form and IV Estimates: Automation-Potential Dummies} 
    {fig:rf_coef_automation}                              
    {\includegraphics[width=0.9\textwidth]{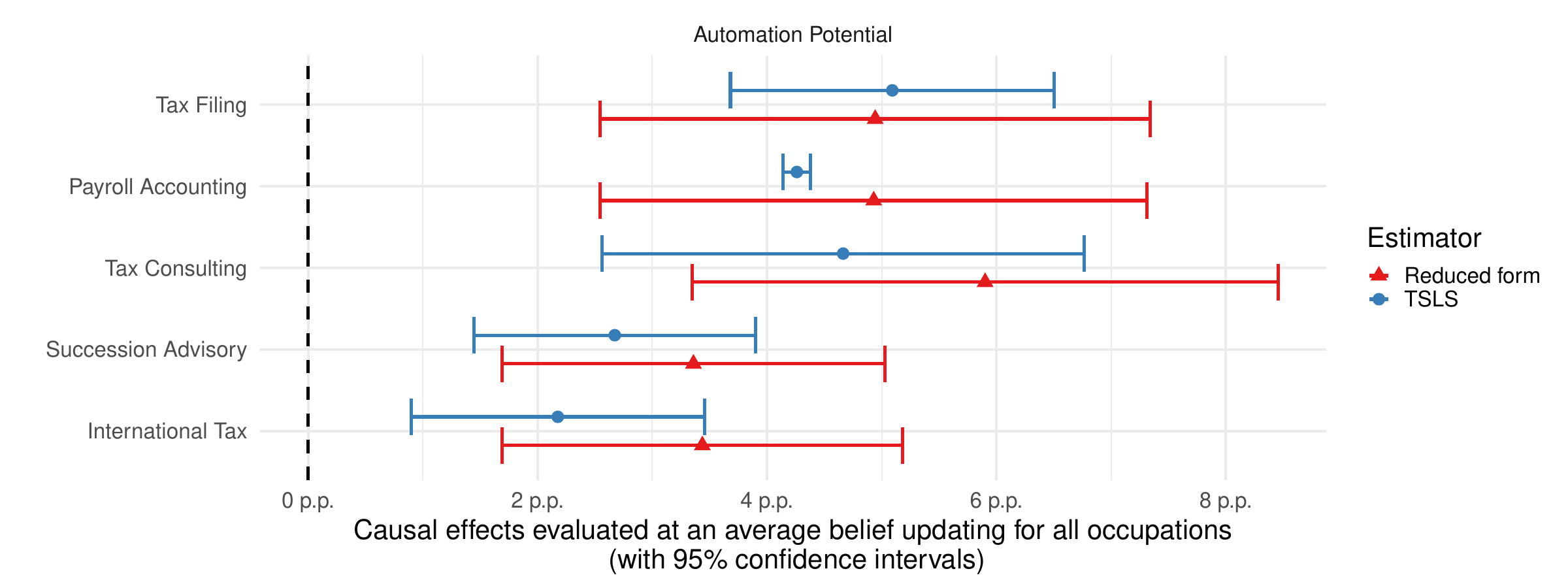}} 
    {Comparison of reduced-form and 2SLS (IV) effects of belief updating on automation potential dummies. \redformIV} 
    {German Business Panel Tax Advisor Survey 2025.}        
    {H}

\customfigure
    {Comparing Reduced Form and IV Estimates: New Task Dummies} 
    {fig:rf_coef_new_tasks}                               
    {\includegraphics[width=0.9\textwidth]{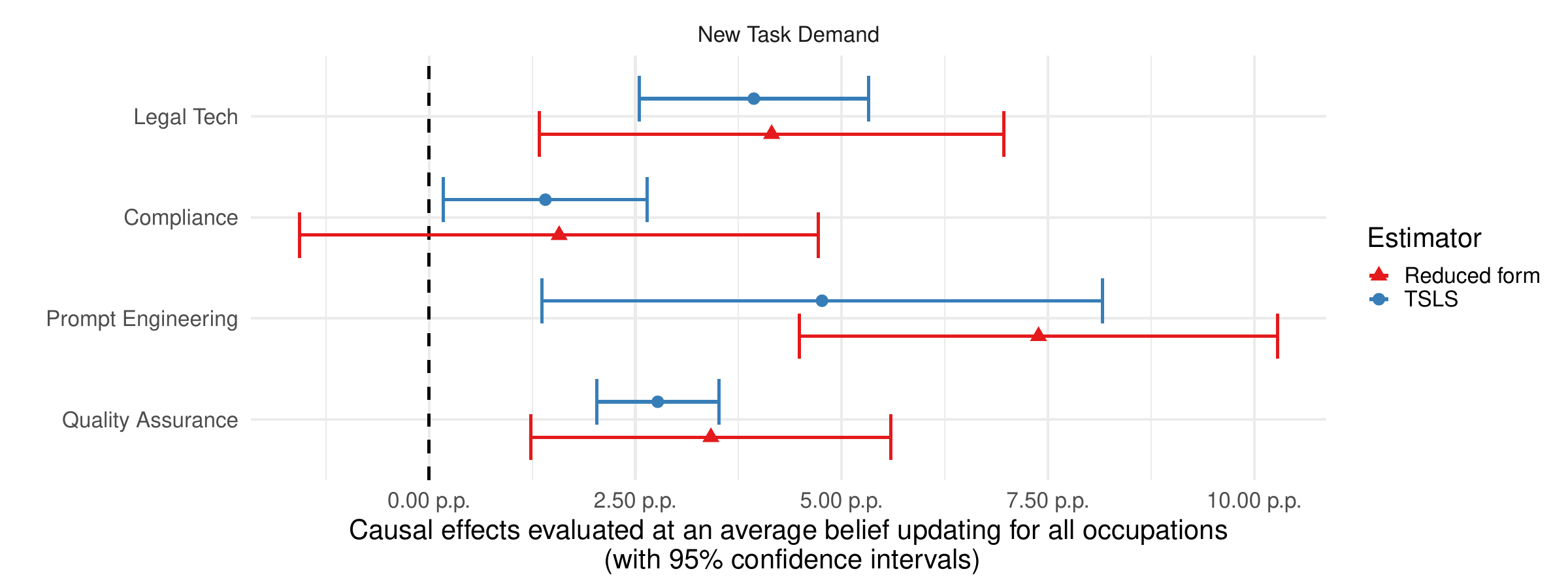}} 
    {Reduced-form and 2SLS (IV) estimates for outcomes related to the introduction of new tasks due to automation. \redformIV} 
    {German Business Panel Tax Advisor Survey 2025.}        
    {H}

\customfigure
    {Comparing Reduced Form and IV Estimates: Training and Skill Investment} 
    {fig:rf_coef_training}                                
    {\includegraphics[width=0.9\textwidth]{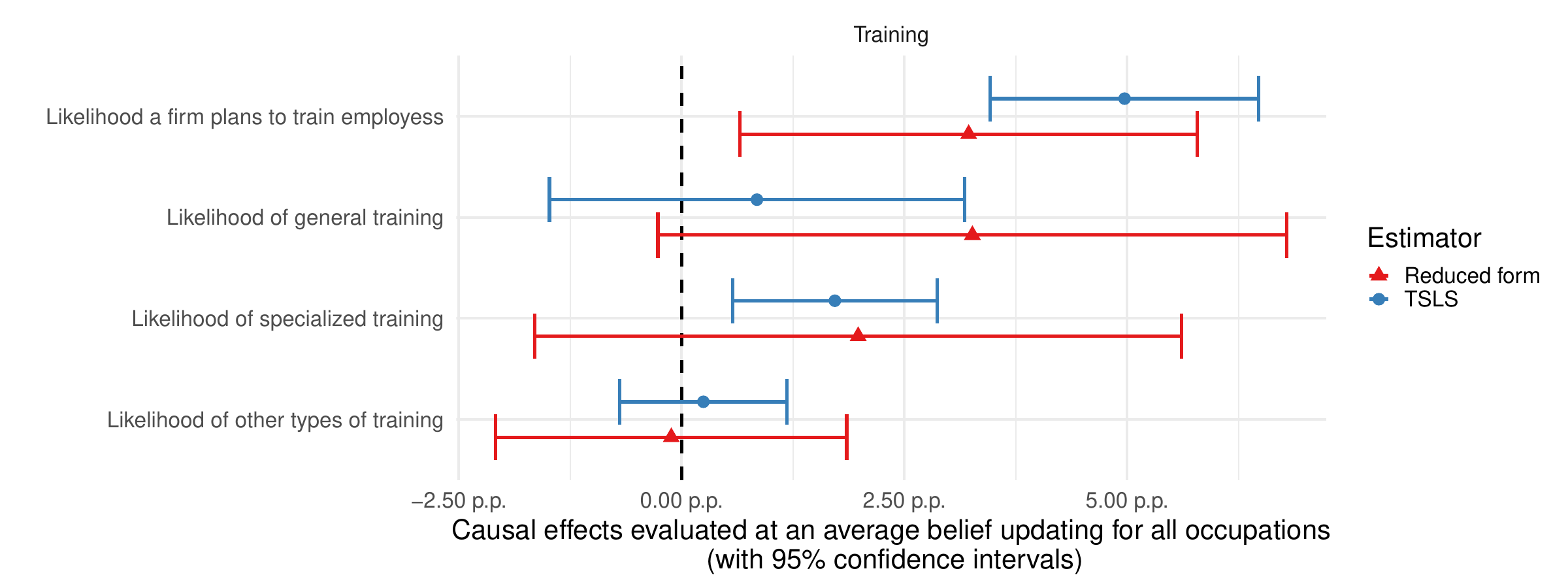}} 
    {Coefficient plots comparing reduced-form and 2SLS (IV) effects of belief updating on training and upskilling outcomes. \redformIV} 
    {German Business Panel Tax Advisor Survey 2025.}        
    {H}

\customfigure
    {Comparing Reduced Form and IV Estimates: Attitudes and Career Plans} 
    {fig:rf_coef_attitude}                                
    {\includegraphics[width=0.9\textwidth]{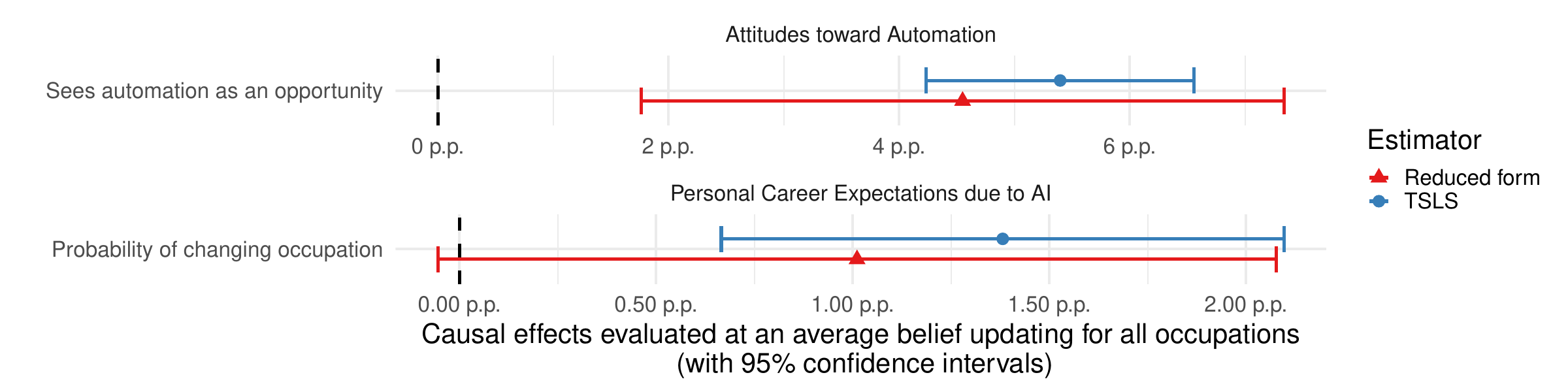}} 
    {Coefficient plots comparing reduced-form and 2SLS (IV) effects of belief updating on attitudes and career plans. \redformIV} 
    {German Business Panel Tax Advisor Survey 2025.}        
    {H}
\customfigure
    {Comparing Reduced Form and IV Estimates: AI Adoption Intentions and Actions} 
    {fig:rf_coef_action}                                
    {\includegraphics[width=0.9\textwidth]{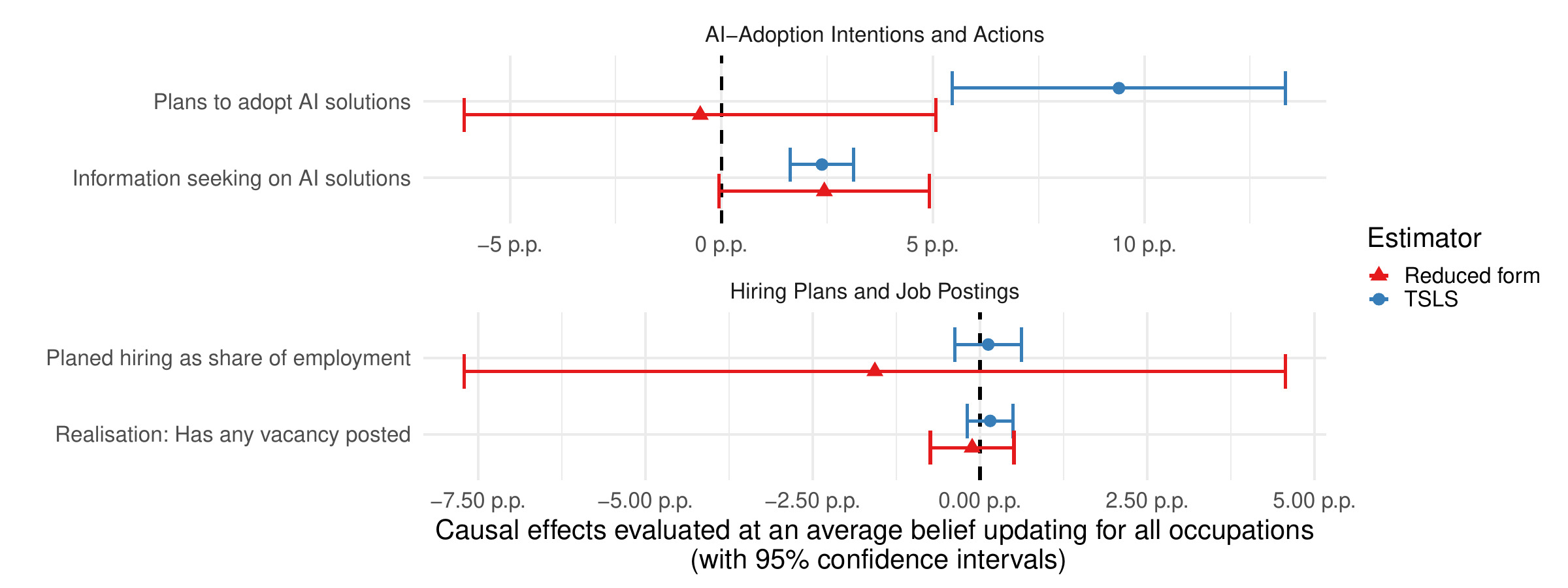}} 
    {Coefficient plots comparing reduced-form and 2SLS (IV) effects of belief updating on AI adoption intentions an actions. \redformIV} 
    {German Business Panel Tax Advisor Survey 2025.}        
    {H}

\FloatBarrier
\newpage
\subsection{Data Collection, Filtering, and Cleaning}
\label{appendix:filtering}

The survey was distributed via the survey software provider Qualtrics through the German Business Panel infrastructure \citep{Bischof2025}. Upon the survey invitation, participants are asked to answer the survey questions carefully and are assured that their participation is voluntary. We inform them that the survey should take (on average) 10 minutes.

In order to facilitate broad participation, the survey was conducted online and designed to be accessible across multiple devices, including desktop computers, tablets, and smartphones. The interface was optimized for both large and small screens to ensure a high user experience across different devices.

To manage outreach effectively, we distributed the survey in weekly batches starting in November 2024. Each respondent received up to three reminders if they had not completed the survey: the first sent one week after the initial invitation, the second two weeks later and the third one month later. This was followed by a thank you email, when finishing the survey. The median survey completion time was 693 seconds (approximately 11.55 minutes), which aligns well with the expected 10-minute duration.

While the overall data quality is high and the survey is representative of the target population we intend to capture, some filtering and cleaning was still needed. Across all filtering steps, our goal was to ensure that only active tax advisory firms remained in the dataset, while excluding respondents who are in the professional register but do not operate in the relevant business segment. 

First, we screened for information in open occupation and legal form fields in our survey to excluded respondents whose occupations (e.g., retirees, university professors) or organizational roles (e.g., heads of large corporate tax departments) did not match with the target population.

Second, we applied revenue plausibility checks. Firms reporting revenue below 25,000 EUR were excluded, as such values indicate economic inactivity.\footnote{These respondents were typically beyond the typical retirement age for tax advisors, reflecting the ability of registered advisors to maintain small advisory roles past retirement.} Likewise, firms with revenues exceeding 15 million EUR or unusually high revenue per employee were flagged for manual review. Many of these cases involved corporate tax departments of firms in other economic sectors rather than independent tax firms, introducing potential bias. Where open responses or contact details confirmed this, we excluded them from the sample.

In addition, respondents who reported employment figures above 150 or revenues above 10 million Euros at their firms were reviewed, as they likely represented outliers compared to typical tax firms. Interestingly, most of these respondents are working at large international auditing firms. While we include these observations in the survey, we checked all of the analyses only for smaller firms, since we only have 148 observations for these larger firms. 

Finally, we excluded a small number observations with obviously erroneous answers. These included only a small number respondents who stated that they fire more than 100\% of their workforce, expect revenue and cost decreases of more than 100\% or cross-referenced appointment year data with the appointment years stated in the register data, removing cases where reported appointment years deviated by more than five years. 

\subsection{Robustness}\label{appendix:robusntness}

\customfigure
{Effects on Employment Plans and Financial Expectations: Sole Proprietorship vs. Other Legal Forms} 
{fig:coefplot_heterogeneity_sp} 
{\includegraphics[width=\textwidth]{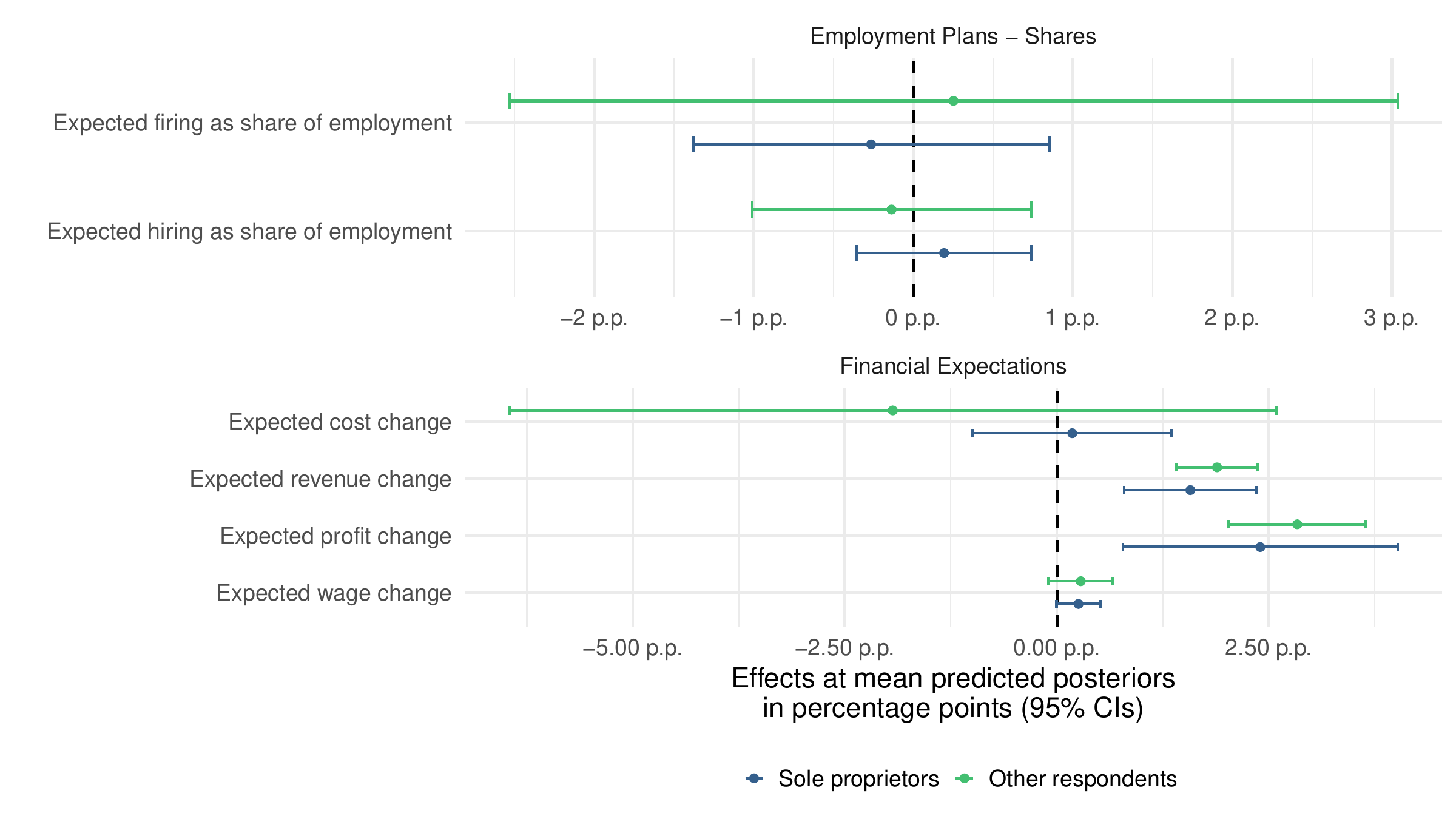}} 
{\secstageupdatepredsub Outcome variables are i) employment plans and ii) financial expectations, estimated separately for sole proprietors and other respondents.} 
{German Business Panel Tax Advisor Survey 2025.} 
{H} 

\customfigure
{Effects on Employment Plans and Financial Expectations: Only Early Responses vs. Responses after Reminder} 
{fig:coefplot_heterogeneity_reminder} 
{\includegraphics[width=\textwidth]{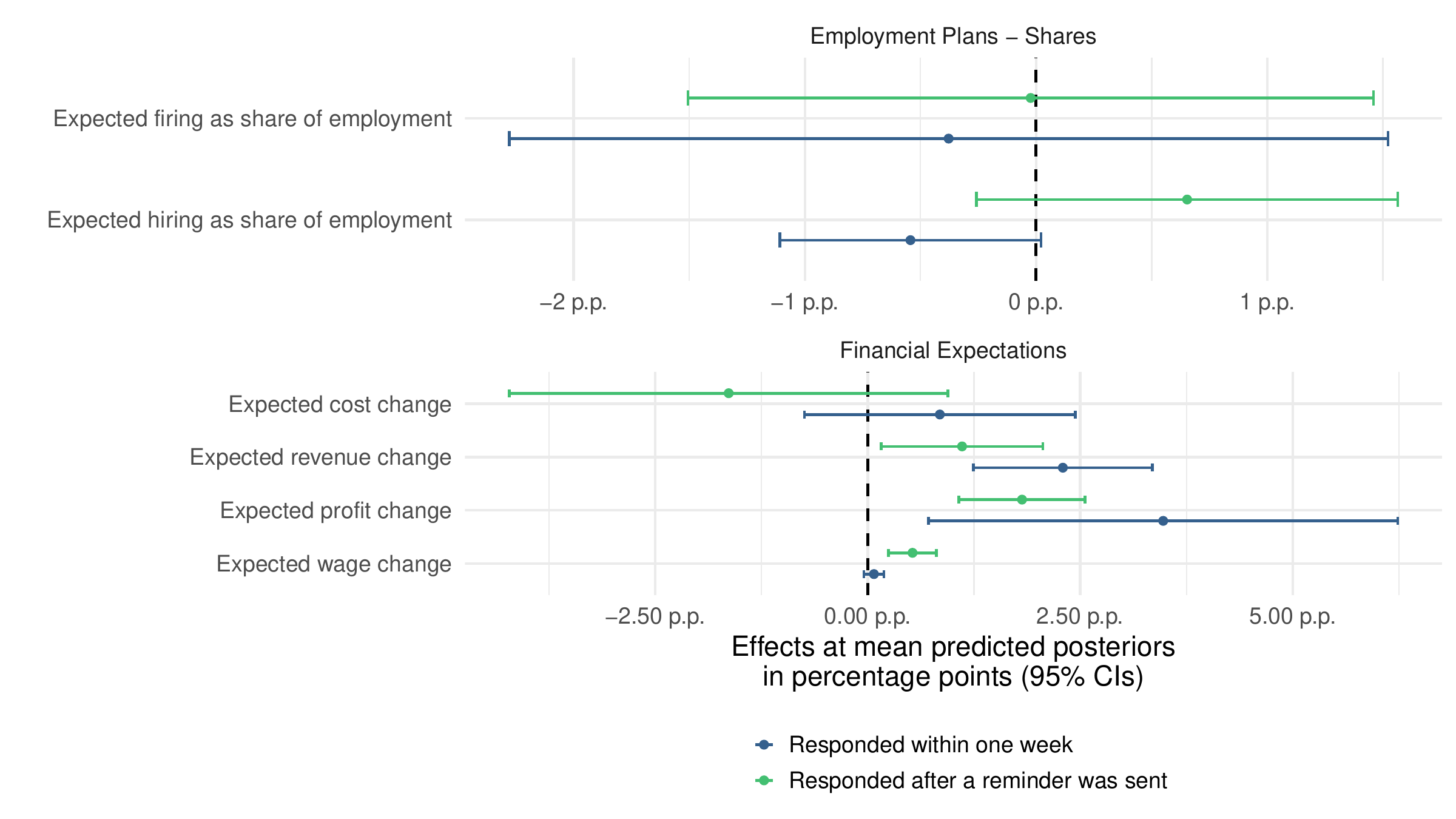}} 
{\secstageupdatepredsub Outcome variables are i) employment plans and ii) financial expectations, estimated separately for respondents who answered within one week after the initial invitation and those who responded after receiving a reminder.} 
{German Business Panel Tax Advisor Survey 2025.} 
{H} 

\customfigure
{Effects on Employment Plans and Financial Expectations: Only Fast Survey Completion vs. Slow Completion} 
{fig:coefplot_heterogeneity_duration} 
{\includegraphics[width=\textwidth]{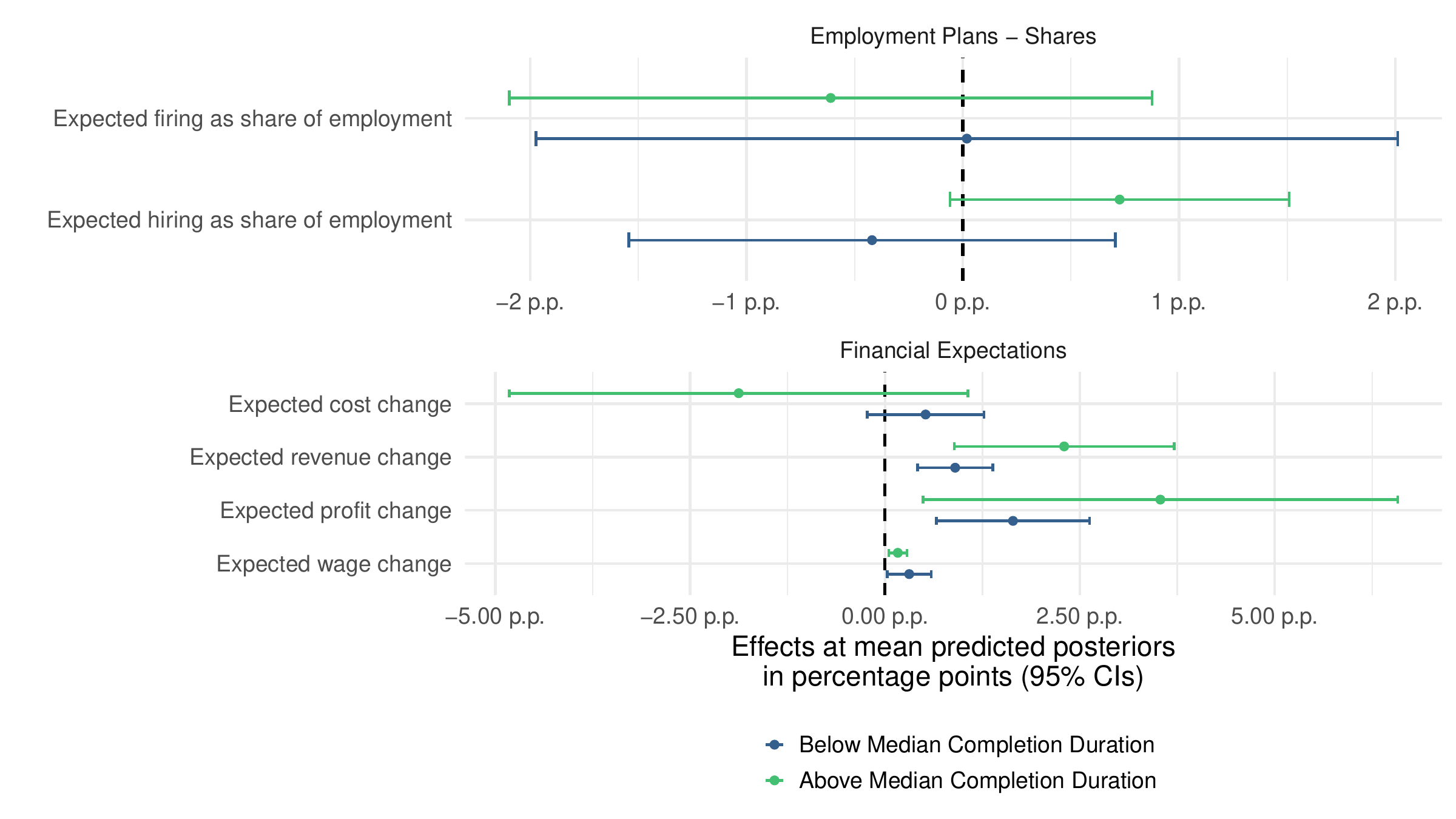}} 
{\secstageupdatepredsub Outcome variables are i) employment plans and ii) financial expectations, estimated separately for respondents with below- and above-median survey completion duration.} 
{German Business Panel Tax Advisor Survey 2025.} 
{H} 

\end{appendix}
\begin{appendices}

\newpage
\vspace*{\fill}
\begingroup
\begin{center}
{\textbf{\LARGE Supplementary Appendix}}
\end{center}
\endgroup
\vspace*{\fill}
\setcounter{page}{0}
\thispagestyle{empty} 

\numberwithin{equation}{section}\setcounter{equation}{0}
\numberwithin{figure}{section}\setcounter{figure}{0}
\numberwithin{table}{section}\setcounter{table}{0}

\newpage 

\clearpage
\newpage
\section{The German Tax Advisory Industry}

\customfigure
    {Labor Market Trends in the Tax Advisory Industry} 
    {fig:trends} 
    {\includegraphics[width=\textwidth]{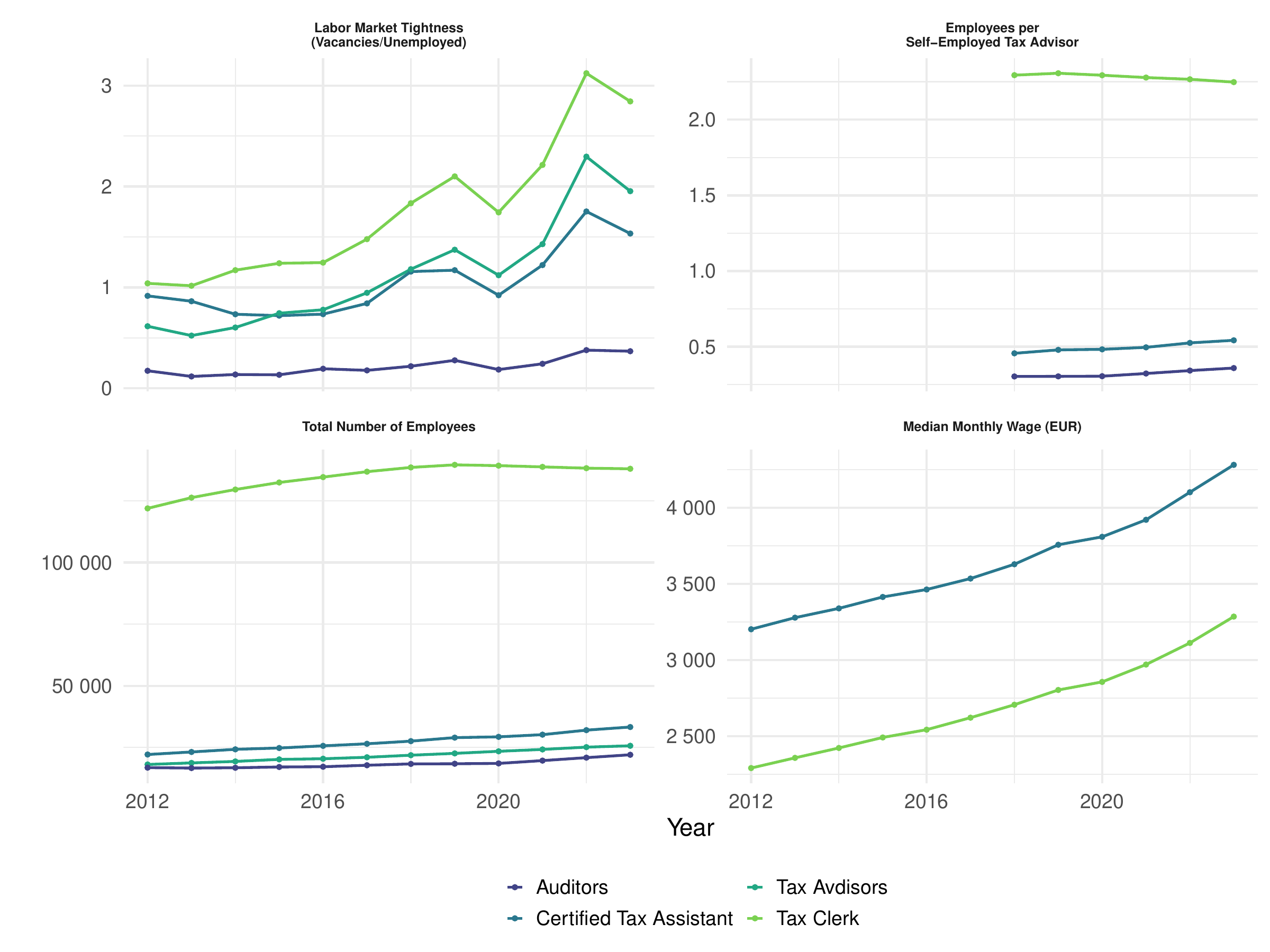}} 
    {Labor market indicators for selected occupations in the German tax advisory sector, 2012–2023. The total number of employees does not include self-employed tax advisors or auditors. Employees per self-employed tax advisor is only available from 2018 onward, as the Chamber of Tax Advisors began publishing statistics on the number of self-employed at that time. Median monthly wages (EUR) are based on social security records and may understate actual earnings for tax advisors and auditors, since many employees in these occupations earn above the social security contribution ceiling (Beitragsbemessungsgrenze).} 
    {German Federal Employment Agency; Statistics of the Chamber of Tax Advisors; Own calculations.} 
    {H} 

\customfigure
    {Comparison of Firm Size Measures from Survey Data and Orbis} 
    {fig:compare_orbis_survey} 
    {
    \begin{adjustbox}{center}
        \begin{tabular}{cc}
         (a) Employees in Survey vs. Orbis & (b) Log Revenue in Survey vs. Orbis \\
          \includegraphics[scale=0.17]{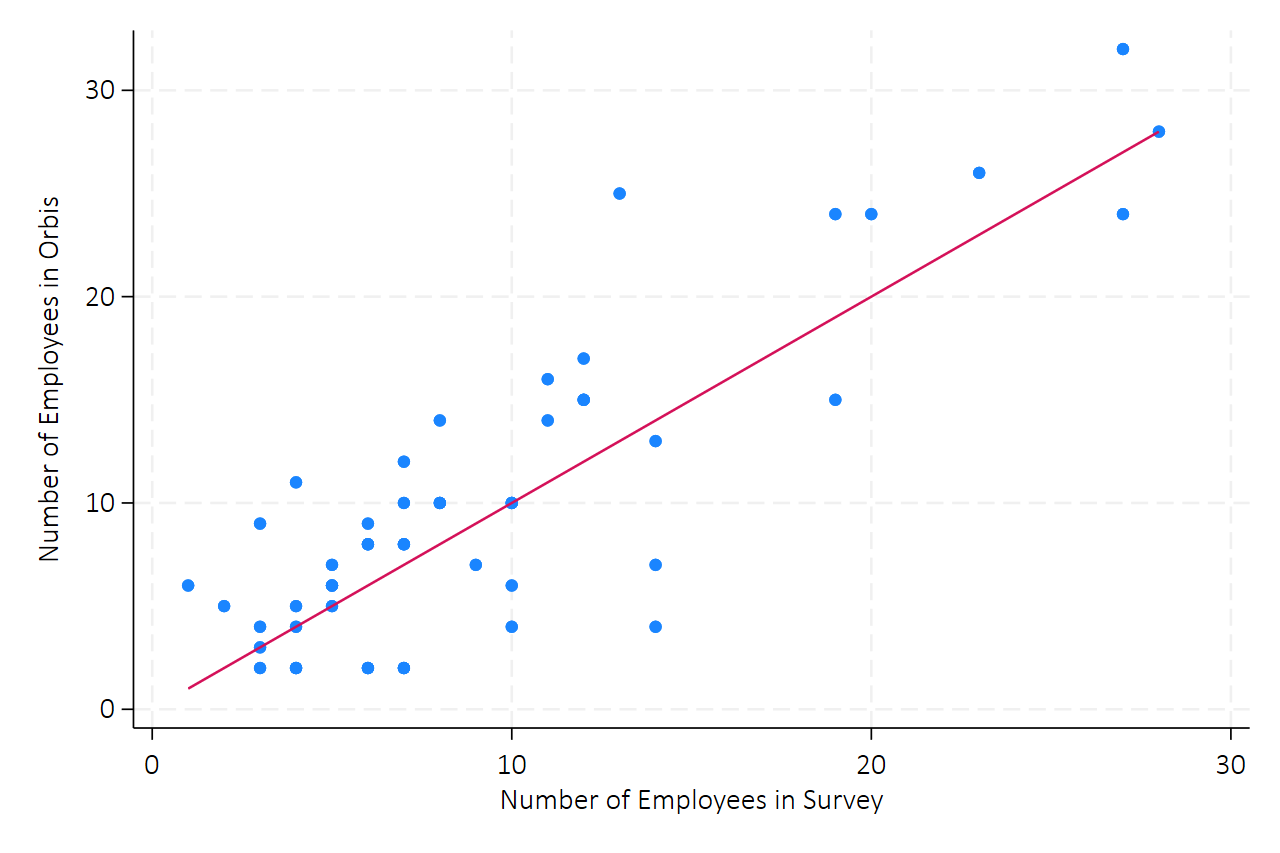} &
          \includegraphics[scale=0.17]{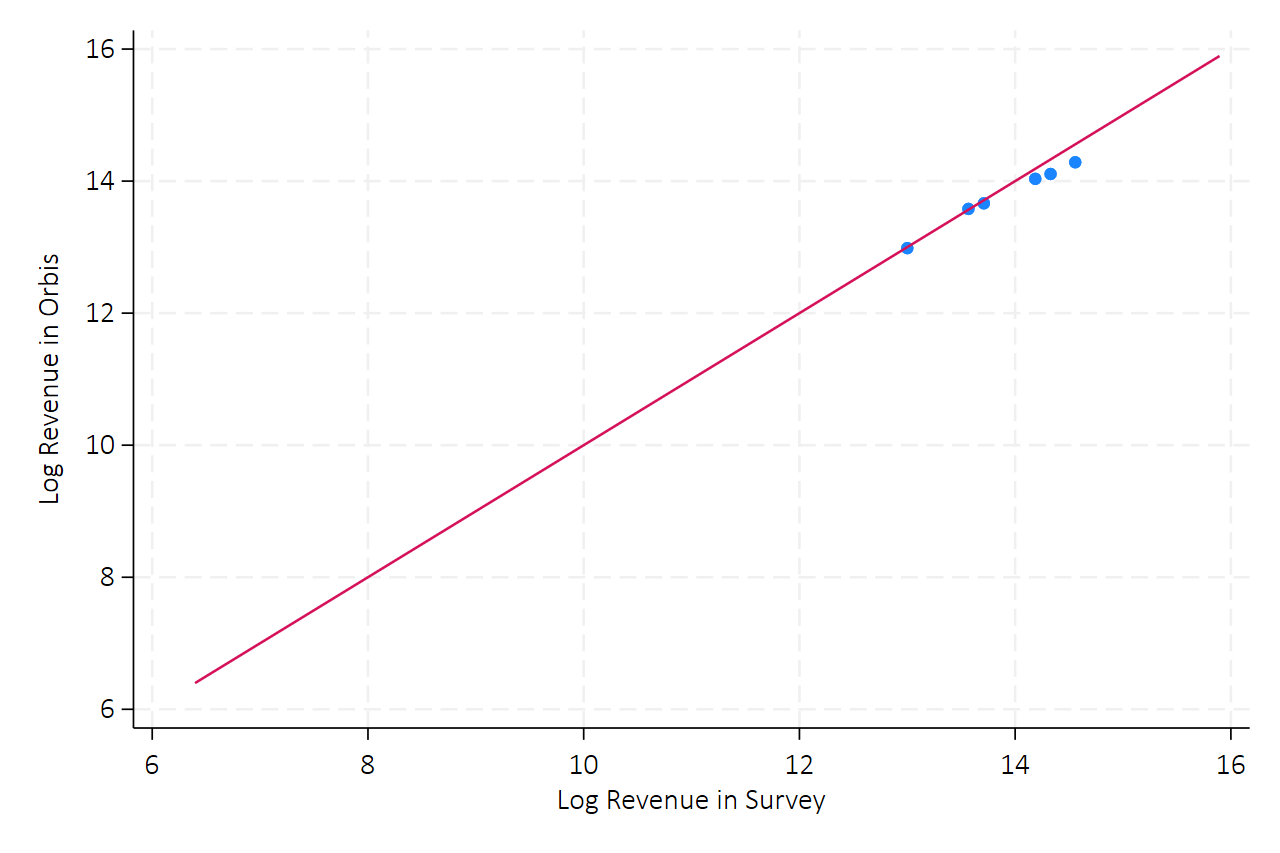} \\
        \end{tabular}
    \end{adjustbox}
    } 
    {Each point represents a firm included in both the survey and the Orbis database. 
    The left panel compares reported employment numbers, while the right panel compares logged revenues. 
    The red line indicates the fitted linear relationship between the survey and Orbis measures, highlighting their close correspondence across firm size dimensions.} 
    {Own calculations based on Orbis data and the German Business Panel Tax Advisor Survey 2025.} 
    {H} 

\newpage

\clearpage

\begin{samepage}

\section{Questionnaire}
\label{sec:quest}

\begin{table}[h!]

\centering\footnotesize
\renewcommand{\arraystretch}{1.5}
\setlength{\tabcolsep}{8pt}
\begin{threeparttable}
\caption{Relevant Questions from the GBP Tax Advisor Survey}
\label{tab:survey}
\begin{tabular}{m{.5cm}m{6cm}m{8cm}}
\toprule
Intro (all) & \multicolumn{2}{m{14cm}}{English translation of transcript of video with the president of the German Federal Chamber of Tax Advisors, Prof. Dr. Hartmut Schwab: Dear colleagues,
ChatGPT solves the tasks of a tax clerk exam in no time at all... and passes it, albeit narrowly.
So what does AI mean for our profession?
Will it soon make us unemployed, or is it the solution to the shortage of skilled workers in our firms?}\\\hline
\textbf{No.} & \textbf{Question} & \textbf{Answer Options} \\ \midrule
Q1 & What is your current employment status? & 
- Employed  \newline 
- Self-employed  \\\hline

Q2 & Which of the following positions best describes your role? & 
- Board Member/Executive Management  \newline 
- Senior Partner  \newline 
- Partner  \newline 
- Director  \newline 
- Senior Manager  \newline 
- Manager  \newline 
- Senior Consultant  \newline 
- Consultant, expert, analyst  \newline 
- student, intern   \\ \hline

Q3 & In your role as \texttt{selected role}: Do you have personnel responsibility? & 
- Yes \newline 
- No  \\ \hline

Q4 & In your professional role: How often do you work with AI-powered tools that generate text independently? For example, ChatGPT, Claude, etc. & 
- Always \newline 
- Often  \newline 
- Sometimes  \newline 
- Rarely  \newline 
- Never \\ \hline

Q5 & In your company: How many employees are working in the following professions? Please provide the number in full-time equivalents. & 
- Tax advisor \hfill [0,100000]\newline
- Auditor \hfill [0,100000] \newline 
- Certified tax assistant \hfill [0,100000] \newline 
- Tax clerk  \hfill [0,100000]  \\ \hline

Q6 & What do you estimate: How much of the core activities in the following professions can be automated by 2024? Please provide a percentage. & 
- Tax advisor \hfill [0,100]\newline
- Auditor \hfill [0,100] \newline 
- Certified tax assistant \hfill [0,100] \newline 
- Tax clerk  \hfill [0,100]  \\ \hline

Q7 & If you think again: What do you think now? Would you like to adjust your information?   & 
- Tax advisor \hfill [0,100]\newline
- Auditor \hfill [0,100] \newline 
- Certified tax assistant \hfill [0,100] \newline 
- Tax clerk  \hfill [0,100]  \\
\bottomrule
\end{tabular}
\end{threeparttable}
\end{table}
\end{samepage}

\begin{table}[h!]
\centering\footnotesize
\renewcommand{\arraystretch}{1.5}
\setlength{\tabcolsep}{8pt}
\begin{threeparttable}
\label{tab:survey2}
\begin{tabular}{m{.5cm}m{6cm}m{8cm}}
\toprule
\textbf{No.} & \textbf{Question} & \textbf{Answer Options} \\ \midrule

Q8 & From your company's perspective: Which of the following areas of responsibility have emerged due to automation in tax consulting? & 
- Quality control of automation results  \newline 
- Data protection and compliance monitoring  \newline 
- Prompt engineering  \newline 
- Application and support of legal tech/large language models (LLMs)  \newline 
- Other areas of responsibility  \\ \hline

Q9 & From your company's perspective: How many new employees do you plan to hire in the coming years? How many of them will be for new areas of responsibility created by automation? Note: Please indicate the number of new hires in each year in full-time equivalents. & 
\begin{tabular}[t]{p{3cm}lll}
& 2025  & 2026 & 2027  \\
Total new hires &&&\\
Of which employees for areas of responsibility created by automation\\
\end{tabular} \newline
 \\ \hline

Q10 & Regarding your personnel planning: How would you proceed if tasks could be replaced by automation? Please provide the number of affected staff. & 
\begin{tabular}[t]{p{3cm}lll}
& 2025  & 2026 & 2027  \\
Assign employees new tasks &&&\\
Dismiss employees\\
\end{tabular} \newline\\ \hline

Q11 & How do you perceive the changes in your profession due to automation? & 
- As a threat  \newline 
- As an opportunity for professional development \newline 
- Neither a threat nor an opportunity  \\ \hline

Q12 & Which profession would you most likely switch to? & 
- Public Accounting  \newline 
- Tax Consulting  \newline 
- Tax Technology Expert  \newline 
- Prompt Engineer  \newline 
- Data Scientist  \newline 
- No change \\ \hline

Q13 & Given the level of automation in your occupation, how likely is it that you would change occupation? Note: 0\% (no career change) - 100\% (career change)& \hfill [0,100]\\

\bottomrule
\end{tabular}
\end{threeparttable}
\end{table}

\begin{table}[h!]
\centering\footnotesize
\renewcommand{\arraystretch}{1.5}
\setlength{\tabcolsep}{8pt}
\begin{threeparttable}
\label{tab:survey3}
\begin{tabular}{m{.5cm}m{6cm}m{8cm}}
\toprule
\textbf{No.} & \textbf{Question} & \textbf{Answer Options} \\ \midrule
Q14 & How many new employees do you plan to hire in the following occupations in total by 2027?    Note: Please indicate the number of new employees in each year in full-time equivalents. &- Tax advisor \hfill [0,1000]\newline
- Auditor \hfill [0,1000] \newline 
- Certified tax assistant \hfill [0,1000] \newline 
- Tax clerk  \hfill [0,1000]  \\ \hline

Q15 & In which area do you see automation potential in your company? & 
- Business consulting \newline
- Financial accounting \newline
- International tax law \newline
- Payroll accounting \newline
- Succession planning \newline
-  Tax consulting \newline
- Tax declaration \newline
 \\ \hline

Q16 & There are now several new AI solutions for tax advisors. Would you like to learn more about examples of such AI solutions? & 
- Yes \newline
- No  \newline
 \\ \hline

Q17 & Have you ever heard of or actively used one of these AI solutions for tax advisors? 
- Taxy.io: A platform that develops AI solutions specifically for tax advisors. This tool analyzes tax questions and provides precise answers based on specialized literature.
 - DATEV LexInform AI Assistant (LEA): An AI solution that supports tax advisors in researching legal documents by providing relevant information and sources (e.g., UStAE, BMF letters).

& 
- Yes \newline
- No  \newline
 \\ \hline

Q18 &  How frequently do you use these AI solutions? & 
- Use them regularly \newline
- Use them irregularly \newline
- Do not use them \\
\bottomrule
\end{tabular}
\end{threeparttable}
\end{table}

\begin{table}[h!]
\centering\footnotesize
\renewcommand{\arraystretch}{1.5}
\setlength{\tabcolsep}{8pt}
\begin{threeparttable}
\label{tab:survey4}
\begin{tabular}{m{.5cm}m{6cm}m{8cm}}
\toprule
\textbf{No.} & \textbf{Question} & \textbf{Answer Options} \\ \midrule
Q19 & Do you plan to use AI solutions in the future? & 
- Yes \newline 
- No  \\
 \hline

Q20 & What increase in revenue per working hour do you expect for the following professions? Note: Please indicate the expected percentage change (positive or negative values). & 
- Tax Advisor \newline
- Auditor \newline
- Tax Clerk \newline
- Tax Assistant
 \\ \hline

Q21 & Compared to today: How does your company plan to adjust the average hourly wage for all employees in the next 12 months? Note: Please enter the change in per cent. You can enter positive or negative values. &
-  Change in hourly wage in per cent
 \\ \hline

Q22 & How much time do you plan to spend on your own digital training in an average week in the future? Note: Please enter the value in hours.
 \\ \hline

Q23 & Is your company planning investments or further training on automation topics for employees? 
&
- Yes \newline
- No 
\\ \hline

Q24 & What kind of investments or further training on automation topics is your company planning?
&
- General further training (e.g. part-time study) \newline
- Specialized further training (e.g. certified fibutronics) \newline
- Investments in hardware and software (e.g. ChatGPT, computer)  \newline
- Other
\\ \hline

Q25 & What do you estimate for your company? By what percentage will the following variables change through the use of AI? Note: Please enter a value in percent. You can enter positive or negative values.
&
- Profit Change \newline
- Revenue change \newline
- Cost change \newline
\\
\bottomrule
\end{tabular}
\end{threeparttable}
\end{table}

\begin{table}[h!]
\centering\footnotesize
\renewcommand{\arraystretch}{1.5}
\setlength{\tabcolsep}{8pt}
\begin{threeparttable}
\label{tab:survey5}
\begin{tabular}{m{.5cm}m{6cm}m{8cm}}
\toprule
\textbf{No.} & \textbf{Question} & \textbf{Answer Options} \\ \midrule
Q26 & When were you appointed as a tax advisor or auditor? Note: Please enter the year of your appointment.
&
- year of appointment \newline
- (not yet) appointed \newline
 \\ \hline
 
Q27 & How would you  like to be addressed in a greeting?
&
- Mr \newline
- Ms \newline
- Not specified \newline
 \\ \hline

Q28 & When were you born? Note: Please enter your year of birth.
 \\ \hline

Q29 & What is the legal form of your company?
&
- Sole proprietorship \newline
- GmbH \newline
- GmbH and Co. KG \newline
- UG \newline
- AG \newline
- oHG \newline
- GbR \newline
- PartG \newline
- KG \newline
- SE \newline
- Verein \newline
- KGaA \newline
- Genossenschaft \newline
- Public-law company \newline
- Other \newline
 \\  \hline
Q30 & Please enter the annual revenue (in EUR) of your company in the previous calendar year. Note: Please enter a whole number without using thousands or decimal separators.
\\ \hline

Q31 & If you could not or did not want to answer our question on revenue, do you think you could at least give us a range in which your revenue lies. Which of the following intervals most closely corresponds to your company's annual revenue in the previous calendar year? 
&
Intervals from less than 50,000 EUR to more than 60,000,000 EUR
 \\ \hline
Q32 & Do you have any comments or questions? Your opinion is important to us! \\
 \bottomrule
\end{tabular}
\begin{tablenotes}
  \footnotesize \item \emph{Note:} The full codebook of the GBP tax advisor survey is available from \href{https://gbpanel.org/}{https://gbpanel.org/}.
\end{tablenotes}
\end{threeparttable}
\end{table}

\clearpage
\section{Second-Stage Regressions}

\input{tab/tab_stage2_main}
\input{tab/tab_stage2_rev_by_occ}
\input{tab/tab_stage2_automation_potential}
\input{tab/tab_stage2_new_tasks}
\input{tab/tab_stage2_training}
\input{tab/tab_stage2_attitudes}
\input{tab/tab_stage2_actions}

\end{appendices}

\end{document}

%% file: gph/experimental_design.tex
\begin{figure}[H]
 \centering
  \caption{Experimental Design}
  \label{fig:Experimental_Design}
\vspace{5mm}
\begin{tikzpicture}[node distance=1.5cm,
    every node/.style={fill=white, font=\sffamily}, align=center]
 
  \node (start)             [activityStarts]              {Start of Survey};
  \node (Prior)     [process, below of=start, yshift=-0.5cm]          {
  \textbf{Prior Beliefs}:\\
  Estimate Share of Automatable Tasks for Different Occupations};
  
  \node (LOW)      [activityRuns, below of=Prior, yshift=-2cm, xshift=-2.5cm,fill=clow!70] {Lower-Skilled};
  \node (HIGH)      [activityRuns, right of=LOW, xshift=3cm,fill=chigh!70] {Higher-Skilled};
  \node (COMBINED)      [activityRuns, right of=HIGH, xshift=3cm,fill=ccombined!60] {Combined};
  \node (CONTROL)      [startstop, left of=LOW, xshift=-3cm,fill=bblue!70]{Control};

\node (Posterior)     [process, below of=LOW, xshift=2.2cm, yshift=-1cm] {
  \textbf{Posterior Beliefs}:\\
  Updated share of Automatable Tasks for Different Occupations};
  
  \node (Outcomes1)     [process, below of=Posterior, xshift=0cm, yshift=-1.5cm] {
  \textbf{Main Post-Treatment Outcomes}: \\
  Hiring and Firing Plans\\
  Cost, Revenue, and Profit Expectations \\
  Wage Expectations\\
  Automation Potential of Tasks\\
  Emergence of New Tasks};
  
  \node (Outcomes2)     [process, below of=Outcomes1, yshift=-2.cm] {
  \textbf{Further Outcomes}: \\
  Investment in Employee Training\\
  Attitudes towards Automation\\
  AI Adoption and Information Acquisition};
  
  \node (Debriefing)     [activityStarts, below of=Outcomes2, yshift=-1cm]          {End of Survey};
  
  \draw[->]             (start) -- (Prior);
  \draw[->]             (Posterior) -- (Outcomes1);
  
  \draw[<-]     (LOW)  -- node [xshift=-0.1cm, text width=3.1cm] {1/4} (Prior);
  \draw[<-]     (HIGH)  -- node [xshift=-0.1cm, text width=3.1cm] {1/4} (Prior);
  \draw[<-]     (CONTROL)  |- node [yshift=-2cm, text width=3.1cm] {1/4} (Prior);
  \draw[<-]     (COMBINED)  |- node [yshift=-2cm, text width=3.1cm] {1/4} (Prior);
  
  \draw[->]     (Outcomes2) -- (Debriefing);
  
  \end{tikzpicture}
  \vspace{5mm}
     \begin{minipage}{1\textwidth}
\scriptsize
{\it Notes:} The figure illustrates the design of our information treatment.
\end{minipage}
 
\end{figure}

%% file: tab/tab_first_stage_by_occ.tex
\begin{table}[ht]
    \centering
    \resizebox{0.9\textwidth}{!}{
    \begin{threeparttable}
    \caption{\centering Occupation-Specific OLS First-Stage Regressions\\ (Excluding Cross-Occupation Effects)}\label{tab:stage1_by_occ}
\begin{tabular}{lcccc}
\toprule
&(1)& (2)& (3)& (4)\\
  & \textbf{Tax}  & \textbf{Certified Tax}  & \textbf{Tax}  &  \\ 
  &  \textbf{Clerk} &   \textbf{Assistant} &  \textbf{Advisor} & \textbf{Auditor} \\ 
\cmidrule(lr){2-2}\cmidrule(lr){3-3}\cmidrule(lr){4-4}\cmidrule(lr){5-5}
Treatment High Skilled & 17.298*** & 12.784*** & 3.949* & 3.372* \\ 
 & (3.259) & (2.450) & (2.301) & (1.791) \\ 
Treatment Low Skilled & 8.701*** & 9.815*** & 8.633*** & 7.169*** \\ 
 & (2.834) & (2.313) & (2.245) & (1.860) \\ 
Treatment All & 17.098*** & 14.841*** & 7.311*** & 10.816*** \\ 
 & (3.017) & (2.338) & (2.163) & (1.882) \\[2ex] 
Prior & 0.886*** & 0.898*** & 0.771*** & 0.902*** \\[2ex]  
 & (0.032) & (0.039) & (0.081) & (0.025) \\ 
Treatment High Skilled $\times$ Prior & -0.113* & -0.074 & 0.048 & 0.011 \\ 
 & (0.057) & (0.058) & (0.106) & (0.040) \\ 
Treatment Low Skilled $\times$ Prior & -0.113** & -0.127** & -0.061 & -0.080* \\ 
 & (0.055) & (0.054) & (0.105) & (0.046) \\ 
Treatment All $\times$ Prior & -0.120** & -0.113** & 0.055 & -0.137*** \\ 
 & (0.054) & (0.053) & (0.099) & (0.042) \\[2ex]  
Constant & 7.481*** & 5.610*** & 6.058*** & 5.042*** \\ 
{} & {(1.580)} & {(1.341)} & {(1.643)} & {(1.130)} \\ 
\midrule
$R^2$ & 0.654 & 0.676 & 0.589 & 0.693 \\ 
N  & 1376 & 1302 & 1410 & 1202 \\ 
F-Statistic & 349 &    428 & 453 & 116 \\
\bottomrule
\end{tabular}
\begin{tablenotes}
  \footnotesize 
  \item \emph{Note:} This table reports the OLS first stage of our IV design, estimated separately for each occupational prior and posterior pair. The dependent variable is the respondent’s posterior belief about the automatability of the occupation named in the column, while the prior variable includes only the prior for this occupation. Treatment main effects capture level shifts in posteriors, while the interaction terms identify learning rates. Standard errors in parentheses. $^{*} p < 0.1$, $^{**} p < 0.05$, $^{***} p < 0.01$.
\end{tablenotes}
    \end{threeparttable}}
\end{table}

%% file: tab/tab_first_stage.tex
\begin{table}[ht]
    \centering
    \resizebox{0.77\textwidth}{!}{
    \begin{threeparttable}
    \caption{First-Stage Regression}\label{tab:first_stage}
\begin{tabular}{lcccc}
\toprule
&(1)& (2)& (3)& (4)\\
&\textbf{Tax}  & \textbf{Cert. Tax} & \textbf{Tax} & \\
&\textbf{Clerk}  & \textbf{Assistant} & \textbf{Advisor} & \textbf{Auditor}\\
\cmidrule(lr){2-2} \cmidrule(lr){3-3} \cmidrule(lr){4-4} \cmidrule(lr){5-5}
Treatment  High Skilled & 17.213*** & 11.036*** & 5.102** & 5.213** \\ 
 & (2.595) & (2.442) & (2.234) & (2.459) \\ 
 Treatment  Low Skilled & 5.206** & 4.523* & 3.774* & 5.104** \\ 
 & (2.607) & (2.461) & (2.247) & (2.480) \\ 
Treatment   All & 14.653*** & 12.154*** & 5.243** & 4.951** \\ 
 & (2.667) & (2.507) & (2.295) & (2.516) \\[2ex]
 Prior  Tax Advisor & -0.122** & -0.038 & 0.701*** & 0.174*** \\ 
 & (0.058) & (0.054) & (0.052) & (0.057) \\ 
Prior  Auditor & 0.102** & 0.069* & 0.146*** & 0.833*** \\ 
 & (0.043) & (0.041) & (0.038) & (0.041) \\ 
Prior  Cert. Tax Assist. & 0.047 & 0.910*** & 0.126** & 0.077 \\ 
 & (0.057) & (0.054) & (0.050) & (0.055) \\ 
Prior  Tax Clerk & 0.847*** & -0.049 & -0.095** & -0.081 \\ 
 & (0.055) & (0.051) & (0.047) & (0.053) \\[2ex]
 High $\times$ Prior Tax Advisor & 0.194** & 0.053 & -0.029 & -0.114 \\ 
 & (0.080) & (0.076) & (0.070) & (0.078) \\ 
Low $\times$  Prior Tax Advisor & 0.422*** & 0.313*** & -0.126* & 0.050 \\ 
 & (0.079) & (0.074) & (0.069) & (0.076) \\ 
All $\times$  Prior Tax Advisor & 0.201** & 0.038 & 0.141* & -0.117 \\ 
 & (0.086) & (0.081) & (0.075) & (0.083) \\[2ex]
High $\times$  Prior Auditor & -0.130* & -0.024 & -0.008 & 0.076 \\ 
 & (0.067) & (0.063) & (0.058) & (0.064) \\ 
Low $\times$  Prior Auditor & -0.232*** & -0.157** & -0.036 & -0.082 \\ 
 & (0.066) & (0.062) & (0.057) & (0.063) \\ 
All $\times$  Prior Auditor & -0.101* & -0.050 & -0.161*** & -0.203*** \\
 & (0.061) & (0.057) & (0.052) & (0.058) \\[2ex] 
High $\times$  Prior Cert. Tax Assist. & 0.055 & -0.160* & -0.066 & -0.099 \\ 
 & (0.093) & (0.090) & (0.080) & (0.091) \\ 
Low $\times$  Prior Cert. Tax Assist. & -0.162* & -0.307*** & -0.227*** & -0.237*** \\ 
 & (0.088) & (0.084) & (0.076) & (0.086) \\ 
All $\times$  Prior Cert. Tax Assist. & -0.016 & -0.186** & -0.103 & -0.009 \\
 & (0.089) & (0.084) & (0.077) & (0.085) \\[2ex] 
High $\times$  Prior Tax Clerk & -0.171** & 0.098 & 0.076 & 0.063 \\ 
 & (0.080) & (0.077) & (0.069) & (0.078) \\ 
Low $\times$  Prior Tax Clerk & -0.008 & 0.173** & 0.335*** & 0.202*** \\ 
 & (0.081) & (0.076) & (0.070) & (0.078) \\ 
All $\times$  Prior Tax Clerk & -0.115 & 0.114 & 0.166** & 0.222*** \\ 
 & (0.080) & (0.076) & (0.069) & (0.077) \\[2ex]

Constant & 8.088*** & 6.846*** & 4.141** & 3.526** \\ 
{} & {(1.886)} & {(1.772)} & {(1.620)} & {(1.775)} \\ 
\midrule
F-Statistic & 186.78 & 284 & 177.32 & 295.27 \\ 
N & 1202 & 1195 & 1200 & 1183 \\ 
$R^2$ & 0.68 & 0.69 & 0.65 & 0.71 \\ 
\bottomrule
\end{tabular}
    \begin{tablenotes}
    \footnotesize
        \item \textsc{Note:}  This table reports the OLS first stage of our IV design based on regression equation~\eqref{eq:belief_update} in which the dependent variable is the respondent’s posterior belief about the automatability of the occupation named in the column. Treatment main effects capture level shifts in posteriors, while the interaction terms identify learning rates; the large first‑-stage F‑-statistics reported at the bottom indicate strong instrument relevance. $^{*} p < 0.1$, $^{**} p < 0.05$, $^{***} p < 0.01$.
    \end{tablenotes}
    \end{threeparttable}}
\end{table}

%% file: tab/tab_stage2_main.tex
\begin{table}[h]
    \centering
    \resizebox{0.9\textwidth}{!}{
    \begin{threeparttable}
    \caption{Second-Stage: Firm-Level Employment and Revenue Outcomes}\label{tab:stage2_main}
\begin{tabular}{lcccccc}
\toprule
&(1)& (2)& (3)& (4)& (5)& (6)\\
& \textbf{Planned} & \textbf{Planned} & \textbf{Expected}& \textbf{Expected} & \textbf{Expected} & \textbf{Expected} \\ 
 &   \textbf{Firing} &   \textbf{Hiring} &   \textbf{Cost} &  \textbf{Revenue} &  \textbf{Profit}  &  \textbf{Wage}  \\ 
 &     \textbf{Share}&    \textbf{Share} & \textbf{Change} & \textbf{Change} & \textbf{Change}  & \textbf{Change} \\ 
\cmidrule(lr){2-2}\cmidrule(lr){3-3}\cmidrule(lr){4-4}\cmidrule(lr){5-5}\cmidrule(lr){6-6}\cmidrule(lr){7-7}
Pred. Posterior Tax advisors & 0.0125 & 0.0059 & 0.0045 & -0.0073 & 0.0031 & -0.0017 \\ 
 & (0.01440) & (0.00849) & (0.01115) & (0.01207) & (0.01181) & (0.00733) \\ 
Pred. Posterior Auditors & 0.0017 & 0.0121 & 0.0004 & 0.0335** & 0.0297*** & 0.0050 \\ 
 & (0.01243) & (0.00773) & (0.00962) & (0.01380) & (0.00942) & (0.00586) \\ 
Pred. Posterior Certified tax assistants & 0.0099 & 0.0048 & -0.0106 & -0.0179 & -0.0023 & 0.0038 \\ 
 & (0.01589) & (0.00793) & (0.01428) & (0.01985) & (0.01197) & (0.00859) \\ 
Pred. Posterior Tax clerks & -0.0237* & -0.0163** & -0.0008 & 0.0274 & 0.0182 & 0.0012 \\ 
 & (0.01436) & (0.00819) & (0.02154) & (0.02263) & (0.01154) & (0.00594) \\ 
Constant & 0.2263*** & 0.1338*** & 0.0171** & 0.1192*** & 0.1015*** & 0.0635*** \\ 
{} & {(0.00774)} & {(0.00485)} & {(0.00865)} & {(0.00774)} & {(0.00548)} & {(0.00202)} \\ 
\midrule
$R^2$ & 0.003 & 0.008 & 0.001 & 0.023 & 0.073 & 0.011 \\ 
N & 1043 & 1089 & 850 & 851 & 843 & 1285 \\ 
\bottomrule
\end{tabular}

\begin{tablenotes}
  \footnotesize 
  \item \emph{Note:}  This table reports second-stage results from an instrumental variables (IV) regression of equation~\eqref{eq:second_stage} estimating the effect of updated automation beliefs on respondents’ expectations about employment, wages, and financial firm outcomes. Each column uses the fitted posterior beliefs as the endogenous regressor, instrumented via the randomized information treatment. The dependent variables are drawn from questions on expected hiring and firing from 2025 to 2027, AI-induced cost, revenue, and profit changes, and expected wage changes. Coefficients are computed for the z-scores of predicted posterior beliefs. Standard errors are reported in parentheses.  $^{*} p < 0.1$, $^{**} p < 0.05$, $^{***} p < 0.01$.
\end{tablenotes}
    \end{threeparttable}}
\end{table}

%% file: tab/tab_stage2_rev_by_occ.tex
\begin{table}[h]
    \centering
    \resizebox{0.9\textwidth}{!}{
    \begin{threeparttable}
    \caption{Second-Stage: Occupation-Level Revenue per Hour}\label{tab:stage2_rev_by_occ}
\begin{tabular}{lcccc}
\toprule
&(1)& (2)& (3)& (4)\\
  & \textbf{Tax}  & \textbf{Certified Tax}  & \textbf{Tax}  &  \\ 
  &  \textbf{Clerk} &   \textbf{Assistant} &  \textbf{Advisor} & \textbf{Auditor} \\ 
\cmidrule(lr){2-2}\cmidrule(lr){3-3}\cmidrule(lr){4-4}\cmidrule(lr){5-5}
Pred. Posterior Tax advisors & -0.0523** & -0.0375** & 0.0086 & -0.0370 \\ 
 & (0.02362) & (0.01636) & (0.01132) & (0.02514) \\ 
Pred. Posterior Auditors & 0.1126** & 0.0698*** & 0.0383*** & 0.1144*** \\ 
 & (0.05148) & (0.02002) & (0.01106) & (0.03703) \\ 
Pred. Posterior Certified tax assistants & -0.0026 & 0.0547*** & 0.0113 & -0.0176 \\ 
 & (0.02827) & (0.01673) & (0.01140) & (0.02726) \\ 
Pred. Posterior Tax clerks & 0.0755*** & 0.0158 & 0.0092 & 0.0258 \\ 
 & (0.02321) & (0.01464) & (0.01078) & (0.02323) \\ 
Constant & 0.2682*** & 0.2135*** & 0.1555*** & 0.1360*** \\ 
{} & {(0.01363)} & {(0.00849)} & {(0.00555)} & {(0.00841)} \\ 
\midrule
$R^2$ & 0.085 & 0.12 & 0.11 & 0.115 \\ 
N & 964 & 956 & 968 & 933 \\ 
\bottomrule
\end{tabular}
\begin{tablenotes}
  \footnotesize 
  \item \emph{Note:} This table reports second-stage results from an instrumental variables (IV) regression estimating the effect of updated automation beliefs on respondents’ expectations about revenues per hour in specific occupations. Each column uses the fitted posterior beliefs as the endogenous regressor, instrumented via the randomized information treatment. Coefficients are computed for the z-scores of predicted posterior beliefs.  Standard errors are reported in parentheses.  $^{*} p < 0.1$, $^{**} p < 0.05$, $^{***} p < 0.01$.
\end{tablenotes}
    \end{threeparttable}}
\end{table}

%% file: tab/tab_stage2_automation_potential.tex
\begin{table}[ht]
    \centering
    \resizebox{0.9\textwidth}{!}{
    \begin{threeparttable}
    \caption{Second-Stage: Firm-Level Automation Potential}\label{tab:stage2_automation_potential}
\begin{tabular}{lccccc}
\toprule
&(1)& (2)& (3)& (4)& (5)\\
  & \textbf{Tax}  & \textbf{Payroll}  & \textbf{Tax}  & \textbf{Succession}  & \textbf{International}  \\ 
  & \textbf{Filing} & \textbf{Accounting} & \textbf{Consulting}  &  \textbf{Advisory} &  \textbf{Tax} \\ 
\cmidrule(lr){2-2}\cmidrule(lr){3-3}\cmidrule(lr){4-4}\cmidrule(lr){5-5}\cmidrule(lr){6-6}
Pred. Posterior Tax advisors & -0.0020 & 0.0156 & 0.0866*** & 0.0244 & 0.0232 \\ 
 & (0.02022) & (0.01746) & (0.02114) & (0.01552) & (0.01509) \\ 
Pred. Posterior Auditors & 0.0394** & 0.0107 & 0.0341* & 0.0241* & 0.0440*** \\ 
 & (0.01630) & (0.01470) & (0.01958) & (0.01424) & (0.01317) \\ 
Pred. Posterior Certified tax assistants & 0.0128 & -0.0008 & -0.0139 & -0.0072 & -0.0081 \\ 
 & (0.02353) & (0.02075) & (0.02372) & (0.01424) & (0.01511) \\ 
Pred. Posterior Tax clerks & 0.0471** & 0.0507*** & -0.0157 & 0.0072 & -0.0112 \\ 
 & (0.01996) & (0.01909) & (0.02061) & (0.01321) & (0.01429) \\ 
Constant & 0.7951*** & 0.8044*** & 0.2443*** & 0.0849*** & 0.0935*** \\ 
{} & {(0.01056)} & {(0.01047)} & {(0.01123)} & {(0.00736)} & {(0.00768)} \\ 
\midrule
$R^2$ & 0.047 & 0.031 & 0.048 & 0.026 & 0.031 \\ 
N & 1398 & 1398 & 1398 & 1398 & 1398 \\ 
\bottomrule
\end{tabular}
\begin{tablenotes}
  \footnotesize 
  \item \emph{Note:} This table reports second-stage results from an instrumental variables (IV) regression estimating the effect of updated automation beliefs on respondents’ expectations about the automation potential of specific tasks. Each column uses the fitted posterior beliefs as the endogenous regressor, instrumented via the randomized information treatment. Coefficients are computed for the z-scores of predicted posterior beliefs.  Standard errors are reported in parentheses.  $^{*} p < 0.1$, $^{**} p < 0.05$, $^{***} p < 0.01$.
\end{tablenotes}
    \end{threeparttable}}
\end{table}

%% file: tab/tab_stage2_new_tasks.tex
\begin{table}[ht]
    \centering
    \resizebox{0.9\textwidth}{!}{
    \begin{threeparttable}
    \caption{Second-Stage: Firm-Level New Tasks}\label{tab:stage2_new_tasks}
\begin{tabular}{lcccc}
\toprule
&(1)& (2)& (3)& (4)\\
& \textbf{Legal} & \textbf{} & \textbf{Prompt}& \textbf{Quality}  \\  
& \textbf{Tech} & \textbf{Compliance} & \textbf{Engineering}& \textbf{Assurance}  \\  
\cmidrule(lr){2-2} \cmidrule(lr){3-3} \cmidrule(lr){4-4} \cmidrule(lr){5-5} 
Pred. Posterior Tax advisors & -0.0031 & 0.0324 & 0.0545** & -0.0022 \\ 
 & (0.02261) & (0.02491) & (0.02385) & (0.01852) \\ 
Pred. Posterior Auditors & 0.0947*** & -0.0030 & 0.0354* & 0.0061 \\ 
 & (0.01954) & (0.02202) & (0.02082) & (0.01506) \\ 
Pred. Posterior Certified tax assistants & -0.0034 & -0.0124 & 0.0250 & 0.0359* \\ 
 & (0.02797) & (0.02911) & (0.02670) & (0.02095) \\ 
Pred. Posterior Tax clerks & -0.0016 & 0.0161 & -0.0243 & 0.0102 \\ 
 & (0.02464) & (0.02630) & (0.02334) & (0.01962) \\ 
Constant & 0.2838*** & 0.5608*** & 0.3184*** & 0.8584*** \\ 
{} & {(0.01226)} & {(0.01378)} & {(0.01269)} & {(0.00967)} \\ 
\midrule
$R^2$ & 0.039 & 0.004 & 0.035 & 0.019 \\ 
N & 1299 & 1299 & 1299 & 1299 \\ 
\bottomrule
\end{tabular}
\begin{tablenotes}
  \footnotesize 
  \item \emph{Note:} This table reports second-stage results from an instrumental variables (IV) regression estimating the effect of updated automation beliefs on respondents’ expectations about new  tasks being required due to automation. Each column uses the fitted posterior beliefs as the endogenous regressor, instrumented via the randomized information treatment. Coefficients are computed for the z-scores of predicted posterior beliefs.  Standard errors are reported in parentheses.  $^{*} p < 0.1$, $^{**} p < 0.05$, $^{***} p < 0.01$.
\end{tablenotes}
    \end{threeparttable}}
\end{table}

%% file: tab/tab_stage2_training.tex
\begin{table}[ht]
    \centering
    \resizebox{0.9\textwidth}{!}{
    \begin{threeparttable}
    \caption{Second-Stage: Firm-Level Training Investments}\label{tab:stage2_training}
\begin{tabular}{lcccc}
\toprule
&(1)& (2)& (3)& (4)\\
  & \textbf{Plans to} & \textbf{Plans} & \textbf{Plans} & \textbf{Plans}\\ 
 & \textbf{train employees} & \textbf{general training} & \textbf{specialized training} & \textbf{other training}\\ 
\cmidrule(lr){2-2}\cmidrule(lr){3-3}\cmidrule(lr){4-4}\cmidrule(lr){5-5}
Pred. Posterior Tax advisors & -0.0171 & 0.0029 & 0.0219 & -0.0195 \\ 
 & (0.02124) & (0.02883) & (0.02920) & (0.01641) \\ 
Pred. Posterior Auditors & 0.0094 & 0.0334 & -0.0128 & 0.0279** \\ 
 & (0.01684) & (0.02319) & (0.02390) & (0.01367) \\ 
Pred. Posterior Certified tax assistants & 0.0666*** & -0.0693** & 0.0319 & 0.0262* \\ 
 & (0.02486) & (0.03370) & (0.03411) & (0.01556) \\ 
Pred. Posterior Tax clerks & 0.0217 & 0.0504* & -0.0061 & -0.0249* \\ 
 & (0.02294) & (0.03016) & (0.03023) & (0.01399) \\ 
Constant & 0.7552*** & 0.3812*** & 0.4632*** & 0.0783*** \\ 
{} & {(0.01136)} & {(0.01509)} & {(0.01548)} & {(0.00831)} \\ 
\midrule
$R^2$ & 0.034 & 0.006 & 0.005 & 0.008 \\ 
N & 1388 & 1046 & 1046 & 1046 \\ 
\bottomrule
\end{tabular}

\begin{tablenotes}
  \footnotesize 
  \item \emph{Note:} This table reports second-stage results from an instrumental variables (IV) regression estimating the effect of updated automation beliefs on respondents’ expectations about intentions to train employees. Each column uses the fitted posterior beliefs as the endogenous regressor, instrumented via the randomized information treatment. Coefficients are computed for the z-scores of predicted posterior beliefs. Standard errors are reported in parentheses.  $^{*} p < 0.1$, $^{**} p < 0.05$, $^{***} p < 0.01$.
\end{tablenotes}
    \end{threeparttable}}
\end{table}

%% file: tab/tab_stage2_attitudes.tex
\begin{table}[ht]
    \centering
    \resizebox{0.9\textwidth}{!}{
    \begin{threeparttable}
    \caption{Second Stage: Attitudes}\label{tab:stage2_attitudes}
\begin{tabular}{lcc}
\toprule
&(1)& (2)\\
  & \textbf{Probability of}  & \textbf{Plans to adopt}  \\ 
& \textbf{changing occupation} & \textbf{AI solutions}  \\ 
\cmidrule(lr){2-2}\cmidrule(lr){3-3}
Pred. Posterior Tax advisors & 0.0393*** & 0.0193 \\ 
 & (0.01274) & (0.02143) \\ 
Pred. Posterior Auditors & -0.0036 & 0.0104 \\ 
 & (0.00802) & (0.01821) \\ 
Pred. Posterior Certified tax assistants & -0.0119 & 0.0312 \\ 
 & (0.01385) & (0.02574) \\ 
Pred. Posterior Tax clerks & -0.0003 & 0.0238 \\ 
 & (0.01095) & (0.02336) \\ 
Constant & 0.1048*** & 0.6776*** \\ 
{} & {(0.00468)} & {(0.01231)} \\ 

\midrule
$R^2$ & 0.026 & 0.027 \\ 
N & 1391 & 1408 \\ 
\bottomrule
\end{tabular}

\begin{tablenotes}
  \footnotesize 
  \item \emph{Note:} This table reports second-stage results from an instrumental variables (IV) regression estimating the effect of updated automation beliefs on respondents’ expectations about attitudes to job change and automation. Each column uses the fitted posterior beliefs as the endogenous regressor, instrumented via the randomized information treatment. Coefficients are computed for the z-scores of predicted posterior beliefs. Standard errors are reported in parentheses.  $^{*} p < 0.1$, $^{**} p < 0.05$, $^{***} p < 0.01$.
\end{tablenotes}
    \end{threeparttable}}
\end{table}

%% file: tab/tab_stage2_actions.tex
\begin{table}[ht]
    \centering
    \resizebox{0.9\textwidth}{!}{
    \begin{threeparttable}
    \caption{Second-Stage: Actions}\label{tab:stage2_action}
\begin{tabular}{lcccc}
\toprule
&(1)& (2)& (3)& (4)\\
  & \textbf{Plans to} & \textbf{Information} & \textbf{Hiring} & \textbf{Has}\\ 
 & \textbf{adopt AI} & \textbf{Seeking} & \textbf{Plans} & \textbf{Vacancy}\\ 
 & \textbf{solutions} & \textbf{on AI} & \textbf{(Survey)} & \textbf{posted}\\

\cmidrule(lr){2-2}\cmidrule(lr){3-3}\cmidrule(lr){4-4}\cmidrule(lr){5-5}
Pred. Posterior Tax advisors & 0.0244 & 0.0038 & 0.0059 & -0.0041 \\ 
 & (0.05102) & (0.02012) & (0.00849) & (0.00394) \\ 
Pred. Posterior Auditors & 0.0257 & 0.0168 & 0.0121 & 0.0004 \\ 
 & (0.04025) & (0.01703) & (0.00773) & (0.00542) \\ 
Pred. Posterior Certified tax assistants & 0.0009 & -0.0135 & 0.0048 & 0.0046 \\ 
 & (0.05631) & (0.02307) & (0.00793) & (0.00447) \\ 
Pred. Posterior Tax clerks & 0.0900* & 0.0351 & -0.0163** & 0.0011 \\ 
 & (0.04889) & (0.02156) & (0.00819) & (0.00384) \\ 
Constant & 0.6804*** & 0.7837*** & 0.1338*** & 0.0107*** \\ 
{} & {(0.02613)} & {(0.01095)} & {(0.00485)} & {(0.00274)} \\ 
\midrule
$R^2$ & 0.08 & 0.009 & 0.008 & 0.001 \\ 
N & 301 & 1405 & 1089 & 1408 \\ 
\bottomrule
\end{tabular}

\begin{tablenotes}
  \footnotesize 
  \item \emph{Note:} This table reports second-stage results from an instrumental variables (IV) regression estimating the effect of updated automation beliefs on respondents’ AI adoption plans and actions. Each column uses the fitted posterior beliefs as the endogenous regressor, instrumented via the randomized information treatment. Coefficients are computed for the z-scores of predicted posterior beliefs. Standard errors are reported in parentheses.  $^{*} p < 0.1$, $^{**} p < 0.05$, $^{***} p < 0.01$.
\end{tablenotes}
    \end{threeparttable}}
\end{table}